# Current and Future Applications of PVDF-Carbon Nanomaterials in Energy and Sensing

Joanna Kujawa[1)*], Sławomir Boncel[2,3)], Samer Al-Gharabli[4)], Stanisław Koter[1)], Anna Kaczmarek – Kędziera[5)], Emil Korczeniewski[6)], Artur P. Terzyk[6)*]

*1) Faculty of Chemistry, Department of Physical Chemistry and Physical Chemistry of Polymers, Nicolaus Copernicus University, Gagarin Street 7, 87-100 Toruń, Poland*

*2) Faculty of Chemistry, Department of Organic Chemistry, Bioorganic Chemistry and Biotechnology, NanoCarbon Group, Silesian University of Technology, Krzywoustego Street 4, 44-100 Gliwice, Poland*

*3) Centre for Organic and Nanohybrid Electronics (CONE), Silesian University of Technology, Konarskiego 22B, 44-100 Gliwice, Poland*

*4) School of Applied Medical Sciences, Pharmaceutical and Chemical Engineering Department, German Jordanian University, Amman Madaba Street, Amman 11180, Jordan*

*5) Faculty of Chemistry, Department of Materials Chemistry, Adsorption and Catalysis, Nicolaus Copernicus University, Gagarin Street 7, 87-100 Toruń, Poland*

*6) Faculty of Chemistry, Physicochemistry of Carbon Materials Research Group, Nicolaus Copernicus University in Toruń, Gagarin Street 7, 87-100 Toruń, Poland*

## Abstract

The review unveils the diverse applications of concerted polyvinylidene fluoride (PVDF)-carbon nanomaterial (CNM) systems, spanning from electromagnetic interference shielding, including elimination of 5G-interference, to piezoelectrics and a variety of sensing modalities (breathing, movement, health monitoring, structural integrity assessments, home monitoring, and seismic acceleration). These materials also excel in biomaterials with applications like tactile skin and COVID-preventing facemasks through sunlight sterilization. Moreover, PVDF-CNMs demonstrate excellence in radar absorption, solar-assisted electricity generation, triboelectric energy harvesting, 3D-4D printing materials, anti-icing covers, anti-stealth materials, and heat-dissipating solids in electronics. Across diverse scientific disciplines, the research merges materials chemistry and engineering, yielding materials with multimodal functionalities. The demand for a comprehensive review stems from the need to synthesize insights from fundamental sciences and technologies, capturing the cutting-edge nature of these materials. The scientific goals revolve around elucidating the link between PVDF and CNMs' structural attributes and physico-chemical properties. Two key objectives guide this exploration: (a) shedding light on the conversion from PVDF α- to β-phase toward its applicability in EMI shielding, piezoelectrics, sensors, and energy harvesters, and (b) highlighting the simplicity in generating PVDF-CNMs, presenting a vast potential for tuning material features like hydrophilicity, mechanical properties, piezoelectric characteristics, catalytic activity, and bioactivity. This pursuit of scientific excellence indicates new avenues, underscoring significance of the ongoing research and inviting the scientific community to explore uncharted territories, fostering a continuous environment of discovery and innovation in the dynamic landscape of PVDF-CNMs.

## 1. Introduction

Some people say that “carbon is an old but still new material”. This opinion is caused by the discovery of new allotropes (and nanoallotropes), and among them, graphene and graphene oxide (GO) have been most widely used recently. Thus, in one of the latest studies [1], we provided an in-depth introduction to the role of carbon and CNMs in chemistry along with the comprehensive characterization of these materials. Carbon atom hybridization was related to the chemical and physical properties of 0, 1, 2, and 3D CNMs. Next, the synergy between CNMs and PVDF in membrane science was discussed.

In this study we extend discussion about recent applications (2018 – 2023) of PVDF-CNMs systems focusing on energy – related usage. At the beginning , we explain why we need the concerting approach to create advanced materials. Since CNMs play  an important essential role in the PVDF phase conversion, we discuss how they change PVDF properties. The special attention is paid to the α-to-β conversion since the presence of β-phase leads to the more prospective  energy-related applications. Next, we discuss the basics phenomena in which the application of CNMs improves the properties of PVDF. In the subsequent chaptersdevelopmentsparticular, we discuss recent developments in the application of PVDF – CNMs, paying a special attention to electromagnetic interference (EMI) and electrostatic discharge (ESD) shielding, piezoelectric, energy harvesting, triboelectric nanogenerator (TENG), sensors, actuators, transducers, and photodetectors to EMI and ESD shielding, piezoelectric, energy harvesting, TENG, sensors, actuators, transducers, and photodetectors. Finally, we discuss recent results on PVDF-CNM systems modelling, summarizing, concluding, and indicating the most prospective opportunities and the outlook.The motivation to presente the gather data in this review was a fact that PVDF as well as CNMs are collecting more and more attention from researchers, engineers, and scientists working in medical sciences. The PVDF is highly resistant (chemically, thermally, and mechanically), biocompatible, and stable under radiation. Although this is a fluoropolymer, this material is broadly applied in sucha a broad spectrum of application including not only the energy and sensing are but also medicine and medical sciences, particularly cardiology, owing to its bio and hemocompatibility [2].

## 2. Improving fundamental properties and phase conversion in PVDF exhibiting "magical" roles of CMNs

### 2.1. Enhancing fundamental features

PVDF, with the following name approved by IUPAC, i.e., poly(1,1-difluoroethylene), which belong to thermoplastic fluoropolymer is widely used owing to its unique features. The chemical structure, characterized by the repeating unit ($CH_2$-$CF_2$), dictates the final material's properties. PVDF can be tailored to a wide range of applications and characteristics, resulting in the production of either hydrophilic or hydrophobic piezoelectric materials [1].

The most significant physical feature of PVDF is its significant intertness. The simple chemical composition of -$(CH_2CF_2)_n$- monomer unit makes the PVDF tough and strengthen. PVDF materials are characterized by their extraordinary mechanical stability, *i.e.*, high strength, flexibility, high chemical, UV and thermal resistane, as well ashigh dielectric constant (5.6). The PVDF possesses few forms, *i.e.*, alfa (α), beta (β), gamma (γ), delta (δ), and epsilon (ε) which can be transformed between each other [3]. The formation of chosen form can be accomplished via selection of suitable experimental protocol of material preparation or modification. Such unique properties opens even more posibilities for more unique and advanced utilization [4-9].

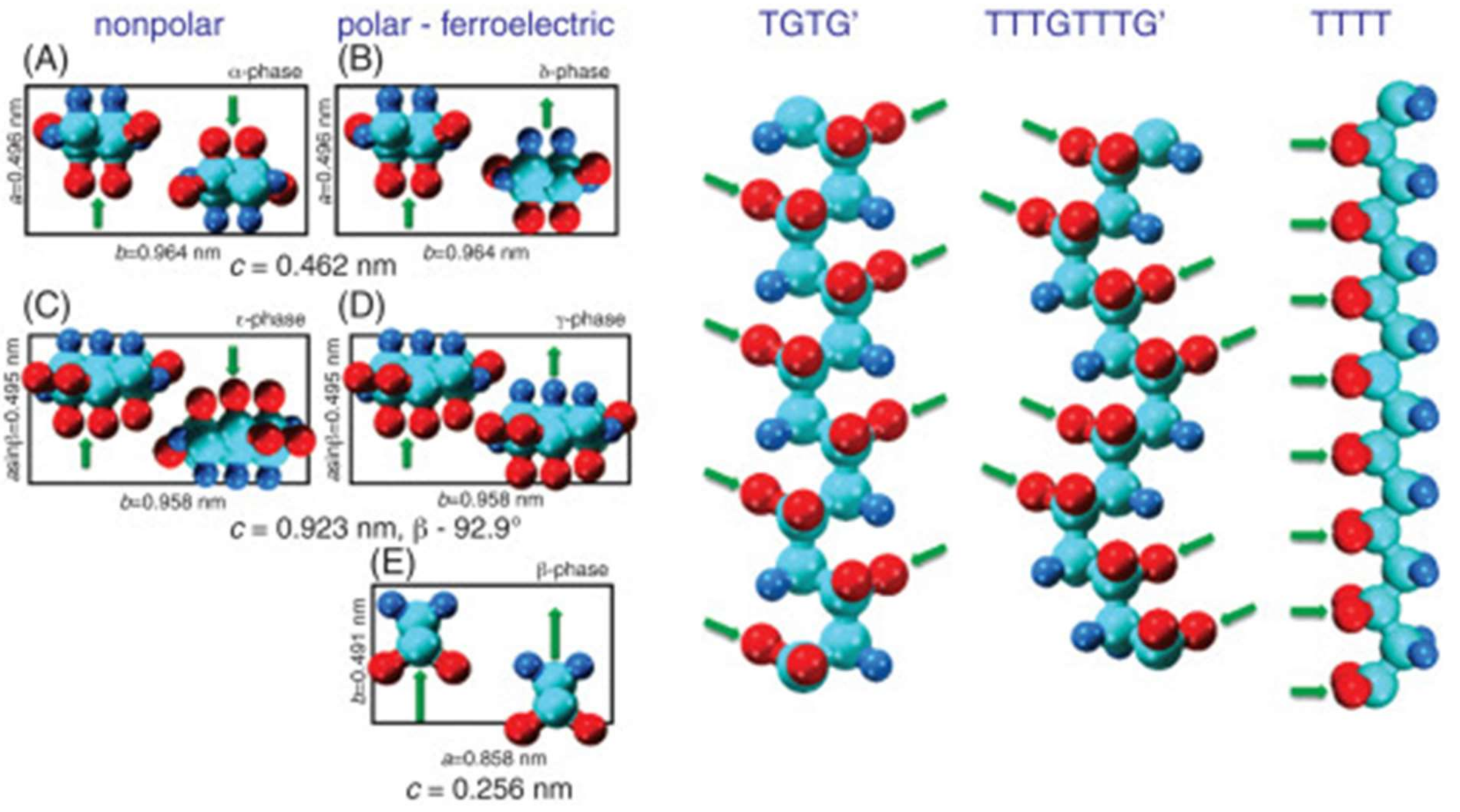

**Figure 1**. The PVDF forms. Adapted with permission from [10]. Copyright 2012 American Chemical Society.

As shown, CNMs are applied for tuning the properties of PVDF. Among them, the most important ones are mechanical, thermal, electric, surface (*e.g.*, roughness, wettability), magnetic, dielectric and wetting [11]. Generally, PVDF-based films are not fully transparent, and this causes a severe problem for the optical applications. The simplest method to make them transparent is the application of ultrasounds and vacuum drying, causing nucleation of PVDF in the electroactive, transparent γ-phase [12]. What is important that CNMs induce the transformation of phases allowing to control because CNMs induce the transformation of phases, allowing control of the conversion process. Blending of PVDF, functionalised CNTs and polyacrylonitrile, with oxygen plasma treatment [13], allow to create hydrophilic centers and, in this way, to control wetting properties. A separate group of materials with improved properties are so called metastable intermixed composites [14] storing large amounts of energy being released violently. Here, GO-doped PVDF/CuO/Al nanocomposites are a good example. By the optimization of parameters of electrospinning, it was possible to improve the density and anti-oxidation properties [14].

## 2.2. Tuning PVDF mechanical properties and improving nanomaterials dispersibility

Due to the potential applications of PVDF, also crucial is the improvement of mechanical PVDF properties. In this case, CNMs are applied, usually graphene (G) and carbon nanotubes (CNTs) (short review on G application as a PVDF nanofiller was recently published by Trivedi and Rachchh [15]). The enhancement of mechanical properties of PVDF-CNM composites stems from the intrinsically high performance of individualized, highly oriented/anisotropic systems of 1D- and 2D-carbon nanoparticles (CNTs and G), while the interface bonding remains challenging. Due to van der Waals forces, CNMs can agglomerate (improved dispersibility assures, for example, the formation of 3D-conductive paths [16]). Recently, Ma et al. [17] reviewed physical and chemical methods of improvement of nanofiller dispersion. Among physical (typically less expensive than the chemical ones) methods, the most widely applied are: ultrasonication, milling, and blending. At the same time, among the chemical methods one can mention CNM surface modifications: grafting, coating, and polymerization. As an example, one can consider an interesting procedure proposed by Yang et al. [18] who by decorating of CNFs with $SiO_2$, introducing this material into PVDF matrix, highly improved CNFs dispersibility (an increase in dielectric constant was simultaneously observed). In this case $SiO_2$@CNFs prevented agglomeration. $SiO_2$ played a similar role (preventing the GO nanosheets from aggregation) in the PVDF/$SiO_2$/GO tri-layered nanocomposites [19]. Ma et al. [17] proposed foaming, using supercritical $CO_2$, to disperse CNFs in PVDF matrix. Among a variety of CNMs, G nanoflakes (and nanosheets [20]) can be applied to improve PVDF mechanical properties and to increase the melting temperature [21]. Those characteristics derive from the enhanced crystallinity of the matrix induced by supramolecular templating  shear stress and quenching of oxygen and thermally generated polymer-based radicals by CNMs. Addition of G also leads to decrease in air permeability, and while this behaviour can be important in EMI shielding [20]. Ca. 1 wt.% of G addition drastically improves Young's Modulus (YM) upon stretching, hardness and yield stress. Also nanodiamonds (NDs) [22, 23] can be applied to improve fundamental PVDF properties. Kausar [23] reported the improvement of mechanical properties (MP) for PVDF/poly(3,4-ethylenedioxythiophene, PEDT) blend after the ND addition. NDs were carboxylated and modified with polyaniline. At the same time, the higher degradation and decomposition temperatures were observed, and the increase of critical conductivity. MP can be also improved by addition of CNFs [24], rGO and expanded G [25-27]. In this case, small amounts (2-4 wt.%) of the above admixed nanoparticles increase YM, and, at the same time, the enhancement in thermal and electrical conductivity, hydrophilicity, and roughness are observed. Friction coefficient of PVDF can decrease after dispersion of carbon nanofiber and nanotubes [28]. Sodagar et al. [22] analysed the improvement of mechanical properties using simulation (*via* a finite element model). Park et al. reported application of $TiO_2$ and carbon black [29] to improve mechanical features of PVDF. The decrease in PVDF friction coefficient and the improvement of wear resistance were additionally reported. What is important, generally, the enhancement in tribological properties can be achieved by the higher contents of the PVDF α phase [29], and this can be induced by the presence of mesoporous $TiO_2$ and carbon black (CB) of a high surface area [29]. As it was proved by Lee et al. [30], the same effect can be achieved by the application of carbon nanorods (CNRs), which increase the contents of α phase, responsible for the friction and interfacial adhesion of the PVDF-CNR composites. Quantum dots (QDs) can be applied to improve mechanical properties of PVDF-caesium lead perovskites (for example $CsPbBr_3$) displaying poor mechanical stability [31]. PVDF was combined with hexafluoropropylene (HFP) (PVDF-HFP) to improve elasticity [32].

### 2.3. Tuning PVDF thermal and electrical properties

Highly thermally conducting PVDF containing composites can be obtained after a simultaneous addition of CNTs and BN [33, 34]. In this case, BN disables aggregation of CNTs. A similar role was played by $MnO_2$ nanowires in PVDF/CNT/ $MnO_2$ composites of an improved electrical conductivity [35]. The comparison of different nanotubes leads to the conclusion about augmented electrical properties of PVDF after addition of branched CNTs [36]. This phenomenon was explained by the highest ability of branched CNTs to create connections as a 3D-conductive network. The enhanced electrical conductivity was obtained for modified poly(methyl methacrylate) and epoxy groups containing CNTs applied as a compatibilizer for two immiscible PVDF/poly(L-lactide) blends [37] (simultaneously, thermal and mechanical properties were improved). Application of ionic liquid (IL-)modified CNT/PVDF composites in fibers assures high conductivity, stretchability and, hence, plausible application in the electronic textiles. Also, addition of di-glycidyl ether of bisphenol-A (DGEBA)-grafted MWCNTs enhanced PVDF electrical conductivity and piezoelectric coefficient [38]. Recent studies [39] proved that application of G led to an improved graphitization of pitch-based carbon in the structure of PVDF, and in this way, an overall improvement in the electrical conductivity.

Addition of CNTs together with poly(methyl methacrylate) can change the ability of PVDF to crystallize [40], while addition of GO and magnetic $Fe_3O_4$ [41] improves thermal stability, mechanical, electric and magnetic properties, providing promising material for magneto applications. Improving thermal conductivity is crucial for the dissipation of heat in the electronic devices [42]. The simplest CNM applied here is conductive CB [43] (no more than 10 wt.%). Also addition of G can leads to tunable electric conductivity [20], reported also (together with improvement of dielectric constant) after addition of carbon coated barium titanate (BT) hybrid particles (BT@C), and thermal conductive silicon carbide nanoparticles (SiC NPs).

Krause et al. [44] presented the results of very extensive research on different polymers (among them PVDF)/ CNT mixed composites. The major purpose was to optimize the Seebeck coefficient and electrical conductivity. It was concluded that for PVDF/MWCNT materials the highest Seebeck coefficients were recorded.

### 2.4. Tuning PVDF dielectric properties

Considering application of polymers in the electronic devices, it is important to obtain the materials displayingthe smallest dielectric losses . Wu et al. [45] reported the application of PEG-grafted GO as a surface layer on PVDF and the obtained material revealed, at the same time, small dielectric losses and high thermal conductivity, which is a rare achievement. Ruiz et al. [46] proposed the application of solution blow spinning method for the preparation of sandwich (layered) PVDF/MWCNTs composites possessing high dielectric permittivity and low dielectric loss. Also, addition of IL to CNTs improves the dielectric constant of PVDF [47]. Dielectric permittivity was likewise drastically increased after adding modified MWCNTs [48] while epoxy modification was applied to increase the dispersibility of MWCNTs in the PVDF matrix. The obtained materials displayed improved thermal and dielectric properties. The method of preparation of PVDF-based dielectric composites, having excellent ductility and high elongation at break, was proposed [49], based on a spinodal phase separation process. Similarly, the increase in dielectric properties was observed during 3D printing process of MWCNTs/$BaTiO_3$/PVDF nanocomposites. Application of 3D printing eliminates cracks, voids, and made CNMs distribution more homogeneous [50]. Dielectric permittivity increase, caused by the more pronounced formation of the β-phase, was also observed after addition of carboxylated multi-walled carbon nanohorns (MWCNH) [51] to PVDF. A similar effect was observed for the polypyrrole-functionalized MWCNTs [52]. Also, CNT/PVDF infiltrated with epoxy resin provided high dielectric constant composites at a relatively low percolation threshold [53]. Mao et al. [54] indicated the importance of crystallization conditions of CNTs containing PVDF/poly(butylene succinate) blends during the manufacture of microcapacitor networks, influencing dielectric constant and decreasing dielectric loss. Recently Zhu et al. [55] and Xia et al. [56] reported the polydopamine @CNT/PVDF composites with an ultralow dielectric loss. Considering CNTs, the highest dielectric permittivity and electrical conductivity are obtained using high aspect ratio CNTs and low-viscosity PVDF [57]. Application of the high aspect ratio CNTs minimizes the contact resistance effects and, in parallel, constitutes an economic approach since the volume fraction of long CNTs can be significantly reduced. Additionally, longer CNTs lead to lower viscosity of the blends due to shear thinning guaranteed by alignment [58].

### 2.5. Phase conversion in PVDF exhibiting "magical" roles of CMNs

One can conclude that the introduction of CNMs onto the surface or into a bulk PVDF triggers "magical" roles of CMNs to be manifested. Here, the phase of PVDF is also crucial. Although the conversion of α- to β-form of PVDF is often reported, the detailed mechanism of this extremely important process still remains unclear. This conversion is highly required – especially in applications of PVDF as a component of the EMI shielding materials and pyro/piezogenerators. Moreover, since the β-phase has a dipole moment, the presence of this phase increases hydrophilicity, and reduces negative permittivity [59] what is crucial, for example, in the separation materials, sensors [1] and optical devices applications (see below).

Usually the degree of phase conversion is calculated from the data obtained by FTIR, XRD [60-68] and, rarely, from Raman [69] spectroscopy data, by using derived experimentally equations. Cai et al. [70] reviewed the FTIR procedures and corrected the widely applied approach, showing that the bands at 763 and/or 614, 1275 and 1234 $cm^{-1}$ should be applied for the identification of α, β and γ PVDF phases, respectively. Considering XRD, a peak at $2\theta = 20.44^{\circ}$ should be analysed. Usually [63], but not always [65], the agreement between FTIR and XRD is observed. Generally different CNMs and fillers can increase the content of β-phase [62, 64, 71-73], and it is accepted that crystallinity of the β-phase depends on the type of CNM and its content (the conversion vanishes at low CNMs contents [22, 74] and decreases at higher CNMs concentrations, due to the aggregation effects). Aggregation and agglomeration of CNMs is often reported in literature in different processes. For example, during the mechanochemical milling of G its surface area rapidly increases. However, after this fast increase, a decrease is observed caused by re-agglomeration [75]. Thus, the amount of studied CNMs is crucial. For instance, CQDs [76] stimulate the formation of the β-phase but at the concentration of ca. 5 wt.%, CQDs aggregate [69, 77]. A similar effect occurs for the application of electrochemically exfoliated G [78] above ca. 1 wt.%. There are also reports showing that the addition of too large amounts of nanofillers can decrease the conversion by disturbing the crystal growth process and the crystallinity, as it was concluded by Ghajar et al. [79] who studied the influence of ionic liquid (IL), G and SWCNT on crystallinity and the formation of the β-phase. It was shown that addition of SWCNTs to the PVDF/IL mixture led to a reduction of β-phase amount, as well as its crystallinity. This is caused by the disturbance of the crystal growth process. In contrast, the presence of IL alone in PVDF, increases the β-phase amount [80].

### 2.5.1. Recent progress in the understating of the α-to-β conversion

Mechanistic aspects of the α-to-β conversion are rarely met in literature. In this field, an interesting paper was recently published by Viswanath et al. [81] The authors paid attention to the similarity between the lattice constant of graphene and distance between fluorine atoms in the β phase of PVDF. Similar aspect was discussed by Ramanujam et al. [82] who, based on the lattice mismatch theory [83], discussed the process of PVDF growth on carbon fiber by the calculation of the lattice mismatch factor. This factor determines the process of a polymer nucleation on CNFs, and, if it is smaller than ca. 10%, the process of conversion takes place. Here the question arises since both PVDF phases show a good fit to the G surface (see Fig. 2). It should be also emphasized that a relatively small number of reports exists [59, 84, 85] pointing out the crucial role of PVDF physical adsorption on a nanomaterial during the conversion (there are discussions on interactions, but not in the context of adsorption process and, for example, its temperature dependence). The results of AFM experimental studies clearly indicated that a thin film of the β-phase, created on a GO, forms a nano-lamellae [86]. Also, the temperature of conversion is crucial, due to the changes in mobility of polymer chains [87]. The influence of parameters of a casting process (also influenced by salt solution and drying) on transformation was also studied [88].

Considering a mechanism of the β-phase formation, Sahoo et al. [60] showed that PVDF chains of α phase created spherules, and CNMs played the role of ’destructors‘ facilitating the phase transformation (see Scheme 3 in [60] and Fig. 2).

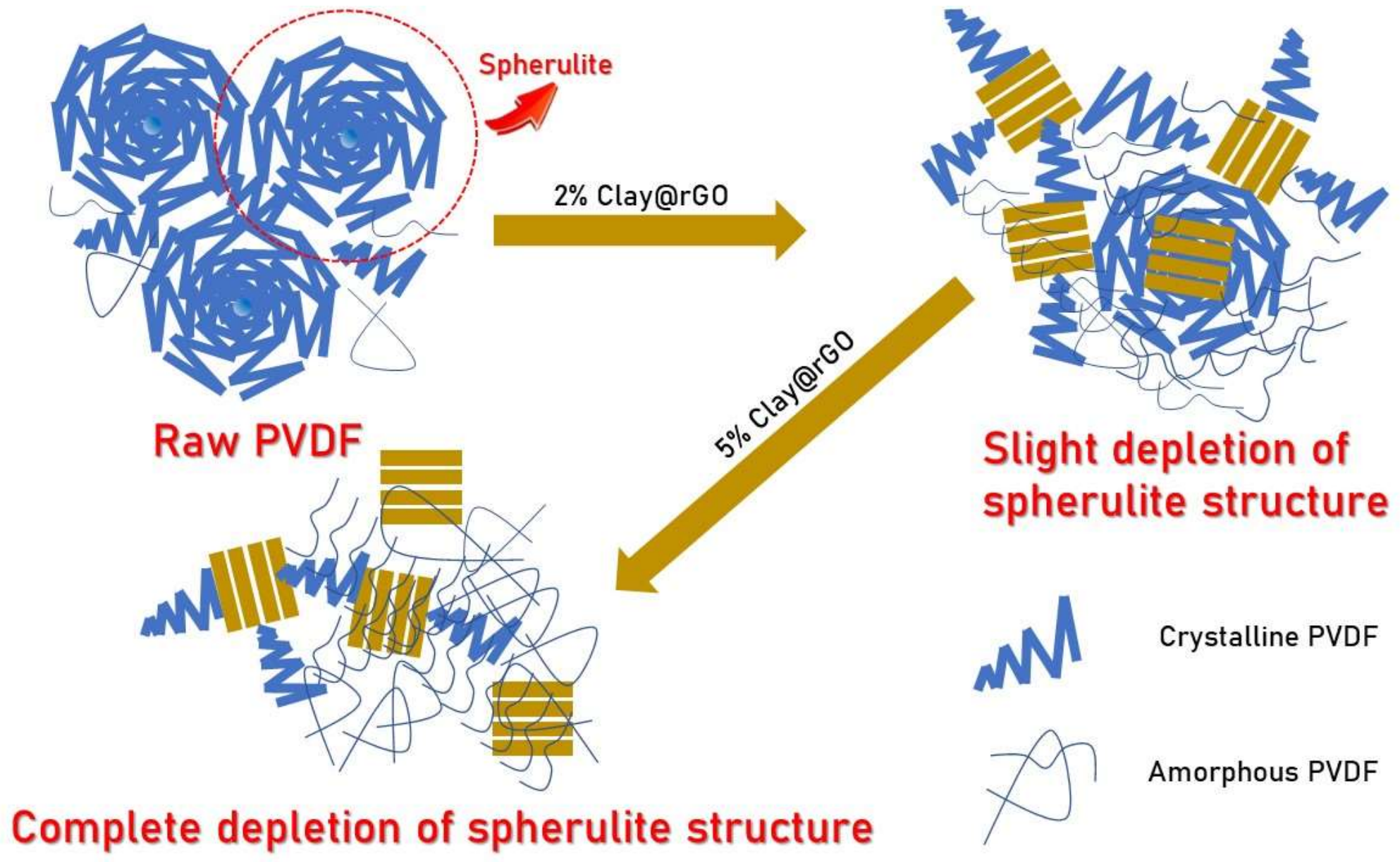

**Figure 2.** Destructive role of clay@rGO in the morphology of PVDF. Adapted with permission from [60] Copyright 2018 Elsevier.

The question arrives now on the nature of CNM-PVDF interactions, since the experimental studies on the interactions between ultrathin PVDF films and CNMs indicate that their nature still remains unknown [86]. In the majority of studies, it was claimed that electrostatic interactions [89, 90] between dipoles were the driving force of the conversion process. However, they can be of a different origin. For example, Choi et al. [69] considered interactions between negatively charged carbon QDs and $CH_2$ groups of PVDF (see Fig. 2 in [69]), and Widakdo et al. [91] interactions between π electrons in G and the cations of a IL (similar interactions were considered by Xu et al. [92]) together with the G-IL π-π interactions. Xu et al. [92] studied the IL-modified GO, concluding about the leading role of ions (forming IL)-$CF_2$ dipole interactions, and the π electrons of GO-ions interactions. Similar conclusion appeared in the study of Wang et al. [47] showing that addition of IL to MWCNTs (and the creation of PVDF/(P[MMA - IL])/MWCNTs system) promotes the formation of β-phase *via* ionic interactions. Hu et al. [93] considered addition of rGO to PVDF-trifluoroethylene (TrFE), and postulated that the improved formation of β-phase was caused by the interactions between polar oxygen groups of rGO and PVDF-TrFE fluorine atoms. They also considered the participation of polarization in the transformation process. Also, Ahmed et al. [62] reported β-phase creation during the study between (PVDF-TrFE-based rGO) and MWCNTs nanofibers obtained by electrospinning process. Begum et al. [48] presented the results of MWCNT oxidation, and then grafting with di-glycidyl ether of bisphenol-A by a linker of hexamethylene diamine. It was concluded that the epoxy from di-glycidyl ether of bisphenol-A moieties interacts electrostatically with the -$CF_2$ group, and plays the role of a nucleation agent helping the β-phase formation. . Also, Essabir et al. [41] concluded that GO nanosheets and $Fe_3O_4$ (added separately or as the hybrid fillers [94] to PVDF) become nucleating centers and thermal stabilizers during the β-phase formation. Similar role of nucleating centers was observed for modified using 3-aminopropyl triethoxysilane) carbon nanotubes, increasing the β-phase content by 103 % [95].

Some authors claim [78] that in the absence of CNMs carbonyl groups-$CF_2$ electrostatic interactions [62, 84, 96, 97], the formation of β-phase is limited– as it was observed after addition of electrochemically exfoliated G to PVDF. An extremely high conversion efficiency (almost a full transformation) was obtained in the study of Song et al. [98]. In this case, the formation of β-phase was attributed to the presence of strong interactions between carbonyl groups of poly(methyl methacrylate) macromolecules covering CNTs and $CF_2$ groups. But not only carbonyls can be responsible for the specific interactions with polymer dipoles. A similar role was assigned to the hydroxylic groups of CQDs [99], MWCNTs [51], NDs [100] or G (GO) aerogel [101] (see Fig. 4 in [101]). In the case of GO the preparation methods of a composite with PVDF was recently studied [102] and it was concluded that solution casting led to the higher amount of the β-phase. On the other hand, the -OH groups can be coated by, for example, polypyrrole [103], and the β-phase formation is, in this case, strongly restricted. The results of Ismail et al. [104] confirm the linear dependence between the content of β phase and the content of rGO, proving that not only G but also rGO increase the transformation efficiency [67, 105]. Also Roy and Mandal suggested [106] that the remaining oxygen groups present in rGO could lead to a very high α-to-β phase transformation efficiency (93%). In this case, $CH_2$/-$CF_2$ dipoles interactions with the π-electrons of the oxygen residues additionally supported the process. Also, fluorination of GO accelerated the degree of transformation [61], and this process was explained by the rise in the $CH_2$-F electrostatic interactions, in comparison to the pure GO. It can be also mentioned that the presence of hydroxyl and carboxyl groups in CNMs not always guarantees β-phase formation but can lead only to the restrictions of the α-phase vibrations [107].

Other nanomaterials and compounds, co-existing with CNMs, can improve the transformation process. Shi et al. [108] studied the influence of barium titanate (BT) nanoparticles and G nanosheets on the transformation. They observed the differences in the β-phase formation on the surface of G and BT nanoparticles caused by the β-phase PVDF chains orientation. Namely, it is oriented with F atoms toward BT surface and with H atoms toward the G surface, due to electronegativity of C (see Fig. 2 in [108]; a similar orientation was proposed by Huang et al. [109] - see also Fig. 3). Addition of $Ba_{0.85}Ca_{0.15}Ti_{0.9}Zr_{0.1}O_3$ (BCZT) nanoparticles and CNTs to PVDF promoted the β-phase formation [110, 111]. It was concluded [110] that BCZT nanoparticles interacted electrostatically with the F atoms and nanoparticles were surrounded by the β-phase (see Fig.5 in [110]). A similar situation was observed after addition of rGO and $BaCo_2Fe_{16}O_{27}$ both acting as β-phase nucleating agents (also rGO alone can play this role [63]) *via* interactions with $CH_2$ groups of PVDF [112]. Huang et al. [113] indicated the role of QDs defects as the centers of nucleation, and Ma et al. [114] as the centers of PVDF self-assembly. It was also proved that the addition of IL [47, 91, 115] and poly-acrylamido-methyl-propane-sulfonic acid [116] to carbon nanofibers (CNF) (or CNTs) induced the transformation [115]. Also, the parameters of micro-injection molding process [117] (especially the injection speed) influence the β-phase contents during the process of G-PVDF formation. It was explained on the basis of the influence of a shear force on the stability of PVDF spherulites. Zhou et al. [118] showed that, created from polydopamine by a heat treatment carbon shell encapsulating $BaTiO_3$, increased the β-phase content (similar synergic effect can be obtained by the application of MWCNT [119]). The same authors confirmed the well-known experimental observation of the enhanced β-phase formation by the incorporation of TrFE. The study of Samadi et al. [120] led to the conclusion about the more efficient β-phase formation during electrospinning, after addition of $TiO_2$-$Fe_3O_4$-MWCNT hybrids to PVDF.

Phase separation [109], light [121] (undoped PVDF is light insensitive), poling [122], electrospinning [76, 123-128] (and liquid electrolyte-assisted electrospinning [129]), micro-injection molding [117], ultrasonication [130], mechanical stretching, stress-inducing [131], hot pressing [132, 133], milling [134], incorporation of PTFE [135] and solvent assistance [136] favour the conversion process in the presence of CNMs. Deformation [117] of PVDFchains by the presence of G (see Fig.10 in [117]) is considered as a (increasing with -temperature) process leading, at the first stage, to the fragmentation of α-phase lamellar fragments. At the second stage, recrystallization with the formation of a new α-phase occurs. Next, the processes of α-to-β phase transformation and recrystallisation are observed. Considering ultrasonication, it was concluded [130] that, in some cases (for example in the presence of CNTs), ultrasonication might cause a higher adsorption of the β-phase on a surface of a nanomaterial, and in this way conversion can be adsorption-induced. Badatya et al. [137] showed that after adding polar solvent ( *N*,*N*-dimethylformamide) to conductive CNTs (in the presence of NaCl) the formation of β-phase is observed, caused by the creation of a local electric field and high density of *N*,*N*-dimethylformamide. The influence of poling was studied by Bhunia et al. [122], who used GO and high electric field in vacuum. Authors proposed a possible mechanism of $CH_2$ – π electrons interaction leading to the rise in the β-phase contents (see Fig. 8 in [122], see also Fig. 3). The similar mechanism was proposed in [133]. Kumar et al. [138] reported the study on simultaneous influence of poling, electrospinning and addition of CNMs (carbon fiber in this case) on transition, and determined the optimal parameters of this process. Fortunato et al. [139] used two strategies of CNMs introduction (G was used) – direct multi-layered exfoliation of G in PVDF solution (strategy 1), and incorporation of the so-exfoliated material (strategy 2). Strategy 1 leads to a larger β-phase content. Nasir et al. [67] performed studies on the conversion of phases using CB and electrospinning. They showed that conversion was in fact preferred, but crystallinity decreased with the introduction of CB. CB is well-known in improving the transformation process [140]. Decrease in crystallinity during polar phase creation was also reported by Xu et al. [92].

Recently Zhang et al. [141] reported an interesting theoretical approach to the G-induced conversion process, showing that the rapid conversion is observed around 0.11% vol. of G.

There are also reports on the lack of CNMs (in this case expanded Gr) influencing the transformation [26]. For example, Lu et al. [142] reported the GO induced transformation but only in the presence of ZnO, improving GO exfoliation, and promoting interactions of oxygen groups of GO with PVDF. Considering ZnO, Hasanzadeh et al. [124] concluded that it played a role of a nucleation center for β phase and established (during electrospinning) the stretching of PVDF chains.

#### 2.4.4. The most common PVDF-CNMs configurations

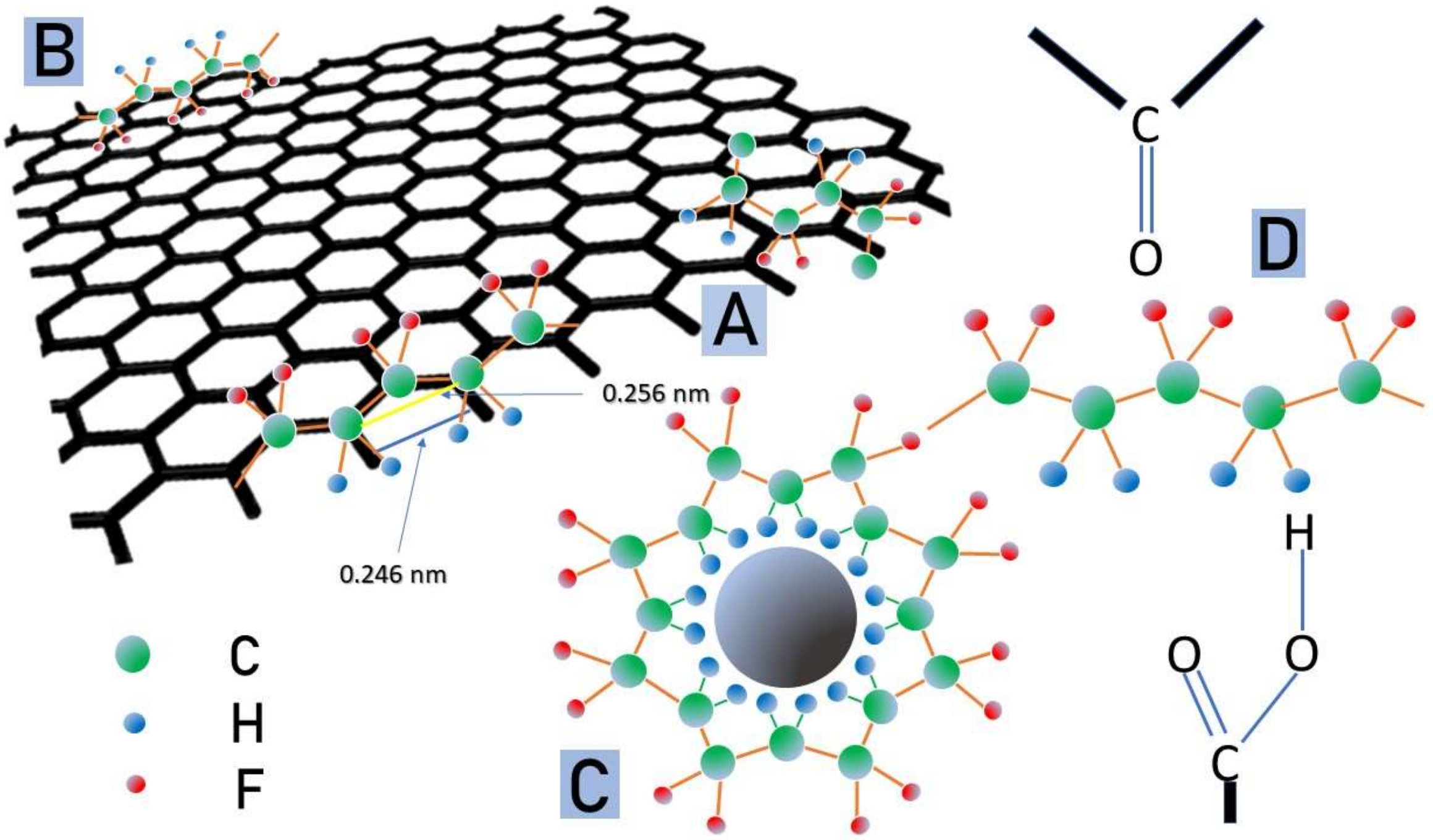

**Figure 3.** Mechanistic approaches to the α-to-β phase conversion. Configurations of α and β PVDF phases on G surface, with characteristic diameters between C atoms (A), PVDF orientation with H atoms towards G surface based on [108] and [122]) (B), PVDF orientation with F atoms towards OH groups of BT surface (based on [108]) (C), interactions between H of surface hydroxylic and carboxylic groups with PVDF F atoms (based on [101]) (D). Adapted

with permission from [108]. Copyright 2018 Elsevier. Adapted with permission from [101]. Copyright 2020 Elsevier. Adapted with permission from [108, 122]. Copyright 2017 John Wiley and Sons.

In Fig. 3, we summarise the most commonly proposed configurations of PVDF-G and GO systems. As one can observe, there are different possibilities, and to discuss details the quantum – type calculations were necessary.

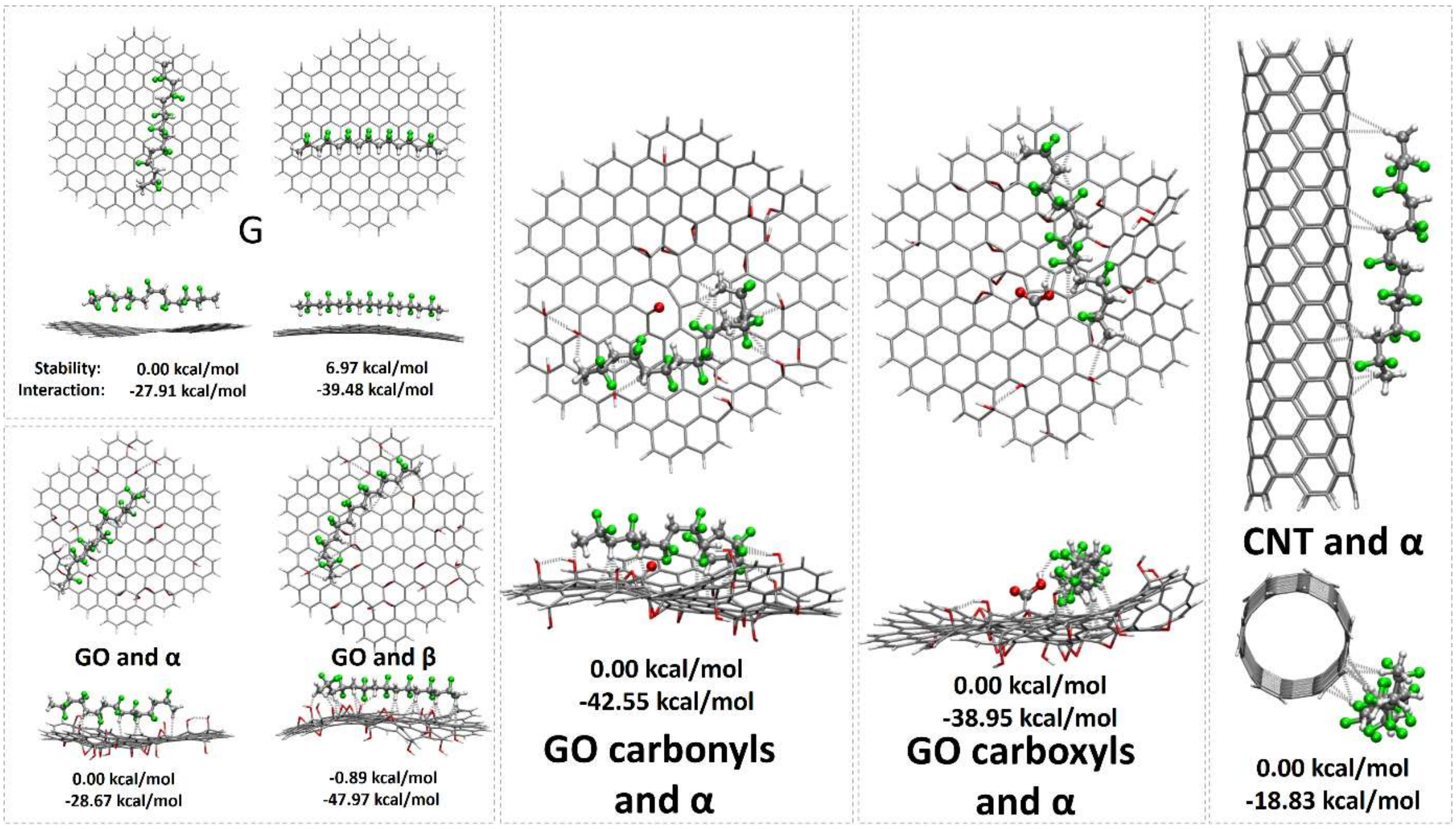

**Figure 4.** DFT snapshots showing the most stable PVDF-CNMs structures.

Thus we performed the DFT calculations (see Supporting Information for details – Fig. S3-S14). The obtained results showed that there was a lack of the correlation between the stability of the complex and the interaction energy, and it comes from the several factors affecting the relative energy of a system (see Supporting Information). It is seen that on G surface, α form is more stable than β, and the most stable configurations are oriented parallelly to G surface. This is caused by a benefit from the dispersive orientations. Considering GO the most stable configuration of the α PVDF phase (Fig. 4), the polymer configuration is determined mostly by the CH…C or CH...O interactions and only scarce CF...HO contacts. In the case of β, apart from a large number of stabilizing contacts with the surface, also the twist of the terminal part of the polymer chain should be noticed. We also considered interactions with surface carbonyl and carboxylic groups. Both types of surface functionalities are most often considered on the surfaces of CNMs (see Fig.3). Our results show, that PVDF interactions with GO surface carbonyls is determined by the minimization of repulsion between partially negatively charged O surface atoms and F atoms of PVDF (see Fig. 4). On the other hand, the carboxyl group is more prone to the mutual attraction with the polymer chain, by forming the double hydrogen bonded quasi-rings C-O-H...F-C-C-H...O(=C), like the one presented on Fig. 4. Considering interactions with a CNT, we observe parallel orientation between PVDF and CNT and the interactions *via* hydrogens with CNT π rings. Summing up, among shown on Fig. 3 orientations all configurations, excepting B, seem to co-exist.

### 2.5.2. The α-to-γ phase conversion

Finally, we should point out that the studies on transformation to the γ phase are rarely met in the literature. In this field, we should mention Roy et al. [12] who studied the application in transparent electroactive films, and Nunes-Pereira et al. [143] who studied a series of CNTs impregnated by metals (Co, Ni, Pd, Pt), and showing that they induced the presence of the γ phase. The largest content of this phase (ca. 90%) was observed for Co and Pt-Co decorated CNTs. Such composites may find application in the materials with tunable crystallinity and optical transmittance. Seena et al. [132] reported hot-pressing induced γ-phase creation induced by the presence of MWCNT modified by $HNO_3$ and $H_2SO_4$. Barrau et al. [144] provided a short review on the influence of different nanofillers on the γ-phase creation, pointing out the importance of this phase due to its ferroelectric properties. A maximum γ-phase content was reached at 0.7% of CNTs. Also addition of MWCNH and IL increases the γ- and β-phase amount [80].

Summing up this part, we can conclude that there are no reports showing the influence of NDs and SWCNHs on the conversion. The number of studies on the γ phase is rather small. Molecular dynamic simulations and quantum DFT calculations of the transformation process are necessary to explain further details of the mechanism (see Supporting Information – Fig. S3-S11).

# 3. Fundamentals of the phenomena in which the application of CNMs improves the properties of PVDF with practical examples

## 3.1. Electromagnetic Interference (EMI) and Electrostatic Charge Dissipative (ESD) Shielding Materials

### 3.1.1. EMI shielding

Nowadays, the number of electronic devices emitting electromagnetic (EM) waves, like mobile phones, computers, TV equipment, radio, cardiac pacemakers, etc., is rapidly increasing rapidly [145]. The EM waves cover a wide range of frequencies – from a few Hz up to several hundred GHz [146] (see Fig. 5). Simultaneously, they interfare with EM waves emitted by other devices. Thus, the problem of a proper shielding of electronic devices appears. From the economic and practical reasons, the housing/casing/cover of these devices/gear/equipment are made of polymer materials, not metal. As the classical polymer materials are insulators, they do not stop the EM waves, they should be modified to fulfil the barrier requirement, *i.e.*, to reduce the influence of the unwanted waves emitted by that device, and to prevent undesirable radiation from penetrating into the interior of the device which would disrupt operation or cause a malfunction of the device. This phenomenon is called electromagnetic interference (EMI).

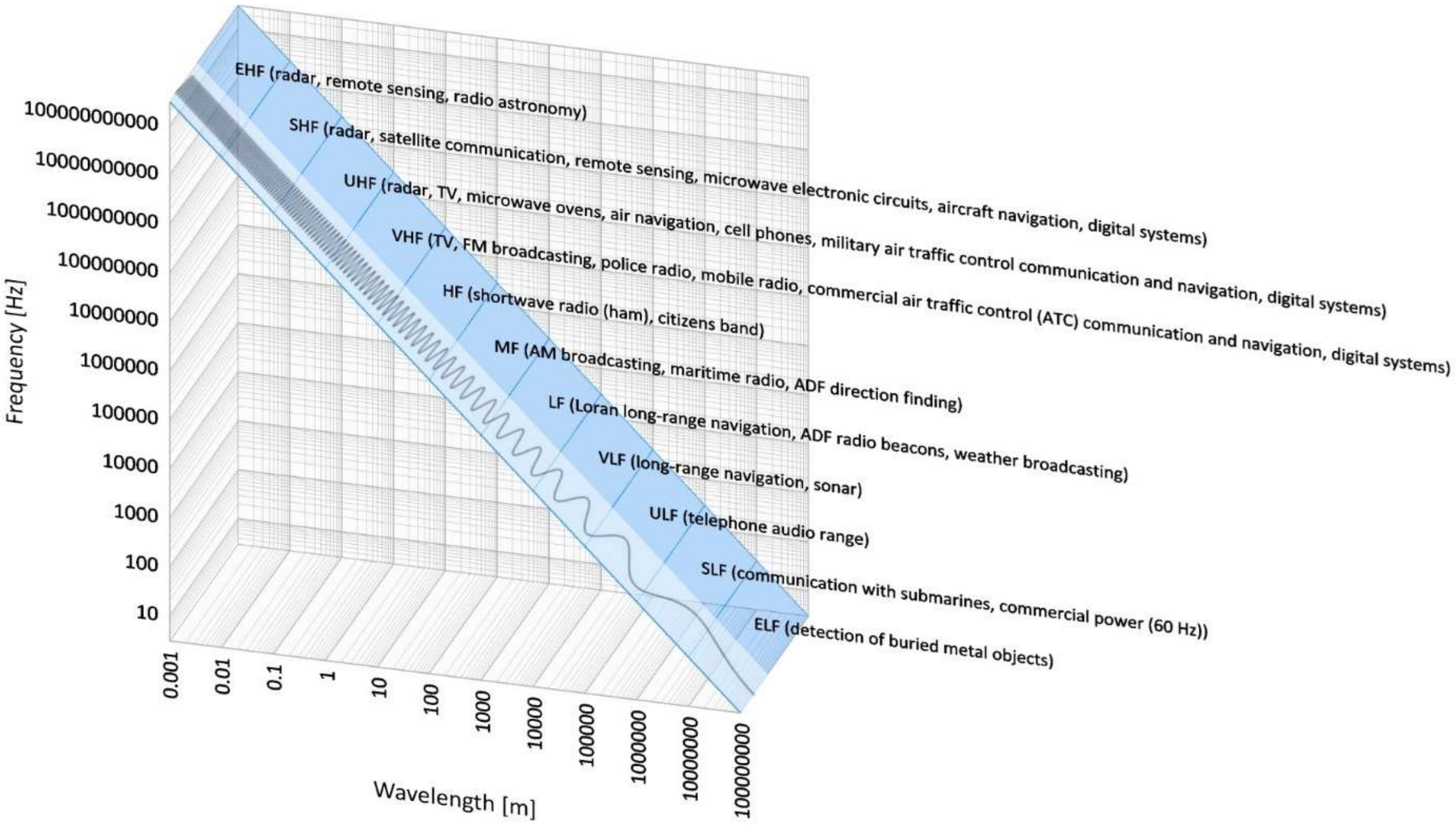

**Figure 5.** Electromagnetic waves frequencies and wavelengths, together with the corresponding applications (E – extra, F – frequency, H – high, L – low, M – medium, S – super, U – ultra, V – very). Adapted with permission from [146]. Copyright 2006 John Wiley and Sons.

The shielding effectiveness of a material, SE, is defined as follows [146-148]:

$$SE = 10\,log\frac{P_i}{P_t}$$

$$SE = 20\,log\frac{E_i}{E_t}$$

$$SE = 20\,log\frac{H_i}{H_t} \qquad (1)$$

where $E = |\mathbf{E}|$, $H = |\mathbf{H}|$; $P$, **E**, **H** are power density, electric, magnetic field respectively, subscripts i and t denote incident and transmitted wave, respectively. $SE$ is expressed in dB; *e.g.* 99.9% reduction of $P$ ($P_i/P_t = 10^3$) gives $SE = 30$ dB.

When the electromagnetic wave passes through a shield, it is subjected to the following actions: 1st reflection (R), 2nd absorption (A), 3rd multiple internal reflections (MR), which are schematically shown in Fig. 6. Thus, SE can be written as the sum of components corresponding to these three mechanisms [146-149]:

$$SE = SE_R + SE_A + SE_{MR} \tag{2}$$

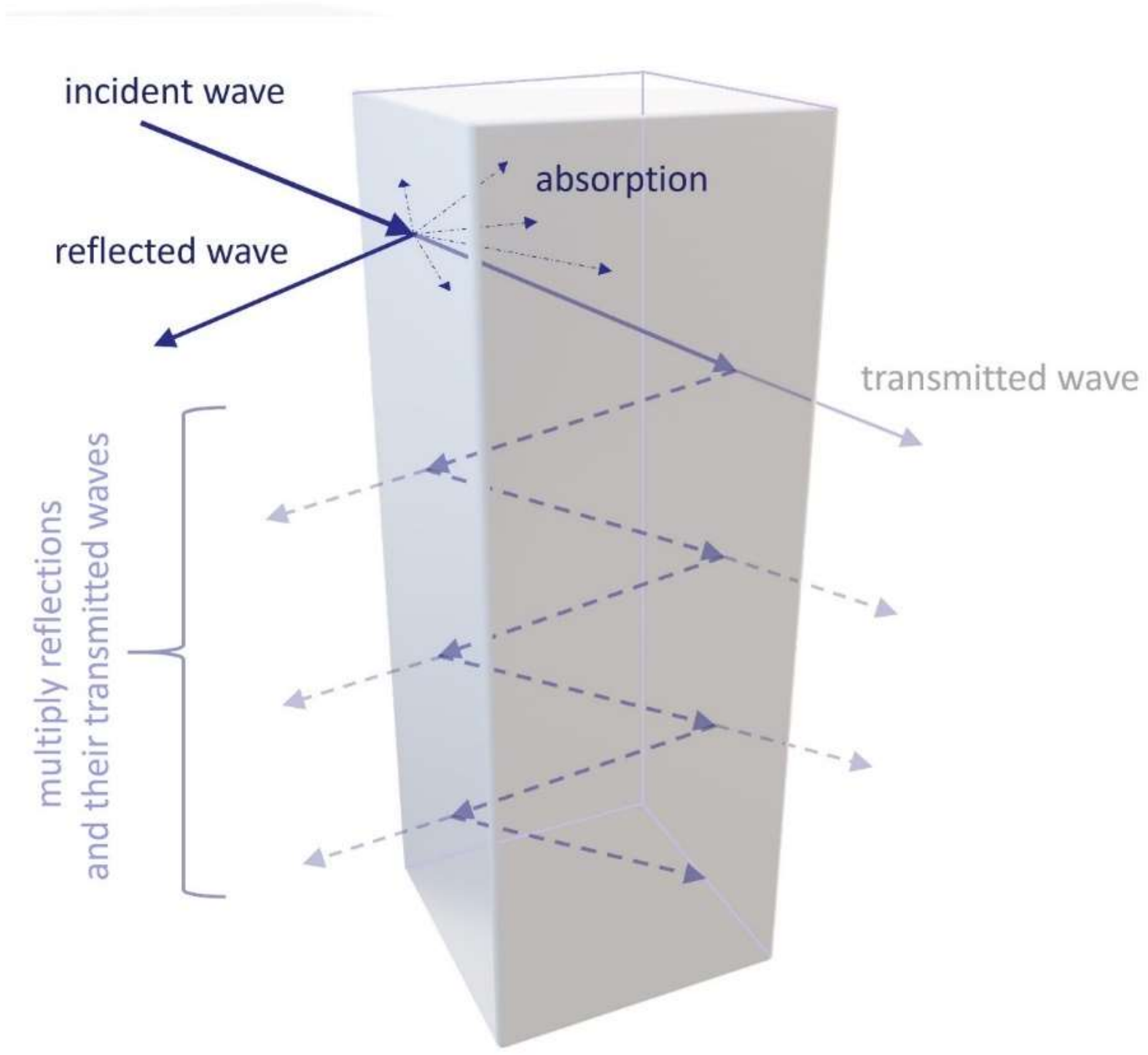

**Figure 6.** Mechanisms of the electromagnetic shielding.

As shown in Fig. 6, the reflection and absorption attenuate the passing wave ($SE_R$, $SE_A$ > 0), whereas the MR augment the transmitted wave ($SE_{MR}$ < 0, see Eq. 2).

For the reflection, the mobile charges (electrons or holes) are needed. The conductivity (connection of conductive domains) is not required for the EMI shielding, however, it augments the shielding.

The reflection of electric and magnetic fields together with the equations describing the reflected and transmitted waves is shown in Fig. S1.

According to these formulae, if the intrinsic impedance of the shield, $\eta_s$, is much lower than the intrinsic impedance for the wave outside the shield, $\eta_w$, $\eta_s << \eta_w$ then $E_1/E_i$ is very small which means that most of electric field is reflected from the first boundary whereas the magnetic field is mostly reflected by the second interface. One can conclude that thin shields are effective for those fields [146]. From the formula (S3), it results that $SE_R$ is high for materials of high conductivity and low magnetic permeability, and low-wave frequency.

Absorption requires the presence of electric and/or magnetic dipoles which interact with EM waves. The absorption loss does not depend on the type of electromagnetic field [148].

The shielding efficiency caused by the absorption loss, $SE_A$, is directly proportional to the shield thickness and the square root of the product of the frequency of the EM wave, $f$, magnetic permeability, $\mu$, and electrical conductivity of the shield, $\sigma$ (Eq.(S6)). It can be noticed that whereas $SE_R$ (far-field, Eq.(S3)) increases with $\sigma/(f\mu)$, $SE_A$ increases with $f\sigma\mu$ (Eq.(S6)).

The multiple reflections (MR) denote reflections at the surfaces and interfaces of the shield material. For this effect to play a role, these surfaces and interfaces should be large (porous materials, composite materials with a very fine filler). Moreover, the shield thickness, $t$, should be lower than the skin depth, $\delta$. If $t > \delta$ the component $SE_{MR}$ in $SE$ can be neglected; *e.g.* for $t = 2\delta$ $SE_{MR} = -0.16$ dB as calculated from Eq.S7 [146, 148-150]. For the conductive shields ($\eta_s << \eta_w$) that are much thinner than the skin depth the contribution of MR may not be negligible.

According to Eq.S7, $SE_{MR}$ is always negative – the lower $t/\delta$, the more negative $SE_{MR}$. For $t >> \delta$, $SE_{MR}$ goes to zero. The negative contribution of $SE_{MR}$ to $SE$ results from the fact that, as a result of MR, the transmitted wave is augmented as indicated in Fig.S1.

MR is not important for electric field because most of the incident wave is reflected from the first boundary (if $\eta_s << \eta_w$). This is not the case with a magnetic field where most of the incident wave passes through the first boundary and is reflected from the second one. If the shield is thin enough to substantially decrease the absorption loss, then MR should be considered.

Summarizing, SE depends on the shield parameters (conductivity, permeability, permittivity, thickness), the electromagnetic field (frequency, electric, magnetic fields), the distance source-shield, and other factors.

### 3.1.2. Practical applications of PVDF-CNMs in EMI shielding

EMI (occurring usually in the range $10^4 – 10^{12}$ Hz of the electromagnetic spectrum [105]) shielding is important due to the need of the interference elimination in everyday live (for example human health), military industry, 5G network [151, 152], aviation industry (here the most interesting is the improvement of resistance against lighting; in Boeing 747 50% of the weight are CF reinforced polymers [153]). The most common EMI is observed between our mobile devices and laptops or radio speakers [105]. Thus, in view of those applications, one can conclude that the major features of the EMI shielding materials used in practice (dielectric as well as magnetic loss) are: high dielectric constant (the materials are called high *k*-constant materials [154]), low dielectric loss, high conductivity and thermal stability as well as low density [155]. This interference can lead to serious problems with malfunctioning of medical or communication devices. An excellent introduction to the EMI shielding concept and importance was given by Choudhary et al. [156], who pointed out that since the majority of devices works using microwaves, the larger part of EMI shielding studies is focused on the GHz range of EM waves (see Fig.5). Here interesting is the so-called $K_u$ band range (12-18 GHz) applied for the satellite communication [157]. Overall, polymers are lighter, more economic due to easier scalable manufacturing and less susceptible to corrosion than metals, and considering CNMs, they are usually applied for this purpose as metal decorated materials (for example PVDF/CNT/rGO/ [158]) and nanocomposites. This stems from the fact that CNMs cause interfacial polarization and this phenomenon leads to the increase of dielectric constant (the simplest material used for this purpose could be CB at the conc. of ca. 3%[159], and/or oxidized carbon nanoparticle from sub-bituminous coal; the maximum dielectric constant was observed for 50% of the nanofiller content [160]). A new method of obtaining new composite fillers of increasing dielectric constant is ball-milling in extensional flow field [161]. EMI shielding mechanism depends on the CNM contents, as it was pointed out by Mei et al. [162] who studied the PVDF/CF/G/biocomponent fibers. Also, other forms of carbon can be used, for example G [105, 163, 164], (as nanoplatelets in foams, enabling MR and scattering – see Fig.9 in [165]), GO (also as 3D porous aerogels [166]), rGO (with Ag) [167], CNFs [17, 151, 168] CNTs [47, 169-174] and carbons obtained from carbonization of a ZIF-67 MOF combined with Co/Ni [175]. Also the mixture of CNMs has been applied sometimes supported by IL [173]. Mei et al. [162] presented CF and polypropylene/polyethylene (PP/PE) core/sheath fibers (CEF-NF) in G/PVDF nanocomposite. No metals were added, however, very efficient EMI shielding (300-1500 MHz) effectiveness, higher than observed for the majority of materials, was achieved (Fig. 8) [162] illustrates the effectiveness in EMI shielding caused mainly by high scattering and absorption of electromagnetic radiation inside the material). Here, SE depends on the amount of G in the matrix, improving, at larger concentrations, the adhesion between GE/PVDF and CEF-NF affecting the propagation path.

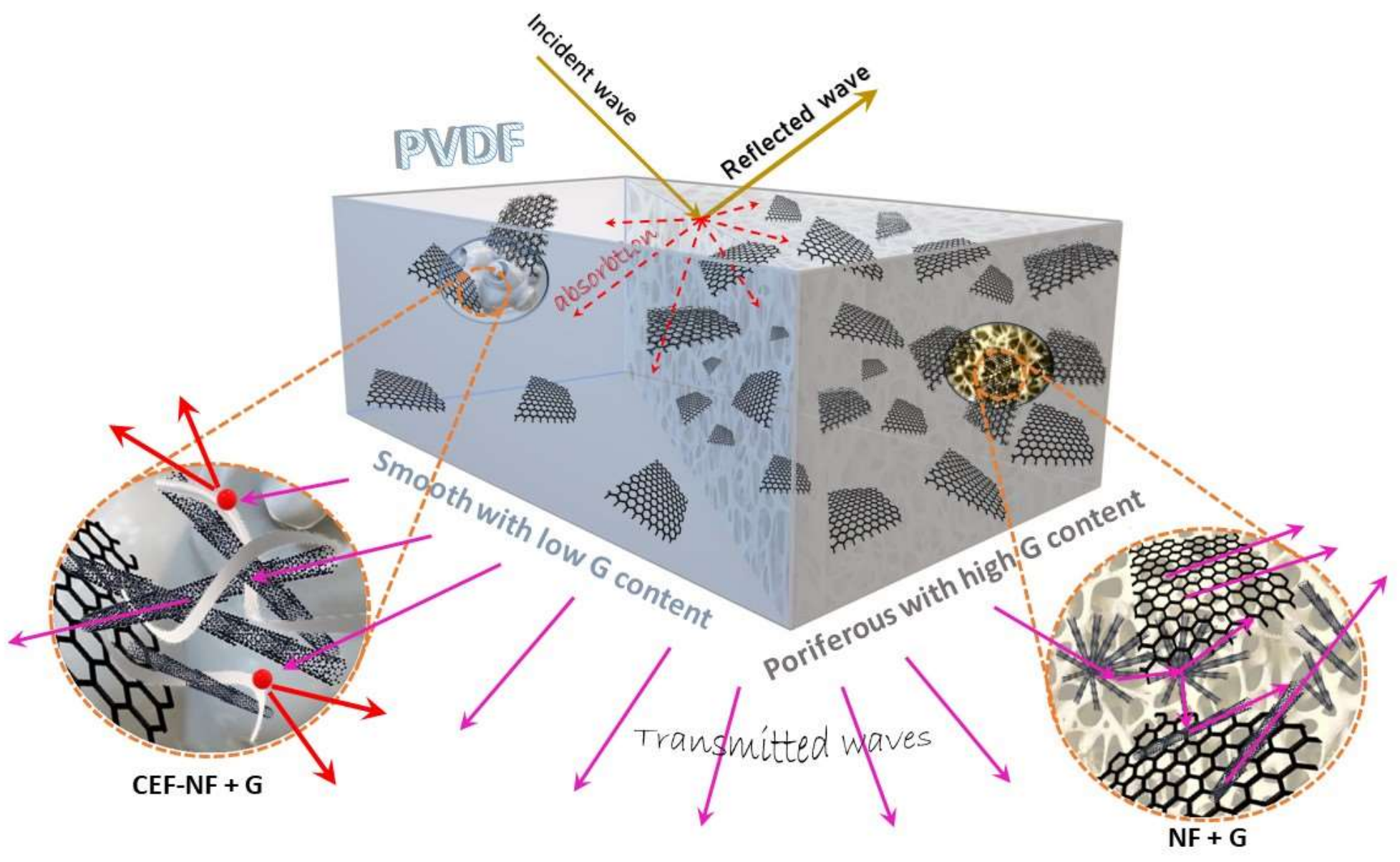

**Figure 8**. EMI shielding by high scattering and absorption of waves inside the material. Adapted with permission from [162]. Copyright 2019 Royal Society of Chemistry.

G was also studied by Dong et al. [176] who showed that ball milling of Gr is the effective method increasing the amount of G (6 wt.% was the most effective amount leading to a highly operational EMI shielding material). An interesting approach was proposed by Jia et al. [177] who used G and CB for the preparation of electrically conductive PVDF – supported, low density, microporous foams (the mechanism of EMI shielding by a continuous reflection was shown in Fig.9 [177]. The cavities present within the structure of microporous material caused a continuous reflection and refraction of EM waves converting them into heat. Here, MR of EM waves is crucial, and it was assured by high specific surface area (and high interfacial area) of the porous material.

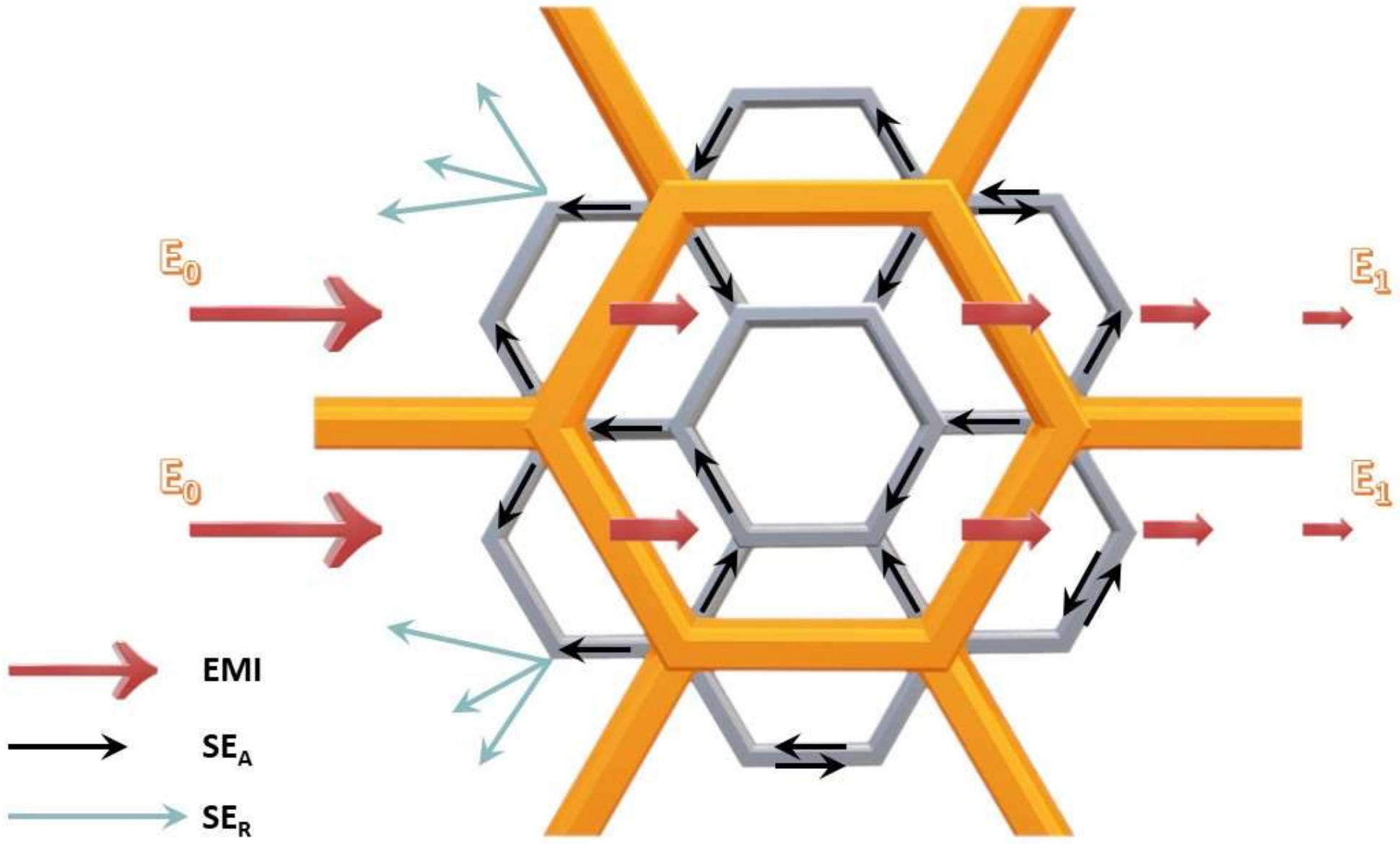

**Figure 9.** The propagation of EM waves in CNMs filled PVDF. Adapted with permission from [177]. Copyright 2021 John Wiley and Sons.

Considering CNTs, an interesting study was published by Biswas et al. [178] who studied the effect of polymer (PP and PVDF), and the structure of CNTs on EMI shielding in the $K_u$ band range. It was concluded that PVDF is a superior dispersant for CNTs, and the most effective shielding was caused by layered materials containing branched CNTs.

Also, an addition of graphitized charcoal to PVDF [179] led to well-performing EMI shielding materials (8.2 - 12 GHz) with a simultaneous improvement of conductivity and mechanical properties, while the addition of CB and expanded G raised EMI shielding in the similar range [180, 181]. However, CB-containing material worked more efficiently. Moreover, the addition of CB widens the efficiency range, which can be crucial in the stealth materials [182]. An excellent study was recently published by Lee and Lim [183] who based on a simple model of polarization (including electronic, ionic and orientational) concluded that addition of RGO to PVDF influenced mainly the orientational polarization of this material, thus changing the dielectric constant. There are also reports on the application of CB and $C_3N_4$/PVDF in the microwave absorption materials [184].

Taking into account that CNMs and metallic cores absorb microwaves, EMI shielding (*via* ohmic loss and polarization) in the range of frequency 8.2 - 12 GHz was observed for Ni and Ni alloys (CoNi, FeNi) combined with CNTs [185], and in a similar range (8 - 12 GHz) for Co nanoparticles embedded in an amorphous carbon matrix [156] (Fig.8) for the model of EMI shielding). In this case, EM waves are the first scattered on carbon defects, and next absorbed by carbon and Co nanoparticles. An increasing scattering was caused by the presence of a large number of interfaces between the components.

Similarly interesting, Ni-containing EMI shielding flexible materials (PVDF/CNTs/Ni@CNTs) with a balanced mechanical properties and temperature-controlled performance, were reported [186]. The temperature rise (by ca. 10%) upon recording of shielding properties was caused by the rise in PVDF matrix crystallinity causing displacement of the conductive fillers inside (see Fig.7 in [186]). Zhao et al. [155] reported PVDF/CNT/Ni-chain and PVDF/G/Ni-chain nanocomposites displaying thickness-tuned EMI shielding properties with a dominating absorption mechanism. The identical mechanism of EMI shielding (8-12 GHz) was also reported for PVDF/RGO/$BaCo_2Fe_{16}O_{27}$ films (Fig.11) [112]. According to the authors [112], surface reflection of EM waves was observed, while still the absorption predominated.

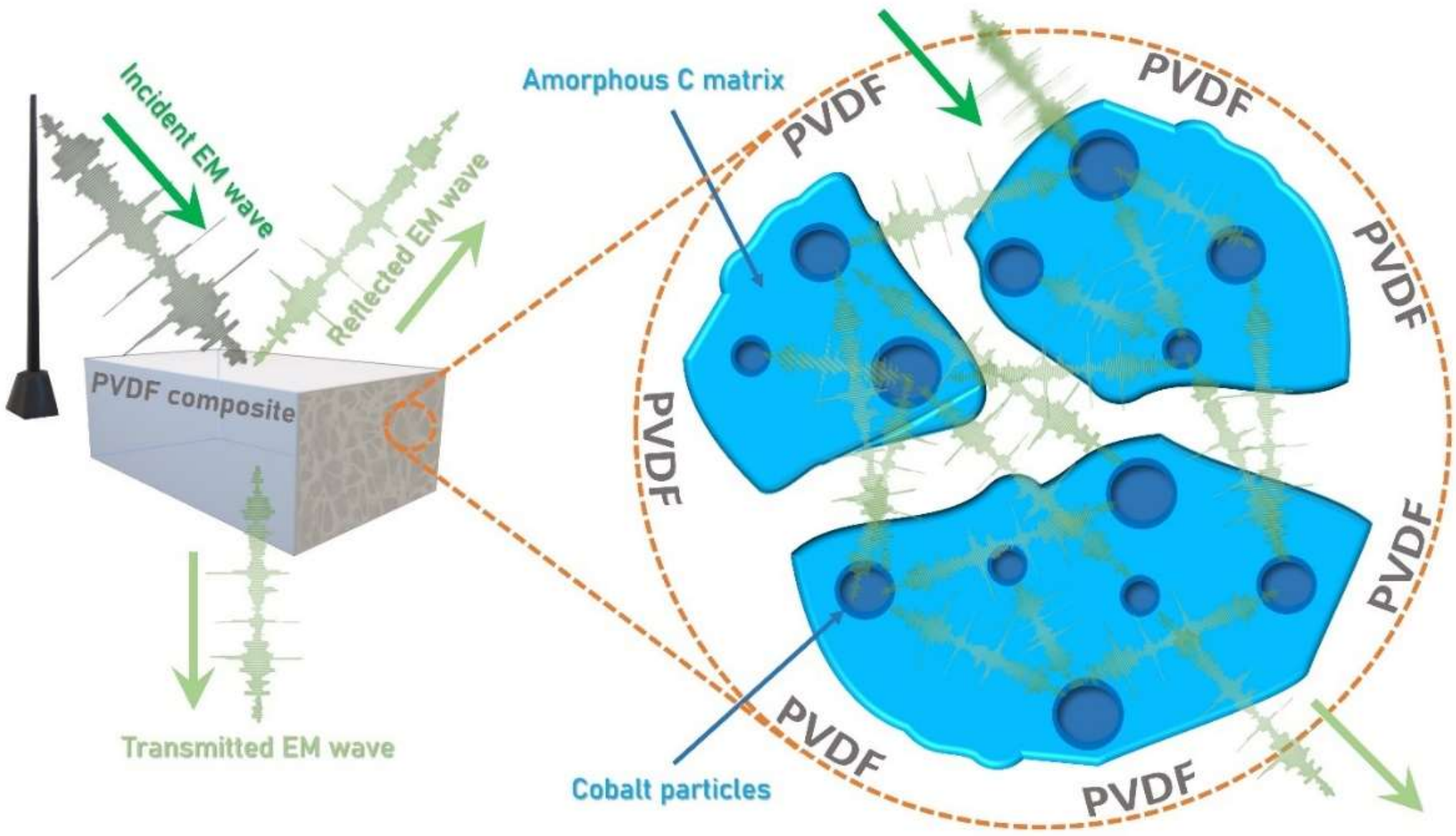

**Figure 10**. The model of EMI shielding for carbon-coated Co NPs. Adapted with permission from [156]. Copyright 2006 John Wiley and Sons.

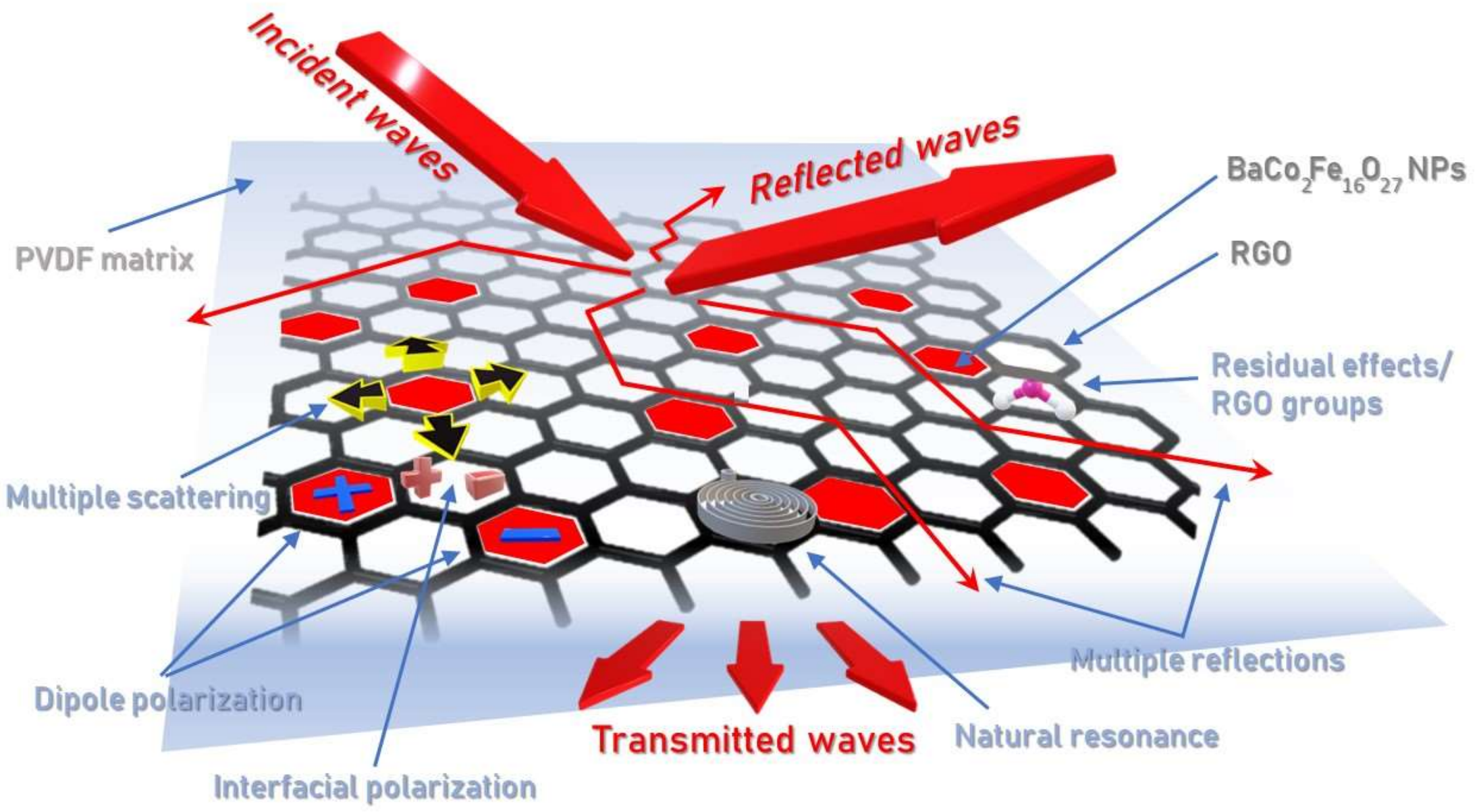

**Figure 11**. The mechanism of EMI shielding for PVDF/RGO/$BaCo_2Fe_{16}O_{27}$ films. Adapted with permission from [112]. Copyright 2020 John Wiley and Sons.

Likewise G/PbS embedded in PVDF showed an effective microwave absorption (9.5 - 12 GHz) [187]. Choudhary et al. [188] reported CNTs and nanoglobules – obtained *via* pyrolysis of different metal containing (Ni and Mn) precursors – blended with PVDF. Ni-containing materials worked more efficiently in the EMI shielding and the differences in shielding efficiency were caused by the nature of nanoparticles: conducting Ni and insulating MnO (this mechanism is shown in Fig. 12 [188]).

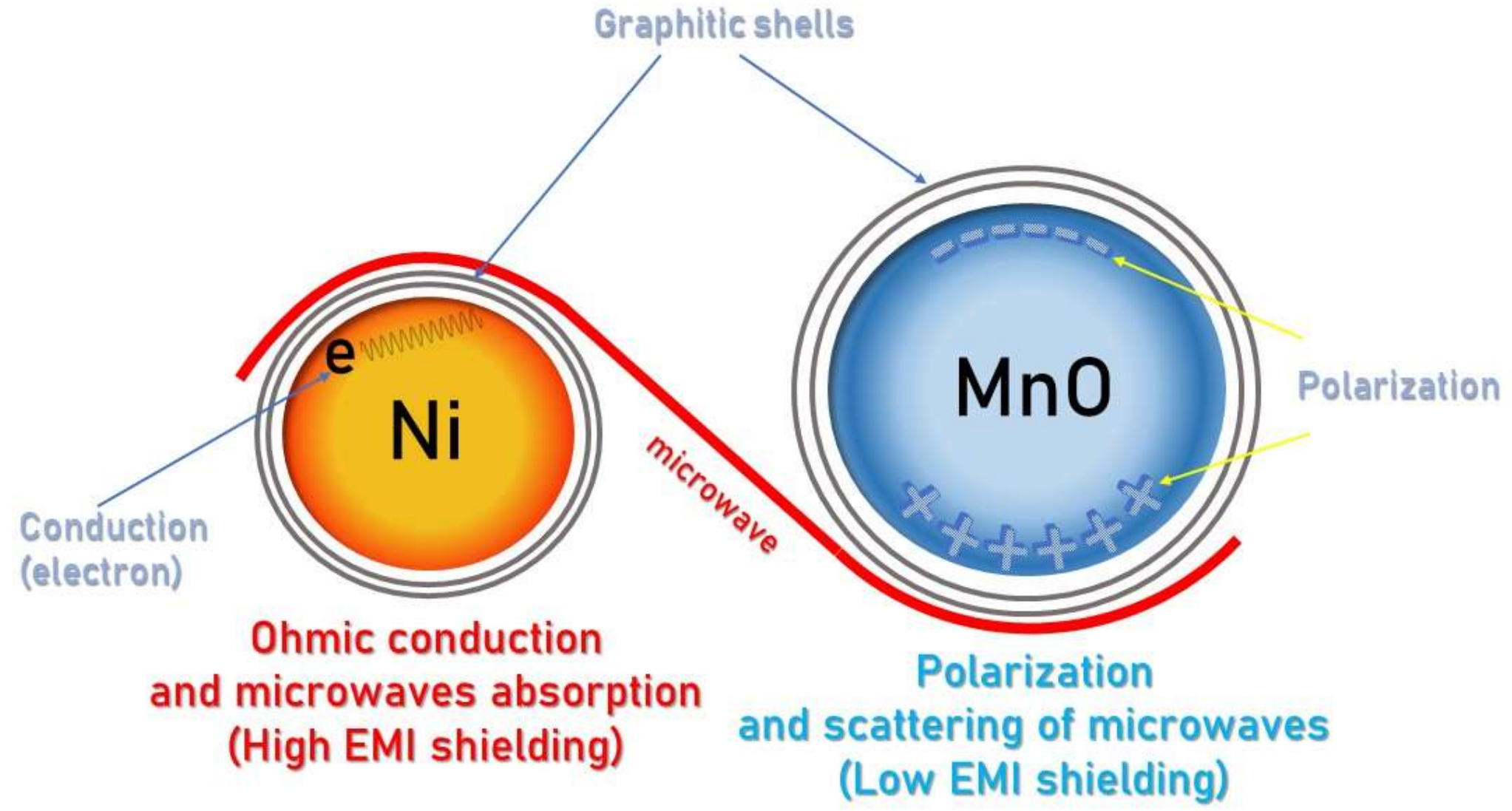

**Figure 12**. The mechanism of EMI shielding for Ni- and MnO-based CNMs containing nanocomposites. The differences in the shielding efficiency were caused by the nature of nanoparticles: conducting Ni and insulating MnO. Adapted with permission from [188]. Copyright 2021 Royal Society of Chemistry.

There is a need to develop EMI shielding and heat dissipating materials, due to the problem of the sustained application of numerous devices. In this field, an attractive material was described by Liang et al. [189] who applied G and Ni nanochains to synthesize Ni@G-PVDF films (Fig. 7 in [361] shows the mechanism of EMI shielding and a simultaneous heat dissipation mechanism). Li et al. [190] reported application of PVDF/CNTs/G/Cu@Ni nanorods (contributing to magnetic and dielectric loss) for EMI shielding flexible films. Very high thermal conductivity allowed to dissipate heat and strong absorption contribution to shielding (see Fig. 5e in [190]). At the same time, a strong synergy between CNMs and Cu@Ni led additionally to the in-plane flexibility and resistance to bending. Magnetic and dielectric loss mechanism was also proposed for $ZnFe_2O_4$@CNT/PVDF composites [191]. Sharma et al. reported rGO/CuS/MWCNT films pointing out the importance of MRs of incoming waves from MWCNTs [192]. The similar scattering loss-based mechanism was presented by Rengaswamy et al. [193] who presented high performance EMI shielding materials (8-12 GHz) based on Ag-Cu/MWCNT/rGO (see Fig. 8 in [193]). The authors also presented the comparison of EMI shielding efficiency for similar materials (see Table 1 in [193]). The importance of MRs in EMI shielding was pointed out by Zhang et al. [194] who reported an interesting approach for the preparation of new, so-called, segregated double networks. Using BN, CNT, and PVDF they obtained a material of a comparable to the earlier reported works efficiency of EMI shielding, electron insulation, and thermally conductivity. Considering the application of CNTs as very important, a complex and comparative study was presented by Sushmita et al. [195] who studied CNTs/polycarbonate/PVDF composites (also GO/$Fe_3O_4$ was used in some cases Considering the EMI shielding efficiency, the authors concluded that the length of nanotubes slightly changed this efficiency while for the case of employing the too short tubes, and the shielding mechanism was indicated as mainly proceeding by absorption.G and Ni spinal ferrites were applied for EMI shielding (by absorption) in the range of 1-12 GHz [196]. Also PVDF-Ni/NiO amorphous carbon (obtained from glucose and urea) [369] were shown as effective microwave absorbers (4.8-18.0 GHz) by polarizations, magnetic loss and dual-configurational mechanisms (see Fig. 10 in [197]) contributions. Ni@C microspheres with a steerable carbon shell thickness emerged as efficient microwave absorbents [198], (Table 1 in [198] compares microwave absorption properties of different Ni-containing composites). Peymanfar et al. [199] proposed the application of carbon microspheres (obtained from glucose) - covered by Ni nanosheets for the absorption of microwave irradiation (Scheme 2 in [199] shows an excellent representation of microwaves absorption by the composite material). Cui et al. [200] demonstrated that porous G/Ni decorated PVDF foam can effectively absorb microwaves (18-26.5 GHz). Lu et al. [201]

reported the synthesis of rGO decorated with octahedral $NiS_2$/NiS nanocrystals as effective materials for microwave absorption by dielectric loss, improved impedance matching, and synergy between PVDF and CNMs. Also, high EMI shielding (in the range of microwaves) properties were reported by PVDF/MWCNT/G/Ni films [202].

$ZnAl_2O_4$/polystyrene and carbon nanospheres (obtained from sucrose and nitric acid) were used for the microwave shielding (the mechanism of shielding is shown in Scheme.2 in [203]), similarly to $ZnFe_2O_4$@$SiO_2$@RGO [204] and $Zn_{0.5}Ni_{0.5}Fe_2O_4$-RGO [205]. Here, the application of a multi-layer strategy (as well as so-called sandwich structures, like in G nanoplatelets-Ni-CNT [206]) improved the EMI shielding by multiple internal reflections mechanism (see Tables. 1 and 2 in [205]). Also, carbon polyhedrons coupled by Co/Ni alloys obtained on a MOF as a template, were shown to be applicable for EMI shielding by the mechanism of dielectric loss *via* dipolar/multi-interfacial polarizations, and magnetic losses.

The latter ones were induced from eddy current loss and a natural resonance (also porosity influence; see Fig. 10 in [207]). Zr-doped $BaZrFe_{11}O_{19}$ and rGO [208] were applied for EMI shielding (*via* the mechanism shown in Fig. 10 in [208]). It was also shown that the application of combining composite Zn-Cu-Ferrite-$C_3N_4$-PVDF leads to a very effective microwave absorber [209], and the combined application of all materials provides high magnetic and dielectric losses. Very effective EMI shielding (10-18.0 GHz) materials, showing also high YM, tensile strength, and UV blocking, were reported by Bhattacharjee et al. [210] Core-shell materials contained carbon nanospheres (CNS, from glucose) covered by $SiO_2$ and $Fe_3O_4$ (CNS@$SiO_2$@$Fe_3O_4$). $Fe_3O_4$ and CB reinforced PVDF were applied for EMI shielding (10 MHz – 1 GHz) *via* dielectric and magnetic loss and the synergy between them [211]. Also G nanoplates, CNTs and $Fe_2O_3$ (*via* the interfacial polarization, magnetic, conduction loss and MR) were shown to be efficient EMI shielding materials [212]. They were applied with MXs (for example $Ti_3C_2T_x$) which yielded materials of high shielding effectiveness and also decent thermal conductivity [213, 214]. MXs were also used with CNTs in composite PVDF based films and foams [215]. Also, $Fe_3O_4$/GO/PVDF materials [216] (as well as Si modified [217]) emerged as effectively EMI shielding solids (8-18 GHz) with improved piezoelectric properties (also in the $K_u$ band) [216]. Here high shielding efficiency in the $K_u$ band range was reported for magnetite-modified G/PVDF [218] (51.74 dB in the range 12-18 GHz). PE/PVDF/$Fe_3O_4$/CNTs nanofoams were active, lightweight, EMI shielding materials (18-26 GHz) operating mainly *via* absorption and reflection mechanisms [219]. Recently Ayub et al. [220] reviewed the G/iron compounds/polymer composites application in EMI shielding, and pointed out the recently observed, rapid increase in the number of published studies in the field. The keyword analysis showed a strong interest of the scientific community in the application of rGO nanocomposites in the microwave absorption.

Efficient heat dissipation and EMI shielding (18-26 GHz) was reported for CNTs and G nanoplates containing Ni-chain composites in PVDF [155]. Effective EMI shielding in the range of 8.2-12.4 GHz was achieved by the application of G/SiC nanowires, and due to formation of conducting G-SiC networks, high thermal conductivity was additionally observed [221]. Also, Ni-ontaining chains and flowers PVDF/MWCNT/G/Ni composite films revealed decent EMI shielding properties under microwave radiation in the range of 18-26.5 GHz [202].

CB and zeolite 13X with PVDF showed an efficient EMI shielding (at the range of 8 - 18 GHz) [157]. The $K_u$ band range (12-18 GHz) shielding materials (*via* the absorption-dominated mechanism) were reported by Rani et al. [157] who used G nanoplatelet, CuO nanoparticles/PVDF/poly(3,4-ethylenedioxythiophene)-*block*-poly (ethylene glycol) blend.

Due to an increasing number of devices working in the THz range (applied in astronomy, biosensing, security screening etc.), Naseer et al. [151] reported application of CNFs for EMI shielding in the range (0.15-1.2 THz).

Radar absorbing materials RAMs are extremely important in military industry: especially materials operating at the X and $K_u$ bands (12-18 GHz) are crucial in the strategic applications [222]. Here, PVDF can be combined with hexaferrite and exfoliated Gr both playing a role of a dielectric and magnetic absorbing material, absorbing more than 99% of the EM radiation [223]. The studies on PVDF reinforced by G nanoplatelets, montmorillonite and $TiO_2$, also revealed suitable EMI shielding properties [224] in the $K_u$ band region (see Table 3 in [224] comparing properties of different nanomaterials in this region). Also rGO-$SrFe_{12}O_{19}$/PVDF was shown as effective [225] in the EMI shielding and RAM applications. Highly absorbing material (90% of efficiency) was reported by Acharya and Datar [222]. The authors applied $CuAl_2Fe_{10}O_{19}$-decorated rGO as a filler in the PVDF matrix (see Tab.II in [222] showing the SE of various materials). D'Aloia et al. [226] discussed new PVDF coatings filled with G nanoplatelets as RAMs, especially in a form of textiles (see Fig. 1 in [226]).

An improvement of PVDF dielectric constant is crucial in the electronic industry, for example in (super)capacitors. This augmentation can be achieved by the addition of dielectric materials such as graphitic carbon (obtained from glucose)- coated barium titanate hybrid particles (BT@C) [32]. In this case, the dielectric constant was increased 158 times, comparing to the pure PVDF-HFP. $Sr_3YCo_4O_{10+\delta}$ (SYCO) combined with CB and PVDF was shown as effective EMI shielding material in the range 8.2-18 GHz (rise in YM with the rise in SYCO content were additionally observed) [140]. Polypyrrolle-decorated GO was introduced into PVDF matrix, as shown by He et al. [136], improving the dielectric properties, and decreasing the dielectric loss of PVDF by interfacial polarization (see Fig. 11 in [136]). G modified with a carboxyl-riched perylene derivative *via* π-π stacking interactions, and PVDF, revealed a high dielectric constant and low dielectric loss [227]. The microcapacitor- based model was proposed (see Fig. 9 in [227]) explaining this behaviour. An interesting approach toward an improvement of dielectric constant was proposed by Li et al. [228] who used 4,4′oxydiphenol-functionalized (4,4′-ODP) G. Upon noncovalent functionalization (adsorption) of 4,4′-ODP on G they observed suppressed electrical conductivity (see Fig. 4a in [228]). Gong et al. [154] showed that large improvement of the dielectric properties of PVDF/GO system could be achieved by the introduction of tannic and Fe-coordination complexesand isolating the GO layers as well as preventing aggregation. which, in total, prevented from the aggregation Likewise, a suppressed aggregation can be achieved by addition of polydopamine and GO [229]. In this case, the rise in PVDF dielectric constant is observed accompanied by a suppressed dielectric loss.

The materials improving X-ray attenuation efficiency are rarely studied. Silva et al. [230] reported the PVDF/$BaSO_4$ nanocomposites filled with GO. Low GO concentration (4%) caused the increase in the attenuation efficiency to 24%. The comparative analysis of different EMI shielding materials can be found in Table 2 in [162]. The comparison of EMI shielding for different polymer-based materials was provided by Mei et al. (see Table 2 in [162]) and Zhao et al. (see Tab.1 in [202]). Cacciotti et al. [231] proposed the application of PVDF/$BaTiO_3$/MWCNTs multi-layered systems for EMI shielding in the microwave range.

PVDF and CNMs are also applied for light-weight electrostatic charge (ESD) shielding materials. Such shielding lead to the avoidance of attraction of dust and different living organisms causing a hygienic hazard [25]. The study on such a type of shielding is rarely met in the literature, and the results published by Peng et al. [25] proved the applicability of expanded Gr (with an interlayer distance 0.8 nm) obtained by intercalation of $H_2SO_4$ between Gr planes.

It can be concluded that various examples given above confirm the flexibility of PVDF, i.e. it can be easily modified using CNMs. In this way PVDF remains major polymer applied in EMI shielding. Since the new CNMs have been still developed we are sure that they will be probed in this field. Although the number of these type of composites is amazing the mechanism of the EMI shielding process strongly depends on the properties of the residual system. In this field RAM are probably the most interesting.

### 3.1.3. Comparison of recent reports on SE

To summarize this chapter, the new results of the studies described above were collected in Table 1.

**Table 1.** The comparison of SE for different recently reported PVDF-based materials.

| Nanofiller | Matrix | Loading | Maximum EMI SE (dB) | Studied Frequency (GHz) | Specific SE (dB $cm^2g^{-1}$) | Ref. |
|---|---|---|---|---|---|---|
| Au, Ag, CNT, rGO | PVDF | 1 wt.% | 28.5 | 12 | 57 | [158] |
| G | PE/PP/PVDF | 40 wt.% | 48.5 | 0.03-1.5 | 1731.4 | [162] |
| G | PVDF | 15 wt.% | 47 | 8-12 | 2350 | [105] |
| G | PVDF | 140 g $m^{-2}$ | 25 | 1-18 | 3048.8 | [164] |
| G | PVDF | 10 wt.% | 37.4 | 40 | 12.5 | [165] |
| CNF | PVDF | 10 %wt. | 41 | 8-12 | 41 | [17] |
| CNF | PVDF/PMM/Cyanoacrylate | 40 %wt. | 20 | 200 - 1200 | 625 | [151] |
| MWCNT | PVDF/OBC | 2.7 vol.% | 33.9 | 8-12 | 16.7 | [169] |
| MWCNT | PVDF | 5 wt. % | 32.5 | 8.2-12.4 | 32.5 | [170] |
| MWCNT | PVDF | 4 % wt. | 32 | 10.3 | 1788 | [171] |
| MWCNT | PVDF/HFP | 10-15% | 36 | 8.2-12.4 | 111.1 | [232] |
| G | PVDF/TPU | 6 % wt. | 21.4 | 2-18 | 23.8 | [176] |
| G | PVDF | 4.0 % wt. | 50 | 8-13 | 5 | [177] |
| Charcoal | PVDF | 80 % wt. | 70.6 | 8.2 | 25.2 | [179] |
| CB and carbonyl iron (CI) | PVDF | 3 % wt. of CB and 20 % wt. of CI | 27 | 8-18 | 13.5 | [181] |
| CNTs, CoNi, FeNi | PVDF | 50 % wt. | 25 | 8-12 | 25 | [185] |
| C/Co | PVDF | 50 % wt. | 25 | 8-12 | 25 | [156] |
| Ni@CNTs | PVDF | 12 % wt. | 51.4 | 18-26 | 102.8 | [186] |
| CNTs, Ni chains | PVDF | 6 % wt. Ni | 57.3 | 18-26 | 95.5 | [155] |
| RGO, $BaCo_2Fe_{16}O_{27}$ | PVDF | 10 % wt. of each | 35.94 | 8-12 | 180 | [155] |
| G, PbS | PVDF | 1 wt. % | 6.9 | 9-12 | 6.9 | [187] |
| CNTs, Ni | PVDF | 50 % wt. | 27 | 8-12 | 27 | [188] |
| G, Ni | PVDF | 10 % wt. of each | 51.4 | 18-26 | 73.4 | [189] |
| CNT, Cu@Ni | PVDF | 8 % wt. of G and 8 % wt. of Cu@Ni | 47.6 | 18-26 | 155.8 | [190] |
| CNT, $ZnFe_2O_4$ | PVDF | 7 % wt. | 54.5 | 8.4-12.4 | 22.7 | [191] |
| rGO, CuS | PVDF | 10 % wt. of rGO/CuS | 25 | 12-18 | 25 | [192] |
| rGO, Ag, Cu | PVDF | 5 % wt. of Ag and 5 % wt. of Cu | 29 | 8-12 | 290 | [193] |
| BN, MWCNT | PVDF | 5 wt. % of MWCNT and 40 % wt. of BN | 8.68 | 8-12.5 | 4.3 | [194] |
| CNT | Polycarbonate/PVDF | 3 wt.% CNT | 23 | 8-12 | 23 | [194] |
| G, nickel spinal ferrites (NSF) | PVDF | 3 wt.% G, 30 % wt. of NSF | 30-53 | 1-12 | 120-212 | [196] |
| MWCNT, G, Ni flowers (NiF) | PVDF | 6 wt. % G, 8 % wt. NiF | 43.7 | 18-26 | 146 | [202] |
| rGO, $Zn_{0.5}Ni_{0.5}Fe_2O_4$ | PVDF | 1 % wt. rGO, 5 % wt. $Zn_{0.5}Ni_{0.5}Fe_2O_4$ | 29.1 | 8.2-12.2 | 16 | [205] |
| CNT, G, Ni | PVDF | 2.75 % wt. CNT, 2.75 % wt. G, 2.75 % wt. Ni | 33.7 | 12-18 | 110 | [206] |
| rGO, $BaZrFe_{11}O_{19}$ | PVDF | 10 % wt. of each | 48.6 | 8.2-12.4 | 243 | [208] |
| NiZn-Cu-Ferrite-$C_3N_4$ | PVDF | 20 % wt. | 71.5, 88 | 8-12, 12-18 | 469, 579 | [209] |
| MWCNT, C spheres, $Fe_3O_4$, $SiO_2$ | PVDF | 3 % wt. of MWCNT | 42 | 12-18 | 70 | [210] |
| CB, $Fe_3O_4$ | PVDF | 40 % wt. | 55.3 | 0.01-1 | 27.6 | [211] |
| CNT, G, $Fe_3O_4$ | PVDF | 8 % wt. of G and CNT | 32.7 | 18-26.5 | 65.4 | [212] |
| G, MXene | PVDF | 2:8 wt. ratio MXene/G | 36.3 | 8-12.5 | 6848 | [213] |
| G, MXene | PVDF | n.a. | 43.4 | 8-12.4 | 35370 | [214] |
| CNT, MXene | PVDF | 12 % wt. of MXene | 65.1 | 22 | 325.5 | [215] |
| CNT, $Fe_3O_4$ | PE/PVDF | 10 % wt. of CNT, 1:1 PE/PVDF, 5 % wt. $Fe_3O_4$, | 26 | 18-26.5 | 110 | [219] |
| G, Ni | PVDF | 8 % wt. of Ni | 57.3 | 18-26.5 | 5 | [155] |
| G, SiC | PVDF | 9.5 % wt. | 32.5 | 8.2-12.4 | 27.1 | [221] |
| G, CuO | (PVDF)/poly(3,4-ethylenedioxythiophene)-block-poly (ethylene glycol) (PEDOT-block-PEG) blend | 2/20 % wt. of GNP/CuO | 17 | 12-18 | 17 | [157] |
| rGO, $CuAl_2Fe_{10}O_{19}$ | PVDF | 1:1 rGO: $CuAl_2Fe_{10}O_{19}$ | 60, 50 | 8-12, 12.4-18 | 40, 33.3 | [222] |
| G, hexaferrite | PVDF | 2.5 % wt. of G, 85 % wt. of hexaferrite | 50 | 1-7 | 7.1 | [223] |
| G, montmorillonite, $TiO_2$ | PVDF | 2.5 % wt. of G, 1 % wt. of montmorillonite, 11.5 % wt. of $TiO_2$ | 12.6 | 8-12 | n.a. | [224] |
| rGO, $SrFe_{12}O_{19}$ | PVDF | 6.5 % wt. of rGO | 33 | 8-13 | 11 | [225] |
| CB, $Sr_3YCo_4O_{10+\delta}$ (SYCO) | PVDF | 30 % wt. of CB, 40 % wt. of SYCO | 50.2 | 8.2-18 | 20.1 | [140] |
| CF, G, polypropylene/polyethylene (PP/PE) | PVDF | 40 % wt. of G | 48.5 | 0.03-1.5 | 1731.4 | [162] |
| G, MWCNT, Ni | PVDF | 3 % wt. of MWCNT, 6 % wt. of G, 8 % wt. of Ni | 27.6 | 18-26.5 | 92 | [202] |

*If a series of materials was studied, the results for the highest-performance ones are only presented*

It should be recollected that the amount of applied CNM, the shielding range as well as mechanical properties should be always considered due to the practical implications. However, based on specific SE values in Table 1, one can conclude that the most effective are G based materials [164], and the most prospective/multi-functional materials are those reinforced by CF [162], MWCNT [160], MXene (MX - a new transition-metal carbides and nitrides, having a general formula $M_{n+1}X_nT_x$) [213, 214]. According to Li et al. [213], MXene reinforced PVDF/G nanocomposites have a high SE due to the synergy between extremely high thermal and electric conductivity of the obtained layered material. This efficiency can be additionally increased by a surface modification of mXenes [233].

### 3.2 Piezoelectrics, Energy Harvesting and TENG

#### 3.2.1. Fundamentals of piezoelectricity

Piezoelectricity is the phenomenon of the accumulation of a charge of the opposite sign on the opposite sides of a material subjected to a mechanical force (see Fig. 13). In this way, an electric potential difference is produced – this phenomenon is called the direct piezoelectric effect. The reverse phenomenon – the production of mechanical stress by applying the electric field – is called the converse (or indirect) piezoelectric effect. Analogous/corresponding coupled effects are common (*e.g.*, electrokinetic phenomena); they are described using the non-equilibrium thermodynamics; at a linear range of the transport equations such effects show symmetry (Onsager's Reciprocal Relations) [234].

Piezoelectricity was discovered by Jacques and Pierre Curie in 1880 [235].

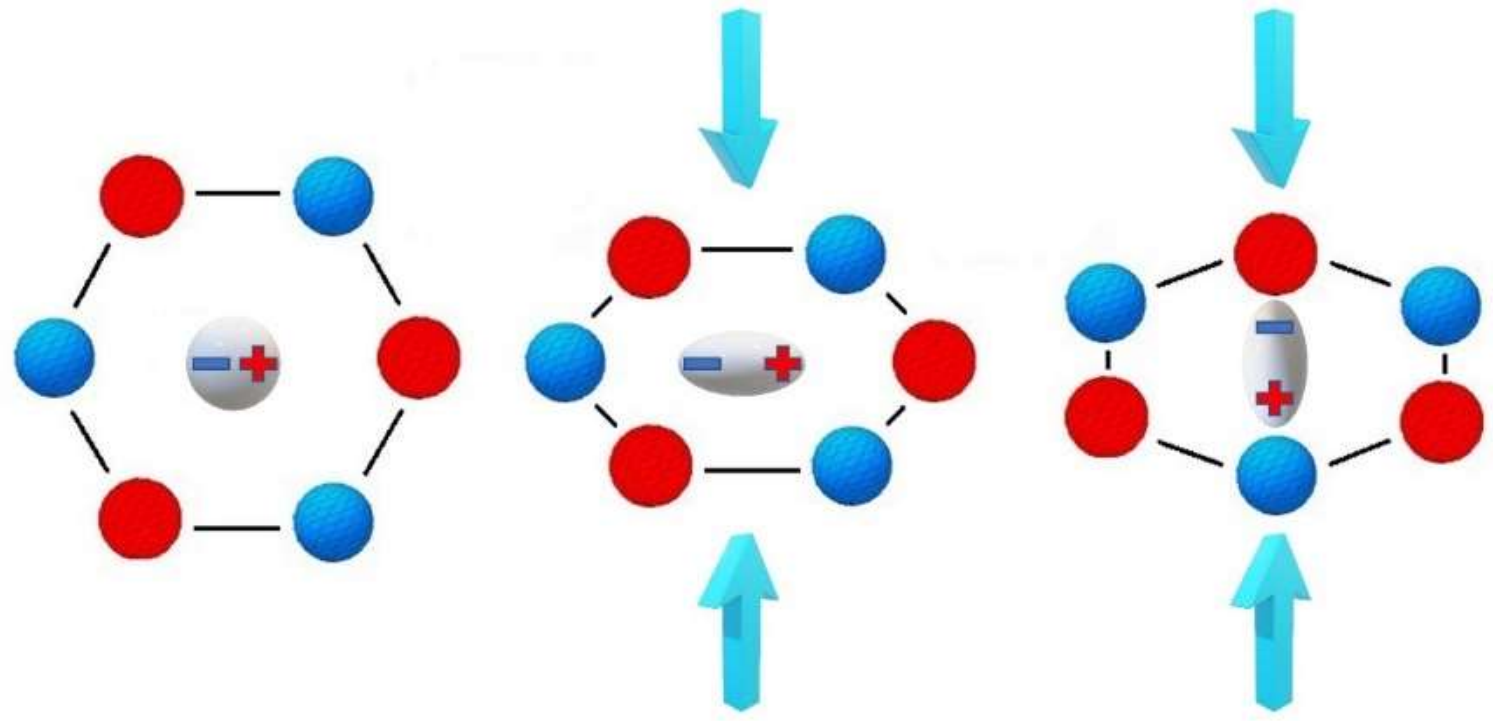

**Figure 13**. Separation of centers of positive and negative charges ($Q_+$, $Q_-$) and generation of electric field under a compressive force; the application of tensile force generates the electric field in opposite direction.

The linear constitutive equations of piezoelectricity can be written in various forms [236-238]. They can be written *e.g.* by expressing the strain, **S**, and electric displacement, **D**, as the sum of two contributions generated by the stress, **T**, and the electric field, **E**.

Graphical relation between these physical quantities is shown in Fig. 14.

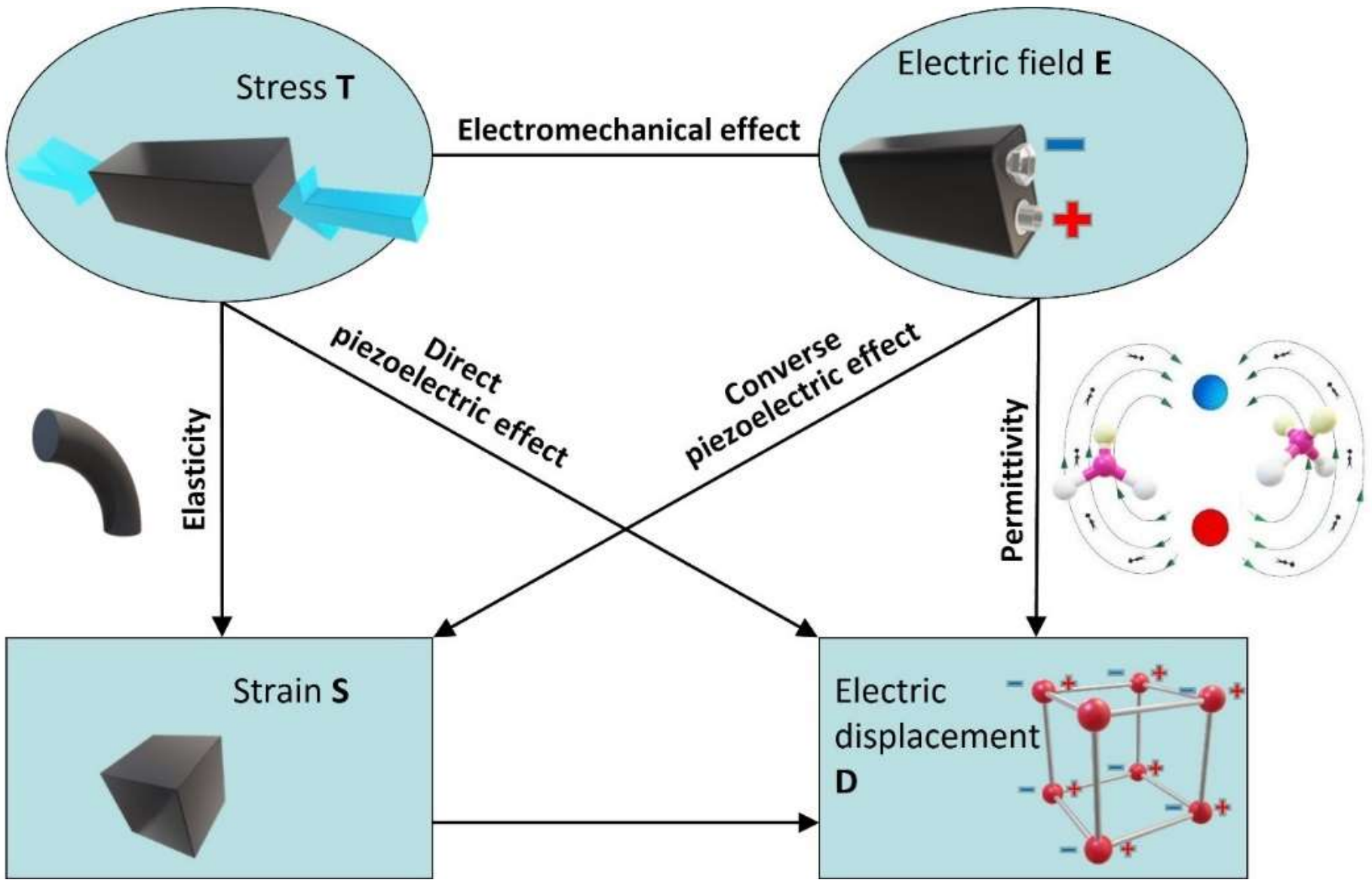

**Figure 14**. Scheme of piezoelectricity phenomena, based on [239]. E and T are in ovals to represent forces, while D and S are in rectangles to point out flows. Copyright Information Springer Science+Business Media Dordrecht 2013.

One of the sets of constitutive equations is given by the following matrix form (Eq. 3,4, S13-S17) [239]:

$$\overline{\boldsymbol{S}} = \underline{\boldsymbol{s}}^{\boldsymbol{E}}\overline{\boldsymbol{T}} + \underline{\boldsymbol{d}}^{\boldsymbol{t}}\overline{\boldsymbol{E}} \quad (3)$$

$$\overline{D} = \underline{d}\,\overline{T} + \underline{\varepsilon}^{t}\overline{E} \quad (4)$$

where

$\overline{T}, \overline{S}$ are the column vectors of stress [N m$^{-2}$], strain [m/m], respectively;

$\overline{E}, \overline{D}$ are the column vectors of electric field [V m$^{-1}$], electric displacement [C m$^{-2}$], respectively;

$\underline{s}^E$ is the matrix of compliance coefficients [$m^2$ $N^{-1}$] (inverse of the matrix of elastic stiffness constants) at the constant electric field;

$\underline{\varepsilon}^T$ is the constant stress permittivity matrix [F $m^{-1}$];

$\underline{d}$ is the matrix of pifdfezoelectric strain coefficients characterizing the direct piezoelectric effect; the transpose of $\underline{d}$, $\underline{d}^t$, characterizes the converse piezoelectric effect; the units of $\underline{d}$ and $\underline{d}^t$ are the same [C $N^{-1}$ = m $V^{-1}$].

In Eqs. (3) and (4), $\overline{S}$ and $\overline{D}$ are expressed as a function of $\overline{T}$and $\overline{E}$. Other relations are also in use: $\overline{T}(\overline{S},\overline{E})$ and $\overline{D}(\overline{S},\overline{E})$, $\overline{S}(\overline{T},\overline{D})$ and $\overline{E}(\overline{T},\overline{D})$, $\overline{T}(\overline{S},\overline{D})$ and $\overline{T}(\overline{S},\overline{D})$. The coefficients of these relations can be related to the coefficients of Eqs.(S13) and (S14) through matrix manipulation [238, 239]. As our discussion will be limited to $d_{ik}$ coefficients, only Eqs. (3 and 4) are shown.

The matrices $\underline{s}^E$, $\underline{d}$, $\underline{\varepsilon}^T$ contain 9 compliance ($s_{ik}^E$), 5 piezelectric ($d_{ik}$) coefficients, and 3 permittivity $\varepsilon_{ii}^T$, respectively; in total – 17 independent coefficients. The most common piezoelectric coefficients are $d_{33}$ and $d_{31}$. The longitudinal coefficient, $d_{33}$, is determined in an experiment where the applied stress and the generated electric displacement are along the same axis, the transverse coefficient, $d_{31}$ – in an experiment where the electric displacement is perpendicular to the applied stress [240], (see Figure 15).

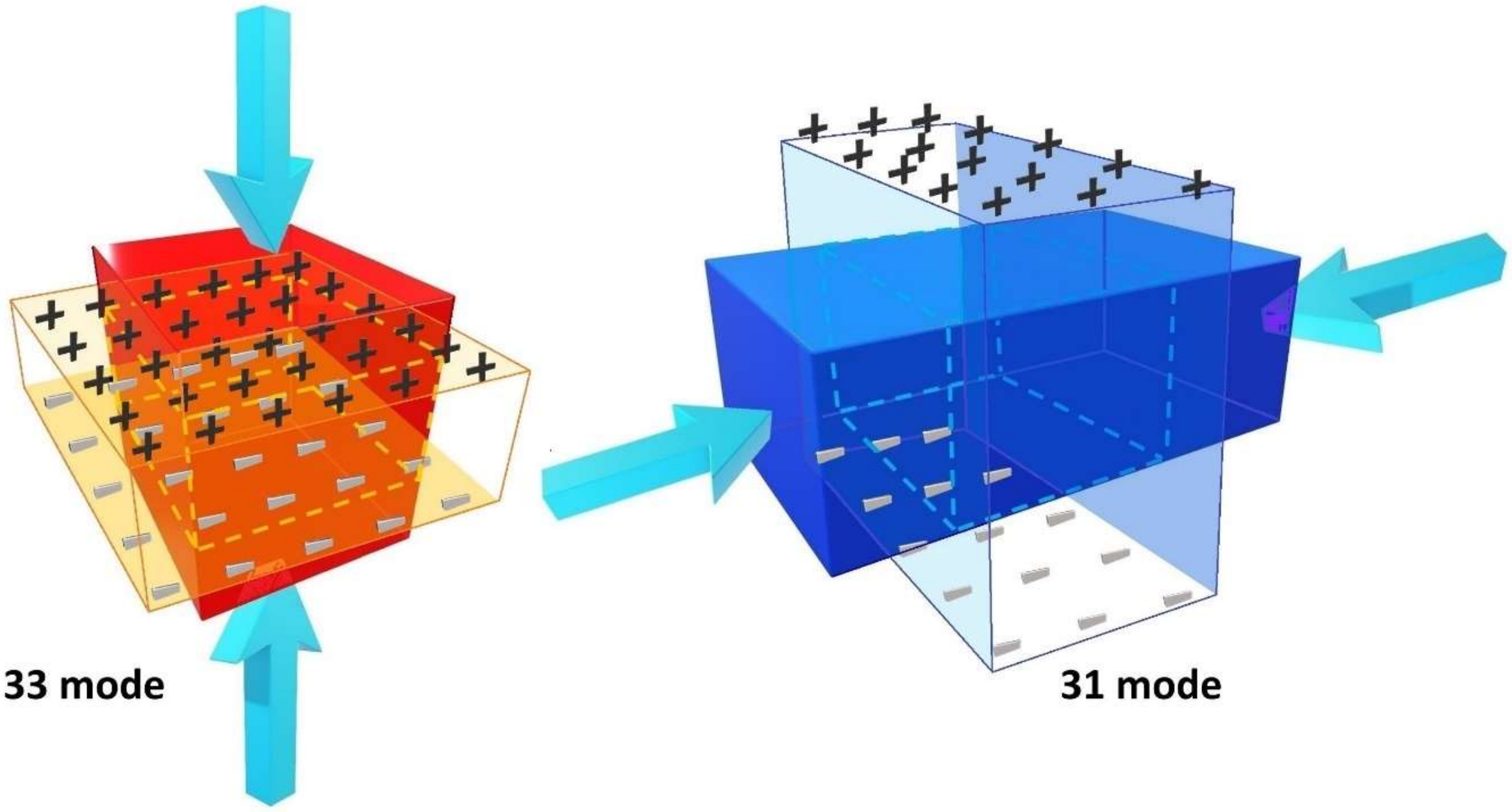

**Figure 15**. Arrangement of mechanical force (F) and electrical displacement for the 33 mode and 31 mode determination.

### 3.2.2. Practical applications of PVDF-CNMs in various generators

The piezoelectric effect can be used for the conversion of mechanical into electric energy in the devices called nanogenerators (NGs) [130]. For the same kind of energy conversion, the triboelectric effect can be used, whereas pyroelectricity allows for conversion of thermal energy into electric energy. The first piezoelectric (PENG), triboelectric (TENG), and pyroelectric nanogenerators were developed by Wang and co-workers [241-243]. The purpose of NGs is a power supply for small, portable electronics, and sensors which are in abundant amounts used by us, and their number is continuously growing. The development of such nanogenerators would significantly reduce the consumption of harmful electrochemical batteries. Recently also fiber-based [244] and hybrid nanogenerators have been intensively studies [245].

In addition to piezoelectric properties and low prices, materials intended for PENGs should be flexible and easy to form. Hence, polymers seem to be the first choice; among them, PVDF and its copolymers show the highest piezoelectricity [246], although not as high-performance as piezolectric, rigid ceramics. One method of improving the properties of PVDF is the introduction of various fillers [247]. As it can be found in [248, 249], the highest power density of PNG based on PVDF was 465 μW $cm^{-2}$ for Vitamin B2/PVDF film, for pure PVDF – only 0.31 μW $cm^{-2}$.

To improve the efficiency of NGs, the hybrid nanogenerators were also investigated with positive results; *e.g.,* the hybrid piezo–triboelectric device [250] generates more power than a nanogenerator working on a single-energy conversion mechanism.

Piezoelectric generators – converting mechanical (see an excellent practical application in motorcycle spring [251]) and radiative [252] energy into electricity – can be widely applied for example in different types of sensors, like for example: movement (finger [64, 88], breathing [253], hand [253], wrist [253], low frequency accelerations in seismic [76]), human health measurement systems, wearable energy harvesters [93], (self-charging) powering nanosystems, piezoresistive strain sensors [24, 254] (applied for example to check the integrity of buildings [255]), sensors for smart home monitoring [256], sensors for monitoring shoe-ball interaction in soccer [257] etc., (Fig. 15 [256]) showing different applications of biomechanical energy harvesters.

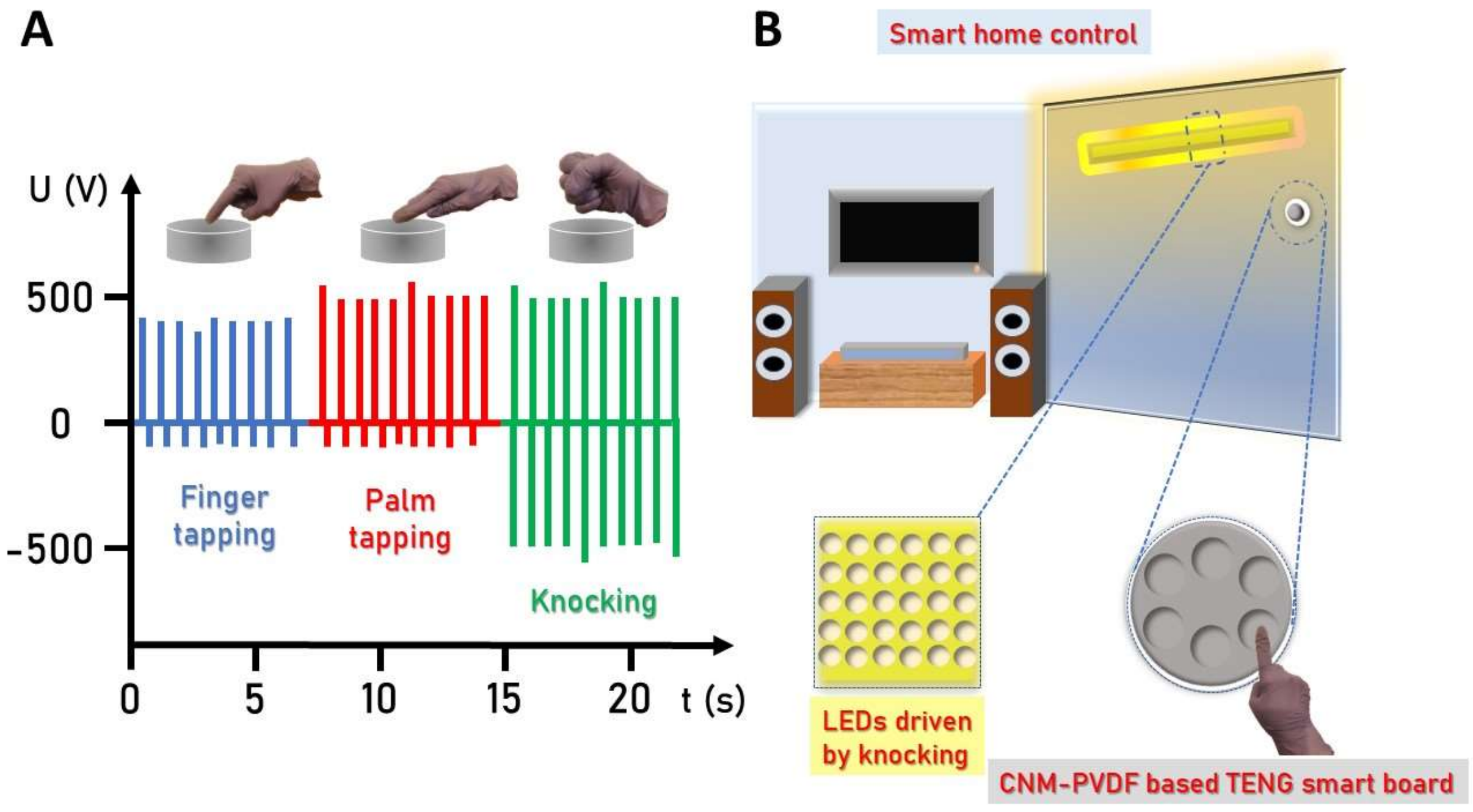

**Figure 16**. Output signals of G-TENGS for fingers activity [256] (A) and smart home monitoring (B). Adapted with permission from [256]. Copyright 2021 Elsevier.

Kadir and Gayen [258] indicated that there was a lot of materials displaying high piezoelectric coefficient ($d_{33}$) (for PVDF it depends on the method of fabrication [259], and in this way being potential candidates for energy harvesting. However, these materials usually are not highly resistant to the external force, thus the application of PVDF-CNMs solids becomes crucial [260]. Also liquid-solid TENG for hydropower energy harvesting are very important [261]. Application of functionalized G with PVDF can provide remarkable solutions particularly considering a high-power density.

Importantly, addition of G improves PVDF piezoelectric properties that can be optimized [262]. In the case of CNMs addition, the dispersion and interfacial adhesion between the components of a composite play a crucial role. Du et al. [78] showed that the application of electrochemically exfoliated G led to a relatively homogeneous G distribution in PVDF matrix, and it is a simple method of a simultaneous increase of both dielectric constant and tensile strength.. Yang et al. [263] proposed to apply Friedel-Crafts reaction to produce G-grafted PVDF. This material was used together with PVDF in electrospinning hot pressing and the obtained composites revealed an enhanced dielectric performance than the neat PVDF. Poly(vinyl alcohol) (PA)-PVDF blends can improve piezoelectric coefficient by ca. 150% comparing to the pure PVDF [264]. The enhancement of PVDF piezoelectric properties can be achieved, for example, by simultaneous addition of G, $MnO_2$ and MWCNTs [265]. In this case, tuning of the properties is possible by changing the contents of $MnO_2$. Addition of rGO [106] can lead to the formation of hybrid films applied as piezoelectric energy harvesters. A similar role can play CNTs applied with $xBa_{0.85}Ca_{0.15}Ti_{0.9}Zr_{0.1}O_3$ in flexible piezoelectric energy harvesters from wind [110]. An improved dispersion of GO in PVDF was reported by Song et al. [84] who applied methyl methacrylate-*co*-glycidyl methacrylate as a compatibilizer during a solution blend process. This approach caused a significant increase in the dielectric properties of a composite by preventing GO nanoparticles from aggregation at the simultaneous improvement of mechanical and thermal properties (see Tabs. 3 and 4 in [84]). To improve the piezoelectric properties of PVDF the amount of β phase should be maximized [266] what can be achieved by using the above-mentioned methods separately and/or combining them [138]. Improving piezoelectric properties of PVDF can be realized after addition of: conducting CB [267, 268], CNTs [251, 269, 270], CNTs and G [134, 271], GO [64, 85, 122, 272-275], CF [276], GO and polydopamine [229] G and GO with halloysite nanotubes [277], rGO [63, 278], fluorinated GO [61], ND [22], QDs [113, 114, 253], rGO and MXene [279, 280], polyaniline functionalized rGO [252], G and $CaCu_3Ti_4O_{12}$ [281], core-shell nanocomposites, *i.e.*, $TiO_2$@C@$SiO_2$ (containing CVD obtained carbon) [282], carbon nanofibers [24] (for conductive thin films), MOF derived Co-enriched carbons [256], CB and $SiO_2$ [283], CB and few layered G [266] etc. The increasing dielectric constant with GO content was explained by Muduli et al. [273] by using the microcapacitor theory (the explanation of dielectric constant decrease after exceeding the threshold GO concentration was provided as the formation of conductive current nanopaths). Considering GO and its different modified and reduced forms, it was proved [284] that addition of reduced forms to PVDF provided higher dielectric constant and lower dielectric loss than the addition of the oxidized ones [285]. Also application of CDs improve the dielectric constant of PVDF [286]. Yang et al. [287] presented the method of dielectric constant adjusting by changing carbon shell thickness in $TiO_2$@Carbon nanowires (thickness 4 - 40 nm; obtained from CVD, leading to even 241 times higher dielectric constant than for PVDF). A similar strategy was proposed by Chen et al. [288] who used graphitic carbon from glucose as a shell for Ag in Ag@C nanocables. The material was proposed as an alternative for ceramics and exhibite a strong adhesion to PVDF, leading the final composite exhibiting low dielectric loss and a high dielectric constant. Tuning of dielectric constant method was also proposed by Patodia et al. [289] who used GO/rGO/MWNT/PVDF composites. Improvement of dielectric constant (11 times larger than for pure PVDF), having crucial role in energy storage flexible piezoelectric films – (Fig. 17A [290]) based on PVDF/PZT@105 (lead zirconate titanate)/GO application materials, was reported by the incorporation of $BiFeO_3$ and CB [291]. The suppressed dielectric loss was additionally reported. Similar properties were observed after introduction of polybenzimidazole functionalized G [292] of prospective applications in robotics and energy harvesting as reported by Zhang et al. [290].

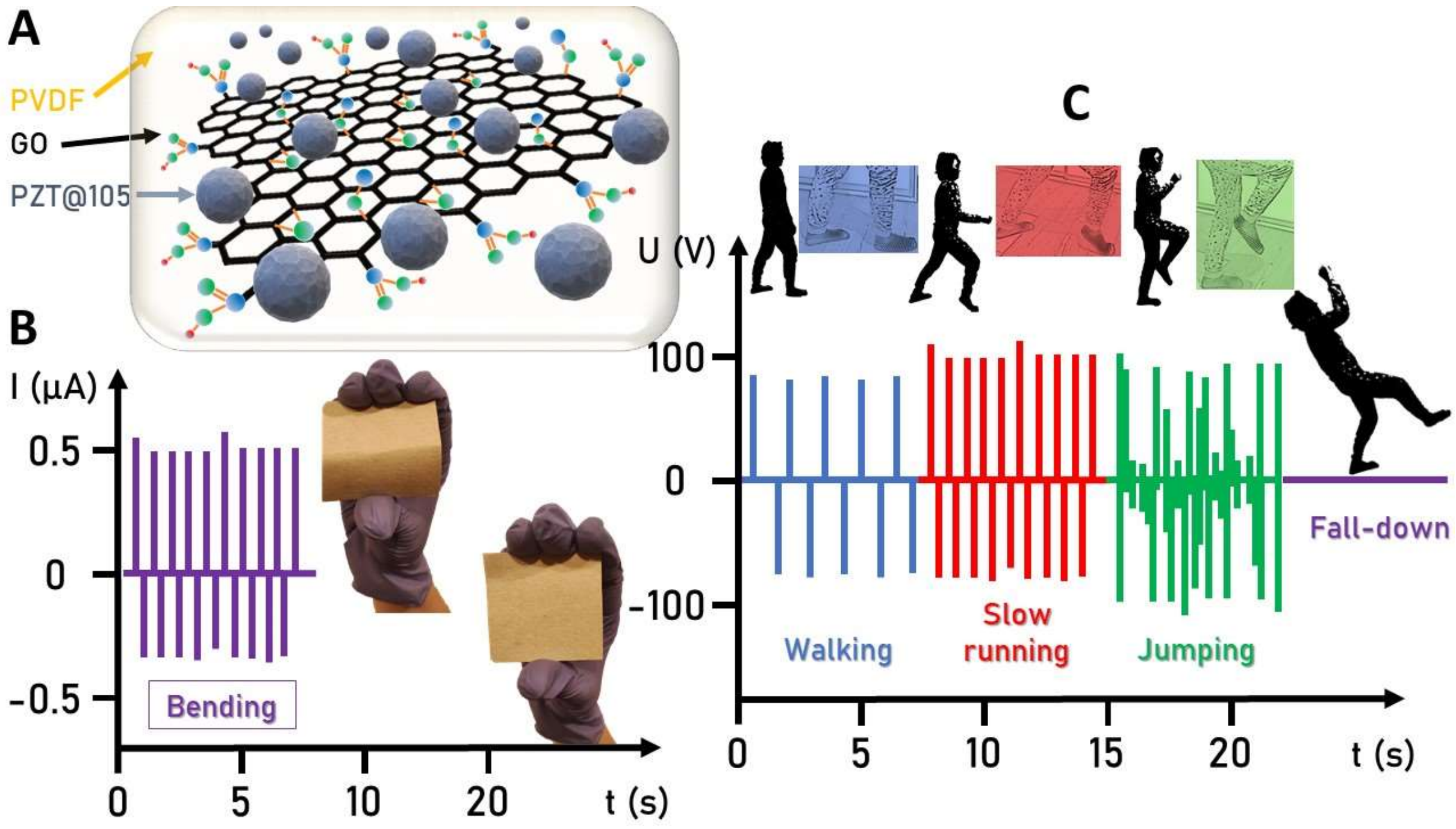

**Figure 17**. Flexible piezoelectric films [290], based on PVDF/PZT@105 (lead zirconate titanate)/GO (A), application of rGO/PVDF-TrFE WNGs for finger bending [93] (B), output signal of a Co-NPC/PVDF NF-based TENG at different stages of human motion for application in self – powered human motion monitoring [256] (C). Adapted with permission from [256]. Copyright 2021 Elsevier. Adapted with permission from [290]. Copyright 2022 Elsevier.

Liu et al. [293] reported the G nanoplatelets/polystyrene/$BaTiO_3$/PVDF blend system of an improved dielectric constant. They showed that the dielectric constant of the blend could be successfully calculated using the proposed mixed dielectric model. $BaTiO_3$ is often applied to increase the piezoelectric properties of polymers because it has high piezoelectric coefficient and is relatively inexpensive [108, 127]. It was applied in electrospinning together with G nanosheets by Shi et al. [108] for the production of PENGs, and with rGO [294] for production of piezoelectric fibers. Abu Bakar et al. [295] applied $BaTiO_3$ and GQDs to prepare PVDF based flexible NG in which poly(styrenesulfonate) replaced ITO bottom electrode. Sridhar et al. [125] reported the NG-based on PVDF/Cloisite-30B and PVDF/$BaTiO_3$/G. Xu et al. [296] studied PVDF/(γ-oxo-pyrenebutyric acid)-modified graphene and $BaTiO_3$ reporting the improvement of dielectric constant and this effect was not only caused by high dielectric constant of $BaTiO_3$ but also by polarization of the nanofillers. Zhou et al. [118] reported $BaTiO_3$@carbon nanoparticle-enabled P(VDF-TrFE) composites for the application in PENGs (GO - P(VDF-TrFE) composite fibers were also studied [297]). Also, rGO-PVDF-TrFE materials emerged as very interesting wearable energy generators (WNGs) [93] – Fig. 17C [93] - showing different application for energy generation from human motions like jogging, walking etc. PVDF-TrFE/SWCNT composites were obtained by Shepelin et al. [298]. Authors, instead of pooling, proposed, for the production of flexible, wearable and recyclable PENGs, extrusion-printing process causing shear-induced alignments of the dipoles. This is important study showing no changes in the β-phase content after introduction of SWCNTs, however, a drastic increase in the $d_{33}$ coefficient. Authors also observed the creation of γ-phase at higher SWCNT concentrations. MDS-supported study delivered conclusions about an almost parallel orientation of PVDF-TrFE dipoles with respect to SWCNT, causing a high volumetric power density of composites. Considering the simplest CNTs – containing NGs Lee and Lim [299] proposed to place the flexible material between two electrodes. CNTs increased the content of β-phase but the electrodes increased additionally the mobility of charges causing higher performance of nanogenerator.

Li et al. [77] reported $BaTiO_3$-rGO piezoelectric films for the monitoring of various human motions in real time - see Fig. 17C (and Fig. 9 in [77] collecting the process of output voltage during monitoring). The comparison with materials without rGO leads to the conclusion about the remarkable improvement of the output voltage and current, and this is mainly attributed to the improved β-phase creation (see above). Increase in the value of $d_{33}$ coefficient (as well as the largest content of β phase) was reported [300] if PVDF/rGO fibers are obtained using dimethylacetamide as a solvent. Also, nitrogen-doped CQDs increase this coefficient value [301]. Polyamide-11/ $BaTiO_3$/G nanocomposite (obtained by solid-state shear milling, coating and laser sintering) was shown to be flexible and having enhanced dielectric and piezoelectric properties (*i.e.*, high $d_{33}$ coefficient) [302]. Yang et al. [103] reported the method of PVDF/$BaTiO_3$@polypyrrole dielectric composites with a low dielectric loss. Lakshmi et al. [99] collected the values of dielectric constants of different polymer nanocomposites (see Tables1 and 4 in [99]) showing that the new PVDF/$BaTiO_3$/G-decorated with GQDs nanocomposites (20% of $BaTiO_3$ and 3% of CNMs) had very large dielectric constant, and the rise in thermal stability was additionally observed. Improvement in the dielectric properties is caused by an increased charge accumulation due to the interfacial polarization and the creation of micropacitors. Silakew and Thongbai [303] published the study devoted to the influence of $BaTiO_3$ crystal size and the volume fraction of SWCNH on dielectric and electric properties of PVDF/$BaTiO_3$/MWCNT composites. Application of $TiO_2$ nanotubes, $SrTiO_3$, rGO and PVDF in piezoelectric compounds was reported by Ponnamma et al. [304] and the composite material had a remarkably higher piezoelectric constant in comparison to the neat polymer. Also $TiO_2$-$Fe_3O_4$-MWCNT nanofiber mats, with promoted β phase creation and piezoelectric properties were reported [120] as well as MWCNTs/$Fe_3O_4$ dispersed into PVDF/polystyrene/polyethylene blend [305]. Moharana and Mahaling [306] proposed the $Nb_2O_5$ reinforced PVDF - GO films with enhanced dielectric properties and low dielectric loss for the energy storage applications.

An interesting study was published by Baek and Yang [307] who reported the application of $CoFe_2O_4$-intercalated GO as –2D magnetostriction (*i.e.* changing shape or dimension during magnetization) PVDF fillers.

Zhang et al. [308] reported the addition of IL and G (2 wt.% of each) to the PVDF improving dielectric constant (by the improvement of β-phase creation - see above). IL also increased G dispersion and plasticity. G nanosheets modified with heptadeca-fluorodecyltrimethoxysilane and γ-methacryloxypropyltrimethoxysilane as coupling agents improved PVDF dielectric constant (by improvement of α to β phase transformation). The dielectric constant was also improved in G/PVDF/polylactic acid nanocomposites [309] also *via* enhanced β phase content. Li et al. [310] reported the improvement of dielectric PVDF properties after addition of hydrophilic fluoro-functionalized GO. It is interesting that crystal PVDF structure did not change after fluoro-functionalized GO addition. Shetty et al. [311] improved the β phase creation in PVDF using addition of benzoic acid to GO, talc nanosheets and electrospinning process. Electroactive mechanically resistant nanofabrics for the energy harvesting were obtained and high piezoelectric response was caused by the synergic interactions between surface groups of GO (and π-electrons), oxygen-containing groups of talc and PVDF (see Fig. 11 in [311]). An improved β-phase content, with the increase in piezoelectric and mechanical properties was observed for PVDF/CNT foams obtained by application of supercritical $CO_2$ foaming process. In this case, ball-milling was used to increase the MWCNT dispersibility [269]. A series of CNT-containing foams with high dielectric constants and low dielectric loss was reported by Zhao et al. [271]. Similar systems but containing additionally $SiO_2$ (MWCNT@$SiO_2$/PVDF) showed enhanced β-phase content, and high dielectric, piezo, ferroelectric and EMI shielding properties [312]. Xu et al. [313] reported a new PVDF/Nafion/GQD dynamic piezoelectric generator (Fig. 1 in [314]) having very high mechanical-to-electrical conversion. Also high dielectric constant and low dielectric losses were reported by Li et al. [315] who introduced rGO and $Fe_3O_4$ (rGO@$Fe_3O_4$) into the PVDF matrix. An interesting method of piezoelectric properties increase was reported by Ramasamy et al. [316] who functionalized GO with phenyl-isocyanate. The nanofibers obtained by electrospinning had an additionally enhanced β-phase content and MP. However, the most interesting property was very high acoustic sensitivity (Fig. 10 in [316]) thus the material can be applied as a piezoelectric acoustic sensor. Kadir and Gayen [258] reported the ZnO/GO based PVDF films with high amount of β phase and its application for mechanical energy harvesting (Fig. 28B, and Fig. 4 in [258] and the application as energy harvesters during finger tapping, stretching and bending). Similar results were reported for G/ZnO nanocomposites [124], and ZnO/CNT [317]. PEG-modified Li-ZnO together with MWCNTs were used for the fabrication of piezoelectric PVDF NGs [318]. In this case PEG enhances the β phase content. Barstugan et al. [319] presented interesting results on the increasing the β phase content in PVDF-G systems by the addition of polybenzoxazoles with four different chemical compositions. In fact, differences in the amount of the β-phase were observed for different polybenzoxazoles compounds leading to increasing piezoelectric properties. Choi et al. [69] reported an application of QDs (5%) in TENG showing that the maximum output power is obtained after using electrospinning.

Improvements in the contents of β-phase CNMs/PVDF and/or reinforced by other polymers (for example QDs/PVDF-*co*-HFP [253]) materials, can be applied for the construction of piezoelectric nanogenerators. Also CDs, improving the β-phase contents in the electrospinned mats, are used for construction of piezoresponse materials [76, 301]. Badatya et al. [137] proposed the PVDF/CNT hydrophobic foam application as PENG. The presence of CNTs assured a high β-phase content while the hydrophobicity of the composite guaranteed working at high humidity. Xu et al. [320] reported optimal parameters of the spin coating during the manufacturing of the PVDF/TrFE/SWCNT films with the optimal energy harvesting properties. Han et al. [321] pointed out the importance of the shape of PVDF/TrFE/CNTs PENGs showing that biomimetically inspired nanohelix was the most promising shape due to the geometrical stress confinement. The enhanced β-phase content was assured by the presence of CNTs. Santos et al. [322] performed comparative analysis of PVDF/MWCNT composites obtained for different MWCNT loadings as by three different methods (solution mixing, melt mixing and electrospinning). It was concluded that the solution mixing led to the most conducting materials, while electrospinning led to the materials possessing the lowest dielectric permittivity.

Also increased β-phase content was reported for PVDF/CNTs decorated with Cu-Ni nanoalloy [323]. In this case, the obtained NG possessed high dielectric constant and tensile strength. Increased polarization effects were attributed to the appearance of more ordered interactions between nanotubes and F atoms of β-PVDF (see Fig. 11 in [323]). The role of CNTs and potassium sodium niobate (KNN in PVDF)/KNN/CNT NG was studied [314]. KNN drastically improved the piezoelectric response. Among well-known properties (increased β-phase content, enhanced electroconductivity etc.), it was concluded that CNTs accelerated the stretching of PVDF during electrospinning process [324]. PVDF-KNN-CNT films can be applied as energy harvesters.

Wu et al. [74] studied an addition of low amount of G into PVDF-TrFE observing an improvement of the piezoelectric properties. Based on scanning electron microscopy (SEM) observations, the authors proved an important role of stretching in the improvement of films crystallinity and piezoelectric properties by rearrangement of polymer chains (see Fig. 6 in [74]). Thin G P(VDF-TrFE) films were studied by Zhao et al. [325]. Some very important aspects were discussed, among them the enhancement of $d_{33}$ to obtain the material for flexible electronics. It was shown that if the TrFE content was larger than 20 %, P(VDF-TrFE) crystallize only as the β phase. Authors also paid attention to the crystallization temperature strongly affecting the coercive voltage and $d_{33}$value. For instance, Fortunato et al. [139] showed that the $d_{33}$ value depended on the method of G preparation and introduction into PVDF matrix. This parameter was larger for the material containing smaller amount of the β-phase due to higher lengths of polymer chain. Also the addition of G (GO) aerogel [101] to PVDF improved piezoelectric properties *via* improvement of the β-phase content, and the material was successfully applied in the field of flexible PENGs (see Fig. 5b in [101]).

Interesting groups of materials were reported by Tu at al. [326] who applied temperature-induced PEG phase transition for tuning the dielectric constant of a PVDF/PEG/rGO material – see Fig. 18A [326]). Recently Taha et al. [327] reported the influence of electron beam on dielectric constant, dielectric loss and conductivity of PVDF/PZT/CNT composites.

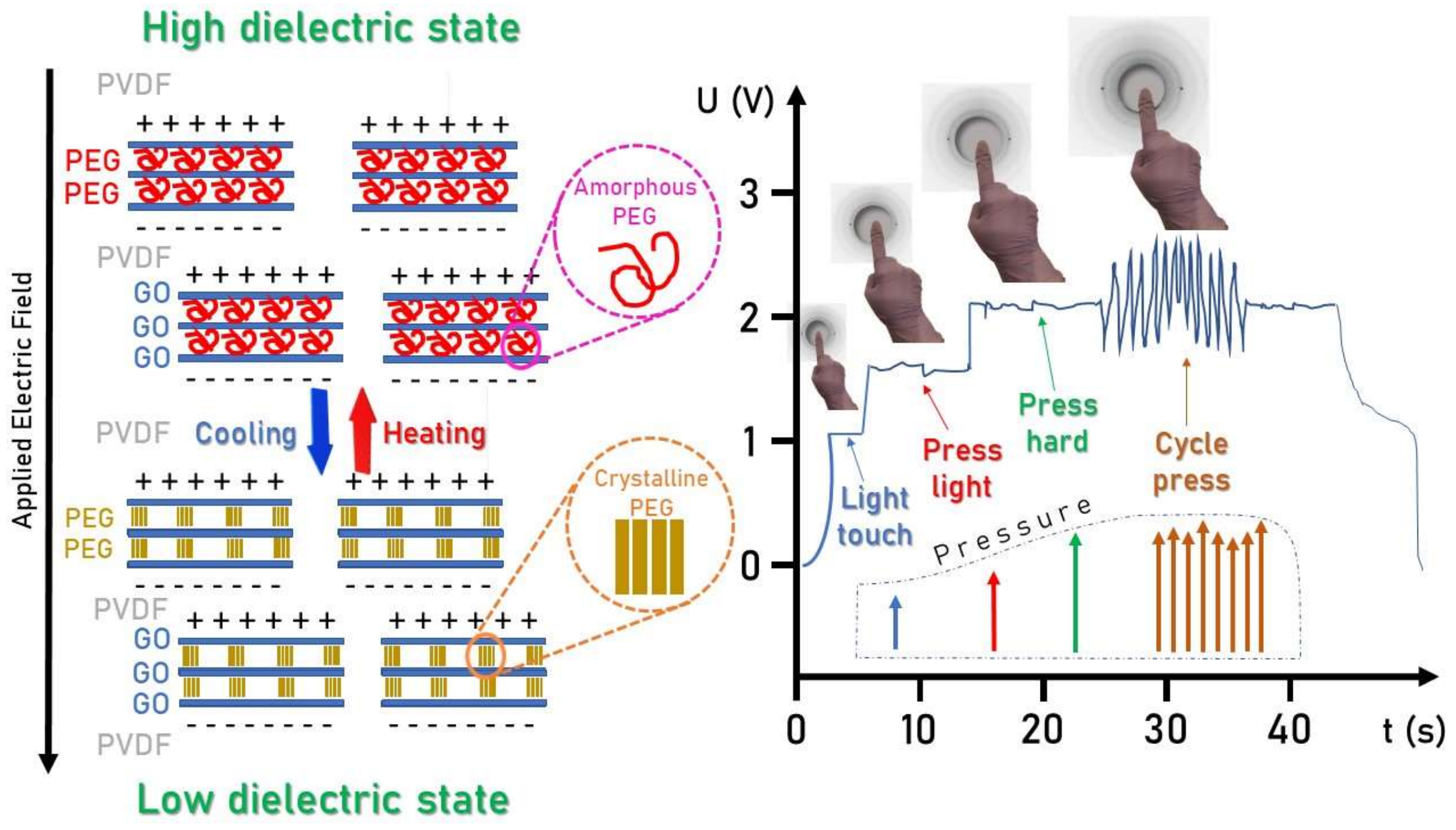

**Figure 18**. Thermal evolution of PVDF/PEG/GO nanocomposites for tuning the dielectric constant (A) and electronic skin application (B). Adapted with permission from [326]). Copyright 2019 Elsevier.

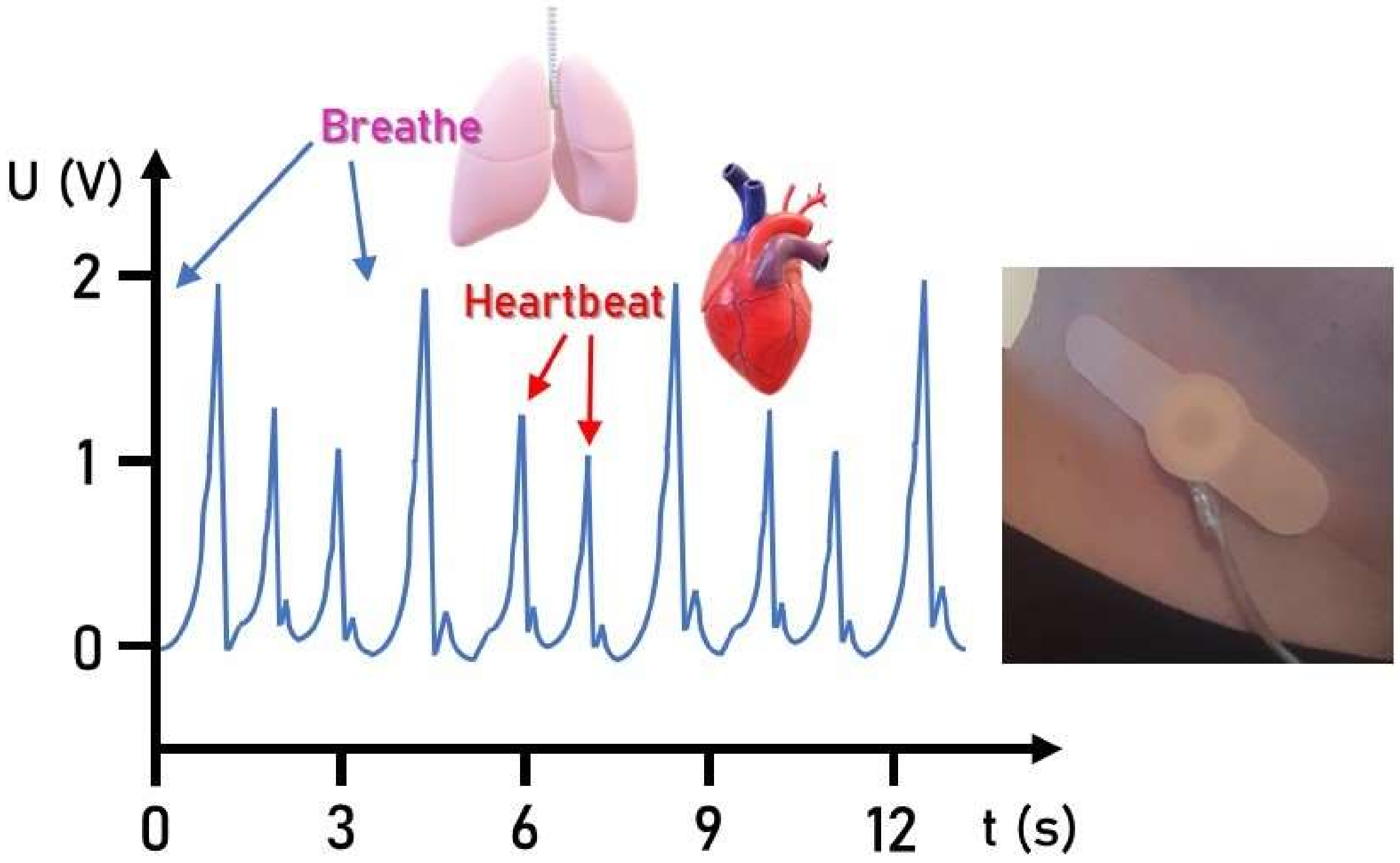

**Figure 19**. Electronic skin on the carotid artery for monitoring of human activity. Adapted with permission from [328]. Copyright 2019 Elsevier.

Also, there is a growing interest in the PVDF-based nanocomposites having negative dielectric permittivity [329, 330].

Future applications of PVDF-CNMs in electronic skin emerge as very promising.  This is not only for the improvement of communication between human and machine, but also for the monitoring of human body (for example breathing, pulse and generally, all-muscles activity) [328]. As it was pointed out by Wang et al. [328], there were four types of electronic skin and among them piezoelectric skin can be very thin (4 mm) having, at the same time, decent flexibility – (see Figs. 19 and 30 in [328]). It can be also mentioned that carbyne-enriched films obtained from PVDF are very promising materials in the field of energy harvesting [331, 332].

The review on the application of G in energy harvesting materials has been published recently by Pusty and Shirage [130]. Zhao et al. [271] tabulated dielectric properties of various polymer – CNMs nanocomposites (see. Table 1 in [271]). Hatta and Mohamed [332] provided a review on the application of G in tribonanogenerators (see Fig. 1 in [332] collecting the application of nanogenerators, and chapter 4.1. reporting the application of G). Yu et al. [333] described the MWCNT-based NG combing tribo- and piezoelectric effects. This flexible and durable NG for sound energy harvesting, based on carbon-containing aerogel, converted sound into mechanical energy – see Fig. 20 [333].

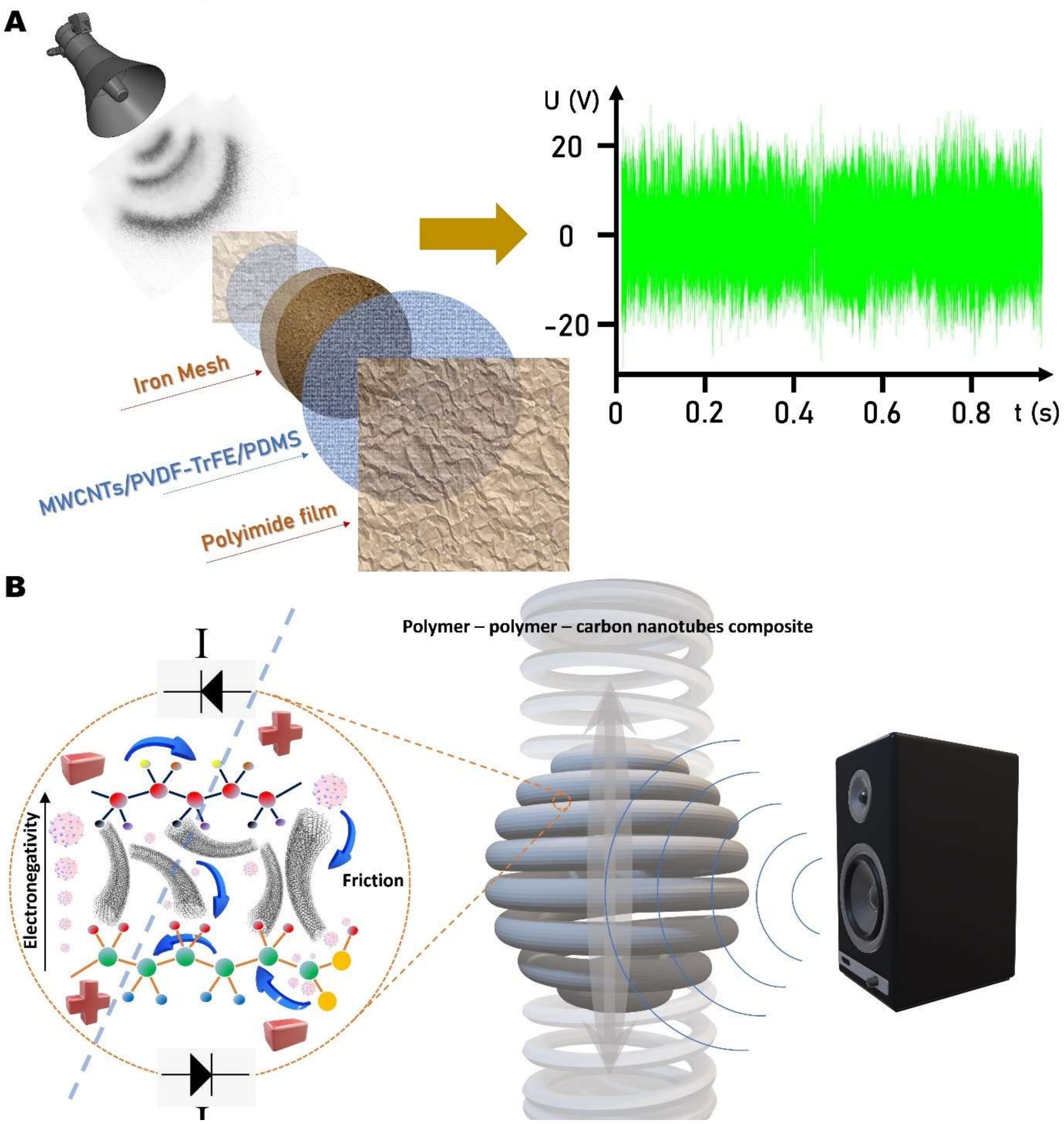

**Figure 20**. The schematic representation of sound conversion into electricity (A), and the structure (B) of PENG. Adapted with permission from [333]. Copyright 2022 Elsevier.

As it can be seen from Fig. 20A, the sound penetrates porous structure of MWCNTs/PVDF-TrFe/PDMS aerogel and, by propagation of the waves, they are converted into mechanical energy, easily convertable into an electric signal. Fig. 20B shows the schematic representation of PENG. The incoming acoustic waves induce structural changes in the composite and the surface charge transfer. The differences in electronegativity between PDMS and PVDF-TrFe cause the electron flow to the more electronegative PDMS. The process is reversible.

It was also proved [334] that PVDF/polyurethane/MWCNT foam absorbed sound in a wide (400 – 1400 Hz) frequency range while conductive MWCNTs increase the interfacial and piezoelectric damping effects.

Summing up, this paragraph presents probably the most interesting, advanced and futuristic applications of PVDF+CNMs system. Due to increasing energy consuming as well as green energy applications the examples presented above confirm important role materials chemistry plays in sustainability.

### 3.3. Sensors, Actuators, Transducers, Photodetectors

The extensive applicability of the PVDF as sensors and actuators has been presented. The requirement addressed to the sensor devices is measurement of a subtle change in a very short period of time and a high sensitivity. The device must sense minor effects to calculate the accurate and subtle sensitivity [335-337]. Owing to the drawbacks of piezoelectric ceramics, *e.g.*, lead zirconate titanate (PZT) which possesses noticeable sensor properties, the small flexibility [338] and significant production costs limit the application of such materials. For that reason, more and more attention were paid to the tactile polymers considered as the most prospective sensors. This application is particularly explored in PVDF owing to the possibility of converting the mechanical deformation into electrical signals. Tactile sensors are primarily designed to detect displacements such as body motion in order to assess physical interactions, and they are commonly employed in robots and touch screen devices to provide a response to the human touch [336, 339, 340].

Tactile sensor is a material more sensitive and exposed to different deformations, including stretching and bending [341, 342]. For this reason, the flexibility of the material is essential from the desig perspective. The applications of the PVDF in sensing is not limited to the tactile ones. There are broadly utilized as nanosensors for humidity [343-345], force [344, 346-348], weather conditions [349], and pressure measurements [350-352]. In the case of the latter one, their significant role has been appreciated in the biomedicine [350, 353, 354], health monitoring [355-358], and acoustics [359-361].

Considering the pressure sensors application, the most important parameter is the level of β-phase in PVDF polymer solution [62]. The high content of β-phase will ensure, high flexibility of the PVDF material. For that purpose, the modification focusing on doping PVDF by different nanofillers is in the spotlight. Thanks to such procedure, it is possible to enhance the sensitivity and efficiency [362, 363]. Merlini and co-workers presented that coating an electron spun PVDF samples with 50 wt.% of polypyrrole exhibit extraordinary pressure sensing features [363].

Many uses of PVDF-based actuators are widely found in acoustic microphones (Fig. 21) [361], robotics, and artificial tissues, paving the door for further research into numerous applications of PVDF as a lightweight pressure sensor. Numerous studies have found that PVDF nanocomposites enhanced the piezoelectric response efficiency for sensors and actuators when compared to pure PVDF [364].

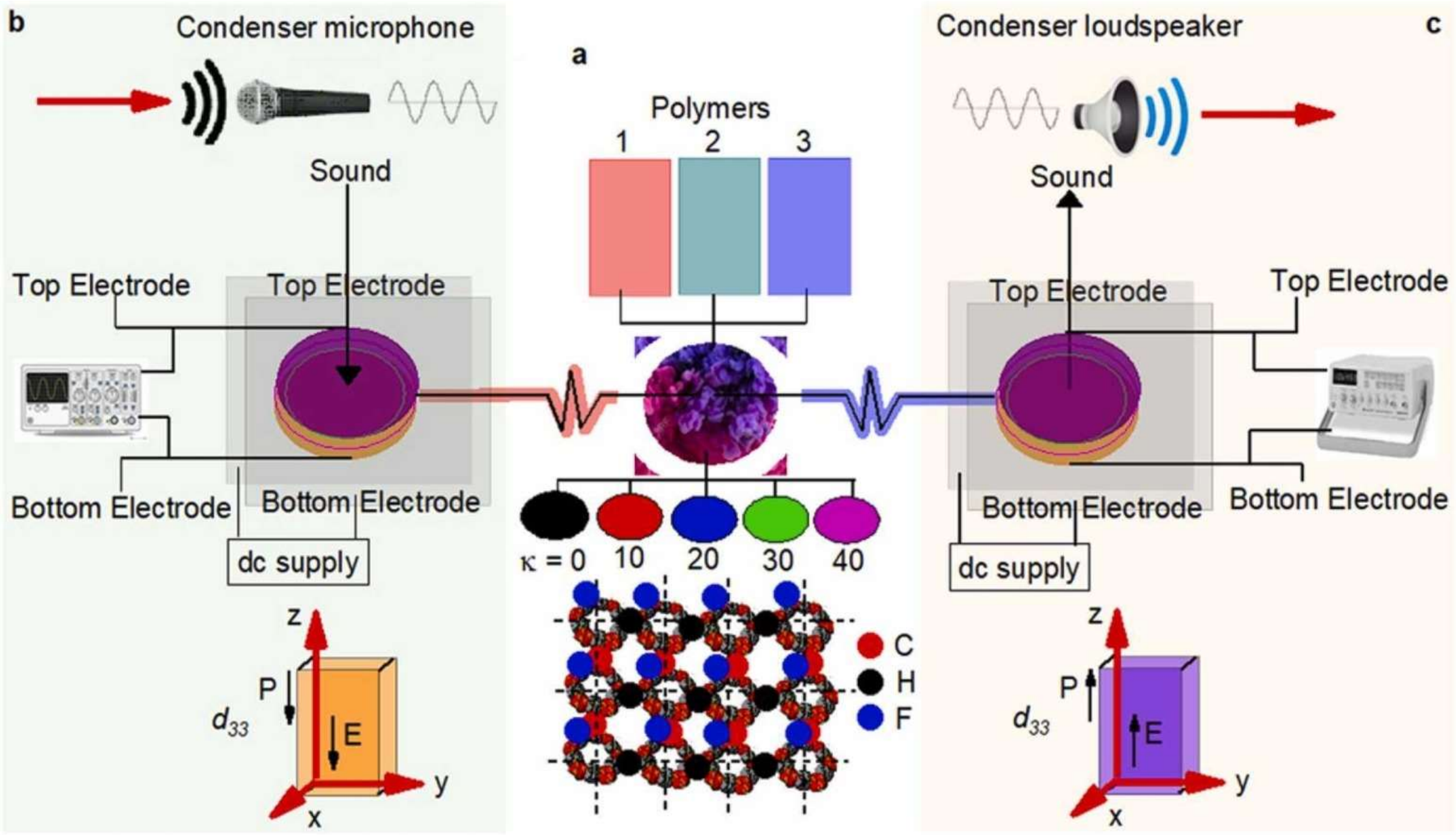

**Figure 21**. (a) Schematic description of triple blend and its chain reaction. (b) Schematic description of Condenser Microphone Unit. (c) Schematic description of Condenser Loudspeaker Unit. Adapted with permission from [361]. Copyright 2022 Elsevier.

Although the neat PVDF possesses the superior features to be applied in sensors and actuators, their composites with CNMs are much more effective [365]. Not only the type of carbon-based filler but also its morphology and orientation could tune the sensing features. Miao et al. pointed out that the thicker porous wall formed by the entanglement of graphene sheets improved the mechanical strength of the sensor [364]. Sharafkhani and co-workers [366] presented that orientation of CNT could change the sensor efficacy significantly. According to the presented research, the key requirments was to generate materials with CNTs orientated along the longitudinal axis of the PVDF nanofibers [367]. Such an achievement was possible by tuning electrospinning conditions and generating the level of β-phase ≥ 90%. By increasing the weight percentage of the well-oriented CNTs from 0 to 1.25, the α- to β-phase transformation was complete, and the output voltage rose by 70% from 4 to 6.8 V, respectively. Because of the abrupt rise in dielectric constant and dielectric loss graphs with increased CNT content, sensing performance immediately deteriorated. The evolution of correct interfacial contacts between the uniaxially oriented CNTs and the PVDF chains, resulting in a perfect coaxial structure, resulted in a 12%-increase in crystallinity, a 100%-increase in elastic modulus, and a 100%-increase in dielectric constant. The piezoelectric sensor also shown outstanding stability over a lengthy duration of 1800 cycles.

Choi and co-workers produced the PVDF-SWCNT composite fiber for the frequency sensing [368]. The active PVDF layer of the piezoelectric device's active layer, PVDF, was synthesized by electrospinning, while the electrodes were made by dip coating in a prepared SWCNT dispersion. A rapid Fourier transform was used to match the output voltage of the external sound with the input frequency, and the frequency matching was achieved even with a mechanical stimulation. The piezoelectric effect and frequency domain peak began to drop dramatically at 300 Hz in a high frequency test, and the limit of the piezoelectric effect and sensing was detected from 800 Hz. The findings of this work offer a strategy for producing flexible piezoelectric-fiber frequency sensors for the acoustic sensor systems based on piezoelectric devices [368].

An interesting case was also to implement carbon nano onions as an example of CNM that increased the efficiency of sensing. Khazani et al. [369] performed piezoelectric fibers-based PVDF-ZnS-carbon nano onions and tested as a flexible nanogenerator for energy harvesting and self-powered pressure sensing. Because of their intrinsic piezoelectric capacity, ZnS nanorods contributed to the alignment of the electric dipoles in PVDF and improve the overall piezoelectric characteristics of the composite nanofibers. Carbon nano onions improved the piezoelectric performance of PVDF nanofibers by enhancing the charge transfer rate and thus enabling the alignment of PVDF's electric dipoles.

Chen et al. [370] formed novel photoelectrochemical sensors for cholesterol which was constructed based on the functionalized perovskites. To build the system, PVDF with the addition of carbon nanodots was used as an additive to perovskite $CH_3NH_3PbI_3$. The produced composite had an improved and mechanical stability as well as water tolerance. Moreover, the molecularly imprinted polymer (MIP), able to recognize specifically molecules of cholesterol, was connected with the composite material by a simple thermal polymerization reaction. The detection limit for cholesterol was $2.1 \times 10^{-14}$ mol $L^{-1}$ – much lower than the existing devices and sensors available to tracking cholesterol [370].

Prasad and co-workers developed a conductive thin film for the application in strain sensor [24]. The conductive nanocomposite material induced thin film of CNT, PVDF and vapor grown carbon nanofibers (VGCNF) as conductive filler at the following concentrations: 5, 10, 15 wt.%. To generate the conductive network, the presence of VGCNF was essential. Moreover, the mentioned materials enhanced the mechanical properties. The conducting behavior was exploited for strain sensing, and the mechanical behavior aided in flexibility toward the strain sensing applications. The sample containing 10% VGCNF produced the most promising results in terms of high conductivity ($1.3\ 10^{-3}$ $Sm^{-1}$) and elongation ($5.7\ 10^{-3}$ $mm^{-1}$). Additionally, the produced material had a gauge factor ca. 2.88, which is greater than metal-based piezoresistive strain sensors and demonstrated a considerable flexibility. The researchers discussed the immediate applicability of thin films in the human health monitoring [24]. The application of CNT-PVDF materials as a strain sensor has been extensively investigated [371-376].

MWCNT-PVDF composite films have been produced by Zhao and co-workers [377]. The purpose of these material was the implementation as composite sensors for a large strain measurement. The researchers studied sensing features of MWCNT-PVDF in direct current (DC) and alternating current circuits. In the case of AC circuits, the change rate of the dielectric loss tangent (DLT) of the generated materials was sensitive to the strain. Moreover, piezoresistivity of the PVDF doped with MWCNT in DC circuits can be successfully applied for the strain sensing. Nevertheless, such application might be incomplete owing to the highly nonlinear strain-resistance relation of the material. Above and beyond, the results revealed that the dependence between the DLT change rate and strain was turning to be linear in the case of suitable selection of the ratio between AC frequency and level of MWCNT content. Another work with CNT and PVDF has been focused on the application of such composite materials in human walking detector [378]. The composite which includes CNT (up to 5 wt.%), PVDF, styrene-b-(ethylene-*co*-butylene)-b-styrene and thermoplastic polyurethane. It was revealed that even though the carbon-based material has been introduced to the matrix of thermoplastics and thermoplastic elastomers, the chemical and thermal properties of the polymers were maintained. The electrical features were directly associated with the property of the matrix. Therefore, PVDF and SEBS exhibited the lowest and largest percolation threshold, at 0.5 and 2 wt.% CNT, accordingly. On the other hand, the TPU showed an intermediate value of the percolation threshold. The highest electrical conductivity has been found for the material with the highest, *i.e.,* 5 wt.% of CNT in composites (0.8 S $m^{-1}$) for the TPU. The highest value of piezoresistive sensibility in 4-point-bending and pressure modes was found for PVDF, for the low deformation bending (GF ≈ 2.8) and pressure tests (PS ≈ 12 $MPa^{-1}$). Based on those findings, it was possible to conclude that PVDF was the most appropriate polymer matrix for the low deformation usage. Furthermore, TPU and SEBS were the most suitable for large deformation application owing to their stretchability.

Another sensor built on the CNT-PVDF for health monitoring has been presented by Aziz and co-workers as a smart fabric sensor [379]. Chaturvedi et al. [380] worked on the tuning of piezoresistive sensitivity implementing polyelectrolyte (poly(4-styrenesulfonic acid) -PSSA) as an interface linker between PVDF and carbon-based material, *i.e.* carbon nanofiber (CNF). It was revealed that by increasing the content of CNF, it was possible to enhance the gauge factor and the piezoresistive sensitivity. The significant shift in the phase-composition of PVDF was noticed. Specifically, the increase in the β-phase and reduction in α-phase was confirmed by XRD and DSC. Furthermore, no differences in the crystallinity of the material were observed. As a consequence of the increasing level of PSSA, an improvement in elasticity has been found. It was pointed out by the researchers that the regulation of PSSA level in carbon-based composite accomplished substantial rise in gauge factor up to 5.07, ascribed to the efficient interfacial linking characteristic of PSSA. The future developments for the produced materials were following: a possible use in biomedicine, prosthetic limbs, and other wearable devices as pressure sensors.

Materials based on PVDF and PANI with the G-addition were used as mats in sensors and nanogenerators [381]. For the production of such materials, Forcespinning$^{TM}$ technique was implemented. The selected methods allowed to produce fiber systems with designed properties for a range of applications. The materials with G-coated nanocomposites exhited a 300%-improvement, referring to pristine mats, in an average output voltage (*i.e.*, 75 mV, peak-to-peak) and an output current of 24 mA (peak-to-peak) by gentle finger pressing. Moreover, graphene-coated materials were characterized by a volume conductivity of $1.2 \times 10^{-7}$ S $cm^{-1}$. It was examined as a promising system for vibration, temperature, and airflow sensing. Moreover, the G-coated material was also analysed as a water-tide energy harvesting piezoelectric nanogenerator. The sample generated ca. ~40 mV for a synthetic ocean wave with a flow rate of 30 mL $min^{-1}$ [381]. The authors pointed out that such materials can be implemented in low-powered sensors and for harvesting blue energy [382]as well as vibration energy [383].

Photodetectors (Fig. 22), as an example of smart flexible electronic devices which can be integrated into wearable products, such as stretchable circuits [384-386], smart clothes [387, 388], flexible displays [389, 390], are in the spotlight of the research for the modern society. The most important applications of photodetector, being the component of the certain device for information transmission, are in-flame sensing [391], missile approach warning system [392], optical communications [393-395], and sensor imaging [396-399].

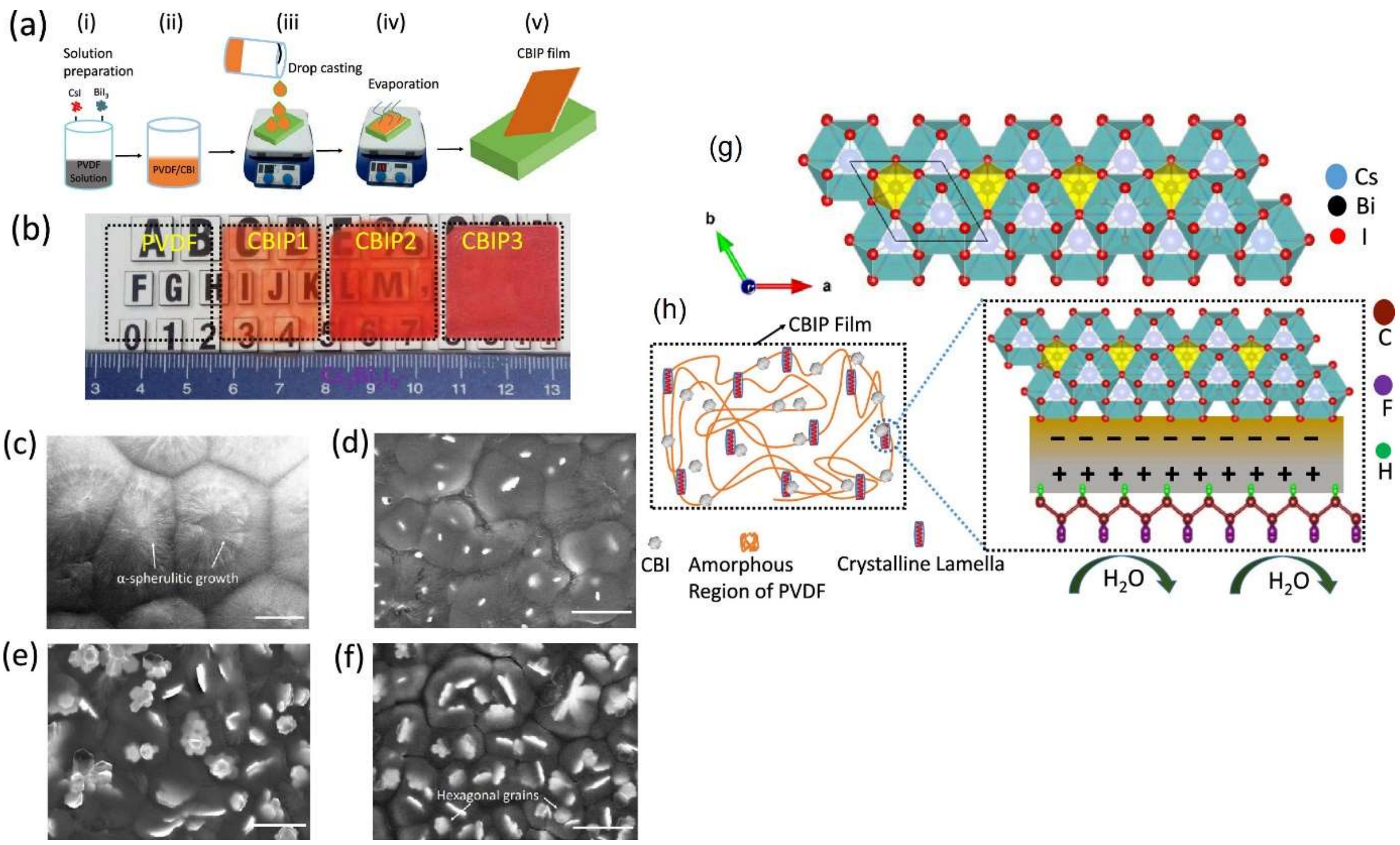

**Figure 22.** (a) Schematic illustration of the *in situ* formation of $Cs_3Bi_2I_9$ in PVDF solution and preparation of CBIP composite films. (b) Digital photographs of CBIP films. FESEM images of (c) neat PVDF, (d) CBIP1, (e) CBIP2, and (f) CBIP3 film. Scale bar is ~10 μm. (g) Model of the hexagonal crystal structure of $Cs_3Bi_2I_9$ perovskite. (h) Schematic presentation of the proposed mechanism for the formation of the electroactive phase in the CBIP films and the stability of the CBI perovskite from an ambient environment [387]. Adapted with permission from [387]. Copyright 2022 American Chemical Society.

Although it is true that photoactive piezoelectric nanogenerators (PENGs) are gaining more and popularity owing to their enormous potential as a self-powered photodetector and pressure sensor, there is a lack of research explaining the mechanism of interactions between photogenerated carriers and piezoelectric charges. Miao et al. [385] tried to explain the mentioned phenomenon. For that purpose, photoactive PENGs based on methylammonium lead halides-PVDF composite were produced. The authors stated that level of methylammonium in the PVDF matrix was essential and played a dual role in improving the nucleation of the electroactive PVDF β-phase and is responsible for an introduction of photoactivity to the composite (Fig. 23) [385]. While irradiated, it had an ideal open-circuit voltage of 4.7 V and a short-circuit current of 0.2 A, which was practically halved in the dark. Additionally, using piezoresponsive scanning force microscopy methods, photoluminescence spectroscopy, and a bandgap analysis, it was fully investigated the interaction mechanism between photogenerated carriers and piezoelectric charges. The findings showed that photogenerated carriers screened a portion of the piezoelectric potential caused by piezoelectric charges, whereas piezoelectric charges stimulated an additional photogenerated recombination of carrier.

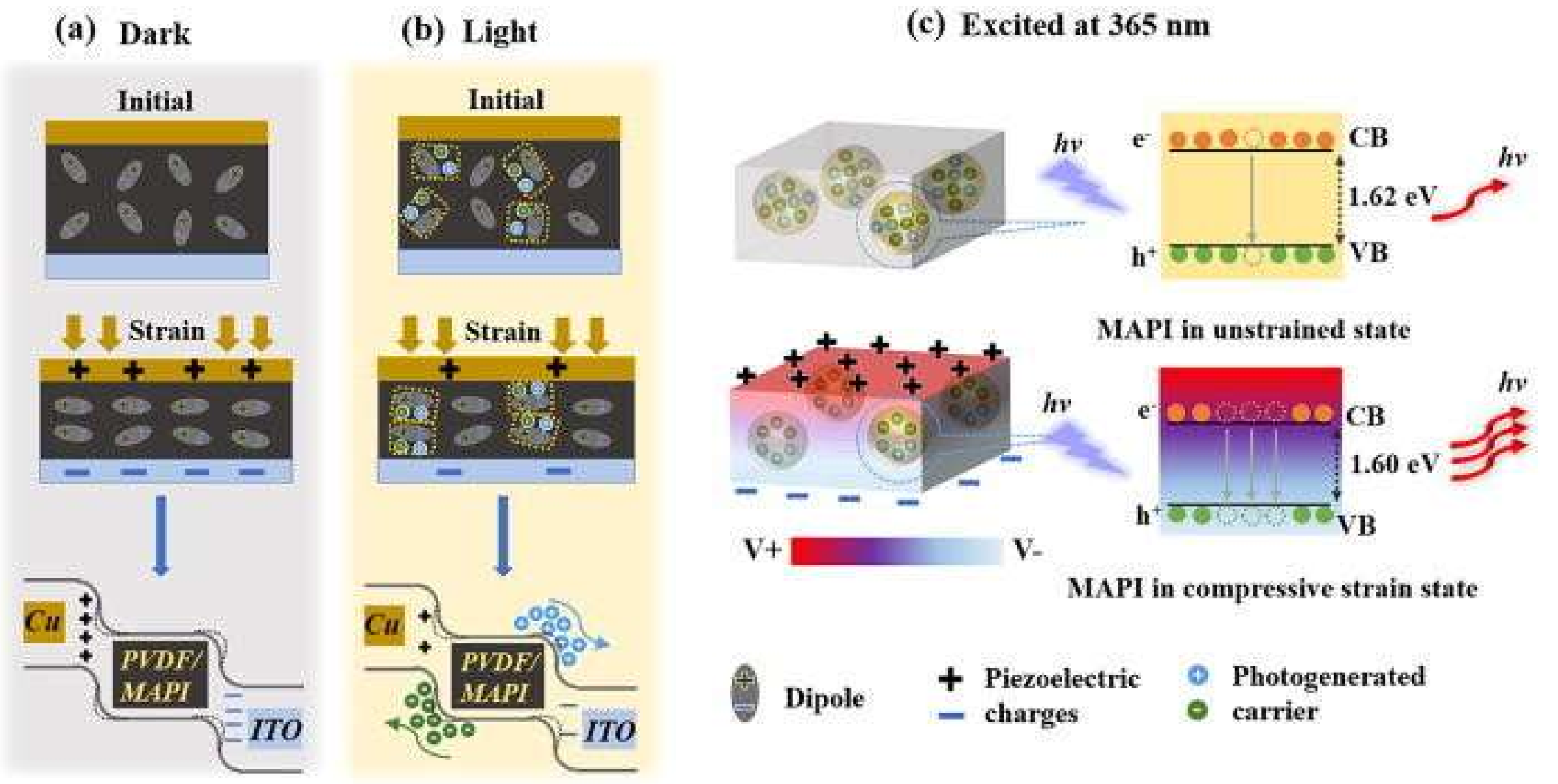

**Figure 23**. A summary diagram of the interaction mechanism of photogenerated carriers and piezoelectric charges in photoactive PENG. Schematic cross section and energy band diagram of the photoactive PENGs (a) in the dark and (b) under illumination. (c) Mechanism of piezoelectric charges tuned PL. Adapted with permission from [385]. Copyright 2022 AIP Publishing.

On and co-workers [389] presented the application of rollable UV in a wireless sensor network technology. The element of novelty in the research was to combine enhanced responsivity of UV photodetectors with rollable and high-performance UV photodetectors. The rollable UV photodetector service has been designed from the ZnAl-layered double hydroxide (LDH) nanosheets on a porous PVDF material. The PVDF with the controlled morphology-porosity was used as a support to grow on vertically aligned ZnAl-LDH nanosheets. An excellent photoresponse of a UV photodetector (for $\lambda<420$ nm) was detected independently on the geometry of the device. The curvature with a 0.6 mm radius was also tested (Fig. 24). The material was highly stable and maintained the responsive performance even after 5 000 rolling cycles, indicating the reasonable durability of the ZnAl-LDH nanosheet. The design of such device and maintain the efficiency was possible thanks to the elasticity of the PVDF as well as the strong interfacial adhesion force between the ZnAl-LDH nanosheets and porous PVDF support [389].

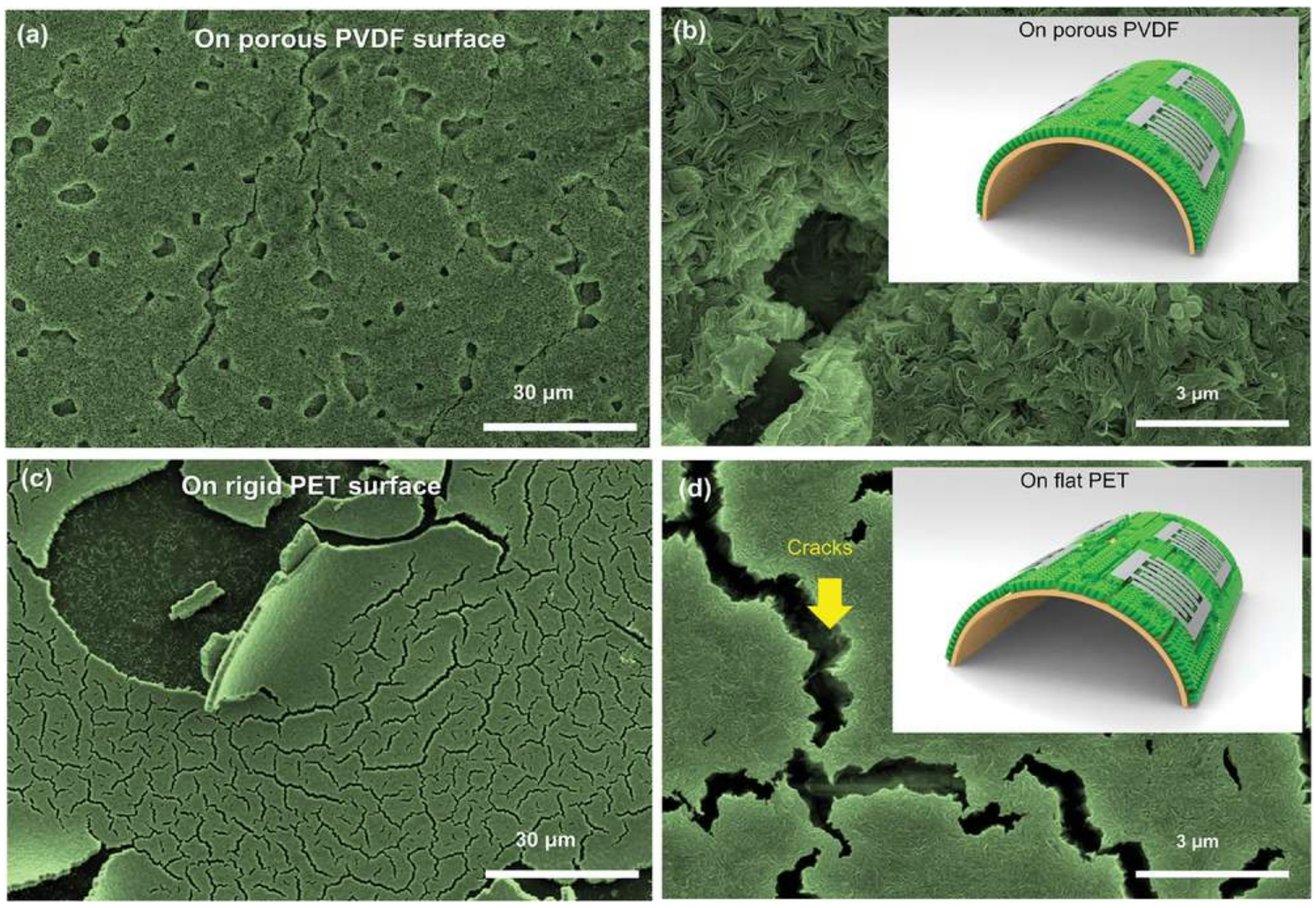

**Figure 24**. Surface images of ZnAl-LDH on the porous PVDF film after the rolling-relief test. c,d) Surface images of ZnAl-LDH on PET substrate after the bending-relief test. Adapted with permission from [389]. Copyright 2022 John Wiley and Sons.

## 4. Other applications of PVDF-CNMs and modeling

Pyroelectric materials are able to convert differences in temperature into electric signal. ND powders supported PVDF was applied for this purpose by Li et al. [100]. The pyroelectric coefficient increases 3.5 times, comparing to pure PVDF, and the optimal loading of NDs was only 0.4 wt.%. Also pyroelectric properties can be strongly improved (even 10 times in comparison to the pure PVDF) after addition of G alone [400] as well as G (and GO) enriched wih halloysite nanotubes [277].

NDs were applied together with TrFe and PVDF for generating ultrashort pulses in Q-switched (*i.e.*, generating high power at a short pulse) lasers [401], and together with silicone carbide (ND@SiC) to improve the thermal conductivity of PVDF – here, an 11.2 times higher thermal conductivity was observed than for pure PVDF [42].

Direct carbonization of PVDF (also combined with other polymers, for example PAN), leads to microporous activated carbon fibers (with specific surface area as high as 1000 $m^2$ $g^{-1}$) used for the enhanced $CO_2$ and $CH_4$ adsorption [402, 403].

Films obtained *via* mixed solvent phase separation of CQD (from citric acid and urea) and PVDF are potential materials for the preparation of sunlight sterilized facemasks for preventing COVID infection [404]. Carbon-reinforced PVDF core-shell nanofibers, obtained using electrospinning, carbon reinforced PVDF core-shell nanofibers were shown to exhibit self-healing properties [405].

CQD synthesized from leaves with PVDF can be successfully applied for UV-A shielding, *i.e.*, applied as a protection against skin cancer and LEDs fabrication [406]. Composite CQD with poly(vinyl alcohol) and hydrophilic PVDF (PVDF-OH), *i.e.*, PVDF-OH@CQDs/ poly(vinyl alcohol) were shown to absorb UV light completely, thus, they could be applied for the effective UV-shielding [407]. PVDF/ thermoplastic polyurethane filled with CB-polypyrrole materials are applicable as materials for 3D printing [408], and due to mechanical and thermal properties are also perspective materials for EMI shielding and flexible sensors.

CNTs, and PVDF/modified by F-containing silane, were used for the preparation of passive anti-icing surfaces. The ice adhesion was low, enabling the removal of frozen droplet by wind [409]. Our group reported the first study on the application of PVDF modified by SWCNH surfaces in the anti-acing applications [410]. A simple thermal feathering of CNMs in PVDF foil produced the surfaces showing the delay in recalescence time, and the extention of the freezing time. Hence, it can be concluded, that PVDF/CNMs represent indeed promising materials in this field.

Molecular simulations of PVDF-CNMs systems were rarely studied. Zeng et al. [411] reported molecular dynamic simulations of GO-PVDF materials to model the most optimal permeable barrier for the water diffusion. Muzzi et al. proved, using Molecular Dynamics Simulations, the creation of hydrogen bonds between F and oxygen-containing groups of GO [412]. Li and Han [413] provided theoretical analysis of the wave propagations in PVDF-G platelets to optimize the mechanical and piezoelectric properties of the material. Krishnaswamy et al. [414] reported the modelling of piezoelectricity of PVDF/G and CNTs - modified matrices. The results are important considering production of piezo-composites with the desired elasticity. It was concluded that due to enhanced mechanical properties of composites G is a more promising nanofiller. Xia et al. [415] reported a new multiscale homogenisation theory for description of CNT-based polymer nanocomposites allowing the calculation, among other things, the dielectric permittivity of a system in the wide range of frequency. A good agreement with the experimental data was observed. The RLC equivalent circuit model describing the characteristics of CNT-containing polymer composites was tested and improved for the series of CNTs [376].

## 5. Summary and Conclusions

### 5.1. Opportunities and outlook

Considering that CNMs can be easily synthesized and introduced into or onto the PVDF matrix, its more advanced application or extended scalability are still challenging. For instance, GO needs to be accumulated in the form of a continuous solid. The drawback is that GO has not the intrinsic mechanical features of a single flake of GO. One should emphasize that GO is susceptible to delaminate or debond. Moreover, this type of materials has weak load transfer features [416].

There are some important issues to be solved considering CNMs. Fo example considering G: [417]: poor electrical output from TENGs, enhancing interactions between interlayers [332] solving of restacking problem [418].

Not only graphene and its derivatives are in the spotlight as nanoenhancers for PVDF-based materials. The growing attention has been paid to MXenes owing to their strong interactions with the polymer chains [419-424].

Another challenge has been identified in the nanogenerators and is related to maintaining their robustness and reliability during the long-lasting processes. The mentioned features of the devices are significantly impacted by the long-term, negative environmental and long-term mechanical use in the course of the operation. To solve the problems, a suitable optimization and design of the materials as well as packaging processes are required [422]. Owing to the introduction of self-healing piezoelectric materials to a composite, the stability still can be boosted. As a consequence, the way towards uninterrupted and stable materials during long-lasting utilization of the nanogenerators will be possible. Based on this inspiration, it is conceivable to join the self-healing piezoelectric bipyrazole organic crystals to the materials having potential in mechanical energy harvesting. The approach can be fulfilled by functionalizing or loading these organic crystals to the graphene-based materials. One interesting solution was the use of vinyl ester resin carrying PVDF green nanofibers for self-healing utilization [425]. The growing interest to introduce self-healing features into the PVDF-CNMs and their further applications in aerospace, defence, food and health industries were proposed [405, 426, 427].

From theselection of the material point-of-view, there is a lot focus on the developing characterization techniques to assess the performances of mechanical features of CNMs-PVDF, particularly for the energy harvesting devices. Nevertheless, a lot of research requires a significant care in this matter.

In the case of nanogenerators, for the scale-up and industrialization and scale-up, it is required to design and elaborate standardized testing conditions. This will help to compare the efficiency of the materials in a clearer way. Moreover, till now it is not fully explained how the introduction of CNMs generates dissimilar output energy values. Henceforth, emerging conditions and testing systems are essential to simplify direct and systematic evaluation [428].

In the design of advanced and materials and miniature devices, the following techniques and methods including microfabrication, can be used, *i.e.*, laser scribing, plasma etching, screen printing, 3D printing, inkjet printing, and mask-assisted methods [418]. Moreover, these techniques can be also applied for the production of PVDF-CNMs materials at the micro-level for energy consumption, storage, and mechanical harvesting [418]. The significant advantages of PVDF-CNMs materials generated by 3D printing technique in comparison to standard solvent casting method have been presented. The problem that can be faced for PVDF-based materials produced by the solvent casting method and having flat block or thin-film structures, is a local residual stress from the substrate. As a consequence, the movement of dipole can be limited, which additionally diminishes the piezoelectric output. However, the size and shape of materials done by 3D printing method can be adjusted and further modification can be introduced. For that reason, owing to the presence of arrays and voids within the structures, the compressive stress is regularly distributed, limiting the strain in the direction of the vertical stress. Additionally, the Young modulus of the material rises which causes in substantial deformation under the applied stress and leads to superior piezoelectric output. In the case of PVDF-CNMs, an interesting solution is to apply the fused deposition modelling (FDM) to 3D printing technique. This method allows stabizing  a higher concent of the $\alpha$-phase crystals. Depending on the final application, the enhanced level of the $\alpha$-phase can be positive as well as very negative issue. In the case of energy harvesting materials, it is a large disadvantage which reduces the efficiency. On the other hand, the utilization of 3D printing in the formation of PVDF composites can help to solve the problem of CNMs filler aggregation by promoting a homogeneous dispersion, decreasing the level of cracks and porosity [429]. However, the low content of $\beta$-phase, the poling methods and introduction of ionic liquid have been presented [430].

Separately from the applicability of PVDF in energy harvesting, nanogenerators and sensing, PVDF is broadly used for the water treatment including radioactive sewage treatment [431]. The highly effective PVDF membranes were produced with the implementation of $\gamma$-ray irradiation (20 kGy) for crosslinking. The generated materials displayed significantly improved mechanical features. However, when the radiation was higher than 200 kGy, the PVDF chains were damaged and the membrane permeability dropped. Moreover, it was pointed out that a higher level of damages was observed when the level of oxygen was substantially reduced in the radioactive waste. The presented outcomes can help to develop new method of the PVDF-based materials modification *via* introduction of CNMs with the control level of oxygen.

According to the presented state-of-the-art, it is suggested that PVDF-based materials with CNMs can be applied to solve the weighty difficulties for the modern society. The spectrum of PVDF-CNMs applicability cover inter alia, the effective separation-based materials. sensors, energy harvesting, and highly effective batteries.

By following the most recent progress in the material science, one can identify , that more and more attention is paid to the Internet of Things and 5G [432, 433]. The main reason of the fast evolution of these emerging technologies is its applicability in smart building and cities (*e.g.*, NEOM in Tabuk, Saudi Arabia) [434-436]. Moreover, the potential utilization in solving problems related to environmental monitoring, overpopulation, and crisis management has been revealed [437].

The advantages of PVDF-CNMs materials used for the devices formation in the micro-scale can be appreciated particularly. It is related to the fact that fuelling the IoT strategies sustainably remains one of the prime challenges, principally owing to the problems with size, accessibility, and threats to the environment. In these points, the efficient energy harvesting materials will be the best possible option to warrant uninterrupted and long-lasting operation. Moreover, one of the most important issues is the operating mechanical energy harvesting is upholding a zero-carbon footprint. Particularly, the nanogenerators own the benefits of possible connecting with power grids and capability to collect energy from renewable sources. In the smart cities, the nanogenerators can gathered the energy from water, sun and wind to "write" letters on smart roads and transportation systems, intelligent healthcare, smart vehicles, and human-machine interface [433, 437-439]. The greatest prospects to tackle the problems of climate change and global warming remain nanogenerators.

### 5.2. Conclusions

The purpose of this review is the explanation of the hybrid approach phenomenon during the preparation of advanced materials. We tried to explain the "magical" roles of CMNs. We explained the EMI shielding mechanisms and we proved that the most effective were G based materials, and the best are the materials reinforced by CF, MWCNT, and MXenes. The magical role of CNMs is in this case the synergy between extremely high thermal and electric conductivity of obtained PVDF-CNMs composites.

In the field of piezoelectric, energy harvesting and TENG the variety of new applications of PVDF – CNMs composites is amazing ( self-powered intelligent agriculture systems [440], smart home monitoring, hybrid films applied as piezoelectric energy harvesters in self-powered human motion monitoring, wearable energy generators and electronic skin on the carotid artery for monitoring of human activity). In this field there are still advances in tuning of the dielectric constant. Here, thermal evolution of PVDF/PEG/GO nanocomposites emerges as particularly interesting.

In the area of sensors, actuators, and photodetectors, the most innovative application of PVDF-CNMs have been presented. The benefits coming from the possible tuning of β-phase and maintaining high level are possibilities to convert the mechanical deformation into electrical signals. For that reason, we exhibit and discuss the purpose of CNMs addition into PVDF in the tactile sensors, robots and touch screen devices, human health monitoring, smart fabric sensor, nanosensors for humidity, force, and pressure measurements, photoelectrochemical sensors, strain sensors, and sensors in biomedical instruments, prosthetic limbs, and other wearable devices. Basing on many applications, the superior features of the CNMs-PVDF in sensors and actuators were presented.

In the majority of discussed applications, the form of PVDF is crucial. Here the magical role of CNMs play in the transformation of PVDF phases. This is why different authors, guided by intuition, propose the lowest PVDF – CNMs interaction energy orientation (see Fig.4).

Performed in this study DFT calculations reveal that among shown on Fig. 2 orientations all configurations, excepting B, seem to co-exist.

There are still emerging fields in the science of PVDF-CNMs. Still little is known on the influence of CNDs and SWCNH on α-to-β phase transformation. There is a small number of studies on the application of SWCNH and new carbon nanoforms in composites with PVDF in anti-icing science.

We demonstrate the realistic portrait of the interactions at the PVDF-MWCNT interface with their (super)structures, physicochemistries, and the molecular structures of the continuous PVDF phases for numerous and frequently multi-task, high-performance applications. We believe control over the molecular structure of the CNM-PVDF interface and the bulk, PVDF matrix at the molecular level, represents the most efficient tool toward functionalities. Our review hence should be a solid point for the further studies on the most pressing and promising PDVF-CNM marriages ‘made in heaven’.

**Abbreviations**

2D – graphene - graphene, graphene oxides, graph(di- and poly)ynes

3D – (nano)carbons - diamond, carbon black / nanospheres / nanoglobules, active(ate)d carbons, graphite (exfoliated / expanded),

BCZT - $Ba_{0.85}Ca_{0.15}Ti_{0.9}Zr_{0.1}O_3$

BT - barium titanate

CB - carbon black

CNFs - carbon nanofibers

CNHs – carbon nanohorns

CNMs - carbon nanomaterials

CNTs - carbon nanotubes

CQDs - carbon quantum dots

CVD - chemical vapour deposition

DC - direct current

DFT - density functional theory

DLT - dielectric loss tangent

EM - electromagnetic

EMI - electromagnetic interference

FTIR – Fourier transform infrared

G - graphene

Gr - graphite

GO - graphene oxide

HFP - hexafluoropropylene

HSP - Hansen Solubility Parameters

IL - ionic liquid

KNN - potassium sodium niobate

MOF - metal organic framework

MP - mechanical properties

MR - multiple reflection

MWCNTs - multi-walled carbon nanotubes

MX – mxene

NDs - nanodiamonds

NGs - nanogenerators

PE - polyethylene

PEG – poly(ethylene glycol)

PENG - piezoelectric nanogenerators

PTFE - polytetrafluoroethylene

PVDF - poly(vinylidene fluoride)

PSSA - poly(4-styrenesulfonic acid)

QDs - quantum dots

R - reflection

rGO - reduced graphene oxide

RLC - a resistor, and inductor and a capacitor circuit

RT - room temperature

SE - shielding efficiency

SWCNTs - single walled carbon nanotubes

SWCNHs - single walled carbon nanohorns

TENG - triboelectric nanogenerator

TrFE - trifluoroethylene

VGCNF - vapor grown carbon nanofibers

WNG - wearable nanogenerator

YM - Young Modulus

XRD - X ray diffraction

**Author Contributions**

**J.K. –** Conceptualization, Writing - Original Draft, Writing - Review & Editing**;** Supervision **S.B. -** Writing - Original Draft, Writing - Review & Editing**; S.A.G –** Writing - Original Draft, Writing - Review & Editing**; S.K.** Writing - Original Draft, Writing - Review & Editing**; A.K.K -** Writing - Original Draft, Data Curation**; E.K.** Writing - Original Draft, Writing - Review & Editing**;** Visualization**; A.P.T -** Conceptualization, Writing - Original Draft, Writing - Review & Editing**;** Supervision, Visualization.

**Conflicts of interest**

There are no conflicts to declare

**Acknowledgements:**

A.P.T gratefully acknowledges the financial support from NCN Opus 22 project: UMO-2021/43/B/ST5/00421.

S.B. greatly acknowledges the supporting action from EU's Horizon 2020 ERA-Chair project ExCEED, grant agreement No. 952008.

References:

[1] J. Kujawa, S. Boncel, S. Al-Gharabli, S. Koter, W. Kujawski, K. Kaneko, K. Li, E. Korczeniewski, A.P. Terzyk, Concerted role of PVDF and carbon nanomaterials for membrane science, Desalination, 574 (2024) 117277.

[2] M. Zięba, T. Rusak, T. Misztal, W. Zięba, N. Marcińczyk, J. Czarnecka, S. Al-Gharabli, J. Kujawa, A.P. Terzyk, Nitrogen plasma modification boosts up the hemocompatibility of new PVDF-carbon nanohorns composite materials with potential cardiological and circulatory system implants application, Biomater. Adv., 138 (2022).

[3] E. Gutiérrez-Fernández, J. Sena-Fernández, E. Rebollar, T.A. Ezquerra, F.J. Hermoso-Pinilla, M. Sanz, O. Gálvez, A. Nogales, Development of polar phases in ferroelectric poly(vinylidene fluoride) (PVDF) nanoparticles, Polymer, 264 (2023) 125540.

[4] H. Park, H. Si, J. Gu, D. Lee, D. Park, Y.-I. Lee, K. Kim, Engineered kirigami design of PVDF-Pt core–shell nanofiber network for flexible transparent electrode, Sci. Rep., 13 (2023) 2582.

[5] J. Dangbegnon, N. Garino, M. Angelozzi, M. Laurenti, F. Seller, M. Serrapede, P. Zaccagnini, P. Moras, M. Cocuzza, T. Ouisse, H. Pazniak, J. Gonzalez-Julian, P.M. Sheverdyaeva, A. Di Vito, A. Pedico, C.F. Pirri, A. Lamberti, High-performance novel asymmetric MXene@CNT//N-doped CNT flexible hybrid device with large working voltage for energy storage, J. Energy Storage, 63 (2023) 106975.

[6] J. Žigon, U.G. Centa, M. Remškar, M. Humar, Application and characterization of a novel PVDF-HFP/PVP polymer composite with MoO3 nanowires as a protective coating for wood, Sci. Rep., 13 (2023) 3429.

[7] Z. Sha, C. Boyer, G. Li, Y. Yu, F.-M. Allioux, K. Kalantar-Zadeh, C.-H. Wang, J. Zhang, Electrospun liquid metal/PVDF-HFP nanofiber membranes with exceptional triboelectric performance, Nano Energy, 92 (2022) 106713.

[8] J.-P. Chen, C.-Y. Guo, Q.-J. Zhang, X.-Q. Wu, L.-B. Zhong, Y.-M. Zheng, Preparation of transparent, amphiphobic and recyclable electrospun window screen air filter for high-efficiency particulate matters capture, J. Membr. Sci., 675 (2023) 121545.

[9] T. Shan, X. Ma, H. Li, C. Liu, C. Shen, P. Yang, S. Li, Z. Wang, Z. Liu, H. Sun, Plant-derived hybrid coatings as adsorption layers for uranium adsorption from seawater with high performance, J. Membr. Sci., 675 (2023) 121547.

[10] L. Zhu, Q. Wang, Novel Ferroelectric Polymers for High Energy Density and Low Loss Dielectrics, Macromolecules, 45 (2012) 2937-2954.

[11] S. Sharma, S.S. Mishra, R. Kumar, R.M. Yadav, Recent progress on polyvinylidene difluoride-based nanocomposites: applications in energy harvesting and sensing, Energy Rep., 46 (2022) 18613-18646.

[12] R.E. Roy, B. Soundiraraju, R.S. Rajeev, Optically transparent nanocomposite films based on poly(vinylidene fluoride) and single walled carbon nanotube: Role of process parameters on polymorphic changes, Polymer Crystallization., 2 (2019) e10074.

[13] E. Dhanumalayan, G.M. Joshi, S. Kaleemulla, M. Teresa Cuberes, R.R. Deshmukh, Studies on the Surface and Wetting Properties of Poly(vinylidene fluoride)/Poly(acrylonitrile)/Multiwalled Carbon Nanotube-NH2 Blends as a Function of Air Plasma Treatment, J. Mater. Eng. Perform., 30 (2021) 7343-7353.

[14] J.-Y. Lyu, S. Chen, W. He, X.-X. Zhang, D.-y. Tang, P.-J. Liu, Q.-L. Yan, Fabrication of high-performance graphene oxide doped PVDF/CuO/Al nanocomposites via electrospinning, Chem. Eng. J., 368 (2019) 129-137.

[15] D.N. Trivedi, N.V. Rachchh, Graphene and its application in thermoplastic polymers as nano-filler-A review, Polymer, 240 (2022) 124486.

[16] M. Kamkar, S. Sadeghi, M. Arjmand, U. Sundararaj, Structural Characterization of CVD Custom-Synthesized Carbon Nanotube/Polymer Nanocomposites in Large-Amplitude Oscillatory Shear (LAOS) Mode: Effect of Dispersion Characteristics in Confined Geometries, Macromolecules, 52 (2019) 1489-1504.

[17] S. Mondal, B. Alke, A.M. de Castro, P. Ortiz-Albo, U.T. Syed, J.G. Crespo, C. Brazinha, Design of Enzyme Loaded W/O Emulsions by Direct Membrane Emulsification for CO2 Capture, Membranes, 12 (2022) 797.
[18] C. Yang, X. Xie, Y. Lu, X.-d. Qi, Y.-z. Lei, J.-h. Yang, Y. Wang, Improving the Performance of Dielectric Nanocomposites by Utilizing Highly Conductive Rigid Core and Extremely Low Loss Shell, J. Phys. Chem. C, 124 (2020) 12883-12896.
[19] C. Chen, R.-P. Nie, S.-C. Shi, L.-C. Jia, Y. Li, X. Li, Y.-C. Huang, D.-L. Han, H.-D. Huang, Z.-M. Li, Simultaneous enhancement of breakdown strength and discharged energy efficiency of tri-layered polymer nanocomposite films by incorporating modified graphene oxide nanosheets, J. Mater. Sci., 56 (2021) 13165-13177.
[20] Z. Fan, R. Liu, X. Cheng, Preparation and Characterization of Electromagnetic Shielding Composites Based on Graphene-Nanosheets-Loaded Nonwoven Fabric, Coatings, 11 (2021) 424.
[21] J. Wang, M. Yi, Z. Shen, L. Liu, X. Zhang, S. Ma, Enhanced thermal and mechanical properties of poly (vinylidene fluoride) nanocomposites reinforced by liquid-exfoliated graphene, J. Polym. Sci. A Polym. Chem., 56 (2019) 733-740.
[22] S. Sodagar, B. Jaleh, P. Fakhri, M. Kashfi, B. Feizi Mohazzab, A. Momeni, Flexible piezoelectric PVDF/NDs nanocomposite films: improved electroactive properties at low concentration of nanofiller and numerical simulation using finite element method, J. Polym. Res., 27 (2020) 203.
[23] A. Kausar, Mechanical, thermal, conductivity, and electrochemical behavior of poly(vinylidene fluoride)/poly(3,4-ethylenedioxythiophene)/polyaniline-grafted-nanodiamond nanocomposite, J. Termoplast. Comp. Mater., 33 (2020) 628-645.
[24] B. Prasad, I.M. Alarifi, F.S. Gill, V. Rathi, V. Panwar, Development of conductive thin films as piezoresistive strain sensor, Mater. Chem. Phys., 276 (2022) 125371.
[25] Q. Peng, X. Tan, M. Venkataraman, J. Militky, Tailored expanded graphite based PVDF porous composites for potential electrostatic dissipation applications, Diam. Relat. Mater., 125 (2022) 108972.
[26] J. Tong, W. Li, H.-C. Chen, L.-C. Tan, Improving Properties of Poly(vinylidene fluoride) by Adding Expanded Graphite without Surface Modification via Water-Assisted Mixing Extrusion, Macromol. Mater. Eng., 305 (2020) 2000270.
[27] W. Deng, W. Yu, W. Cui, H. Guo, Y. Liu, T. Zhao, High dielectric permittivity and low loss of polyvinylidene fluoride filled with carbon additives: Expanded graphite versus reduced graphene oxide, High Perform. Polym., 31 (2019) 778-784.
[28] M.S. Nisha, S. Mullai Venthan, P. Senthil Kumar, D. Singh, Tribological Properties of Carbon Nanotube and Carbon Nanofiber Blended Polyvinylidene Fluoride Sheets Laminated on Steel Substrates, Inter. J. Chem. Eng., 2022 (2022) 3408115.
[29] M.S. Park, C.S. Lee, J.H. Lee, D.Y. Ryu, J.H. Kim, Dissolution–precipitation approach for long-term stable low-friction composites consisting of mesoporous TiO2 nanospheres and carbon black in Poly(Vinylidene fluoride) matrix, Tribol. Int., 145 (2020) 106187.
[30] J.H. Lee, M.S. Park, C.S. Lee, T.-S. Han, J.H. Kim, Wear-resistant carbon nanorod-embedded poly(vinylidene fluoride) composites with excellent tribological performance, Compos. Part A Appl. Sci. Manuf., 129 (2020) 105721.
[31] J. Guan, M. Song, L. Chen, Y. Shu, D. Jin, G. Fan, Q. Xu, X.-Y. Hu, Carbon quantum dots passivated CsPbBr3 film with improved water stability and photocurrent: Preparation, characterization and application, Carbon, 175 (2021) 93-100.
[32] X. Zhang, C. Tan, Y. Ma, F. Wang, W. Yang, BaTiO3@carbon/silicon carbide/poly(vinylidene fluoride-hexafluoropropylene) three-component nanocomposites with high dielectric constant and high thermal conductivity, Compos. Sci. Technol., 162 (2018) 180-187.
[33] X.-d. Qi, W.-y. Wang, Y.-j. Xiao, T. Huang, N. Zhang, J.-h. Yang, Y. Wang, Tailoring the hybrid network structure of boron nitride/carbon nanotube to achieve thermally conductive poly(vinylidene fluoride) composites, Compos. Commun., 13 (2019) 30-36.
[34] Z.-G. Wang, Y.-F. Huang, G.-Q. Zhang, H.-Q. Wang, J.-Z. Xu, J. Lei, L. Zhu, F. Gong, Z.-M. Li, Enhanced Thermal Conductivity of Segregated Poly(vinylidene fluoride) Composites via Forming Hybrid Conductive Network of Boron Nitride and Carbon Nanotubes, Ind. Eng. Chem. Res., 57 (2018) 10391-10397.

[35] M. Kamkar, E. Aliabadian, A. Shayesteh Zeraati, U. Sundararaj, Application of nonlinear rheology to assess the effect of secondary nanofiller on network structure of hybrid polymer nanocomposites, Phys. Fluids, 30 (2018).
[36] B. Krause, C. Barbier, K. Kunz, P. Pötschke, Comparative study of singlewalled, multiwalled, and branched carbon nanotubes melt mixed in different thermoplastic matrices, Polymer, 159 (2018) 75-85.
[37] X. Zhao, H. Wang, Z. Fu, Y. Li, Enhanced Interfacial Adhesion by Reactive Carbon Nanotubes: New Route to High-Performance Immiscible Polymer Blend Nanocomposites with Simultaneously Enhanced Toughness, Tensile Strength, and Electrical Conductivity, ACS Appl. Mater. Interfaces, 10 (2018) 8411-8416.
[38] S. Begum, H. Ullah, I. Ahmed, Y. Zhan, A. Kausar, M.A. Aleem, S. Ahmad, Investigation of morphology, crystallinity, thermal stability, piezoelectricity and conductivity of PVDF nanocomposites reinforced with epoxy functionalized MWCNTs, Compos. Sci. Technol., 211 (2021) 108841.
[39] Z. Lei, Y. Li, Z. Lei, X. Yang, J. Yan, Z. Li, H. Shui, S. Ren, Z. Wang, Y. Kong, S. Kang, Enhanced electrical conductivity of pitch-derived carbon via graphene template effects for high electrically conductive composites, J. Ind. Eng. Chem., 117 (2023) 394-401.
[40] Y.-n. Xie, D.-f. Liu, D.-X. Sun, J.-h. Yang, X.-d. Qi, Y. Wang, Crystallization and concentration fluctuation of miscible poly(vinylidene fluoride)/poly(methyl methacrylate) blends containing carbon nanotubes: Molecular weight dependence of poly(methyl methacrylate), Eur. Pol. J., 105 (2018) 478-490.
[41] H. Essabir, M. Raji, D. Rodrigue, R. Bouhfid, A.e.k. Qaiss, Multifunctional poly(vinylidene fluoride) and styrene butadiene rubber blend magneto-responsive nanocomposites based on hybrid graphene oxide and Fe3O4: synthesis, preparation and characterization, J. Polym. Res., 28 (2021) 437.
[42] C. Guan, Y. Qin, w. Bo, L. Li, M. Wang, C.-T. Lin, X. He, K. Nishimura, J. Yu, J. Yi, N. Jiang, Highly thermally conductive polymer composites with barnacle-like nano-crystalline Diamond@Silicon carbide hybrid architecture, Compos. B. Eng., 198 (2020) 108167.
[43] R. Ram, V. Soni, D. Khastgir, Electrical and thermal conductivity of polyvinylidene fluoride (PVDF) – Conducting Carbon Black (CCB) composites: Validation of various theoretical models, Compos. B. Eng., 185 (2020) 107748.
[44] B. Krause, C. Barbier, J. Levente, M. Klaus, P. Pötschke, Screening of Different Carbon Nanotubes in Melt-Mixed Polymer Composites with Different Polymer Matrices for Their Thermoelectrical Properties, J. Compos. Sci., 3 (2019) 106.
[45] W. Wu, J. Tu, H. Li, Z. Zhan, L. Huang, Z. Cai, Q. Li, M. Jiang, J. Huang, Suppressed dielectric loss and enhanced thermal conductivity in poly(vinylidene fluoride) nanocomposites using polyethylene glycol-grafted graphene oxide, J. Mater. Sci: Mater. Electron., 31 (2020) 807-813.
[46] V.M. Ruiz, R. Sirera, J.M. Martínez, J. González-Benito, Solution blow spun graded dielectrics based on poly(vinylidene fluoride)/multi-walled carbon nanotubes nanocomposites, Eur. Pol. J., 122 (2020) 109397.
[47] P. Wang, P. Xu, Y. Zhou, Y. Yang, Y. Ding, Effect of MWCNTs and P[MMA-IL] on the crystallization and dielectric behavior of PVDF composites, Eur. Pol. J., 99 (2018) 58-64.
[48] S. Begum, H. Ullah, A. Kausar, M. Siddiq, M.A. Aleem, Fabrication of epoxy functionalized MWCNTs reinforced PVDF nanocomposites with high dielectric permittivity, low dielectric loss and high electrical conductivity, Compos. Sci. Technol., 167 (2018) 497-506.
[49] J.-P. Cao, J. Zhao, X. Zhao, Z.-M. Dang, A facile route to prepare high−performance dielectric nanocomposites of poly(methyl methacrylate)/poly(vinylidene fluoride)/carbon nanotubes, Compos. Sci. Technol., 209 (2021) 108792.
[50] H. Kim, J. Johnson, L.A. Chavez, C.A. Garcia Rosales, T.-L.B. Tseng, Y. Lin, Enhanced dielectric properties of three phase dielectric MWCNTs/BaTiO3/PVDF nanocomposites for energy storage using fused deposition modeling 3D printing, Ceram. Int., 44 (2018) 9037-9044.
[51] A.K. Das, C.K. Raul, R. Karmakar, A.K. Meikap, Study of enhanced dielectric permittivity of functionalize multiwall carbon nanotube-based polyvinylidene fluoride free-standing film for flexible storage device, Phys. Lett. A, 407 (2021) 127455.
[52] L. Chen, L. Yan, Y. Guo, H.-C. Liu, H. Huang, H.-L. Lin, J. Bian, Y. Lu, Chemically functionalized multi-walled CNTs induced phase behaviors of poly(vinylidene fluoride) nanocomposites and its dielectric properties, Synth. Met., 260 (2020) 116268.

[53] N. Zhu, L. Yuan, G. Liang, A. Gu, Mechanism of greatly increasing dielectric constant at lower percolation thresholds for epoxy resin composites through building three-dimensional framework from polyvinylidene fluoride and carbon nanotubes, Compos. B. Eng., 171 (2019) 146-153.
[54] H.-j. Mao, D.-f. Liu, N. Zhang, T. Huang, I. Kühnert, J.-h. Yang, Y. Wang, Constructing a Microcapacitor Network of Carbon Nanotubes in Polymer Blends via Crystallization-Induced Phase Separation Toward High Dielectric Constant and Low Loss, ACS Appl. Mater. Interfaces, 12 (2020) 26444-26454.
[55] J. Zhu, J. Shen, S. Guo, C.B. Park, Poly (vinylidene fluoride)-based microcellular dielectrics filled with polydopamine coated carbon nanotubes for achieving high permittivity and ultralow dielectric loss, Compos. - A: Appl. Sci., 163 (2022) 107222.
[56] X. Xia, S. Zhao, J. Wang, H. Du, G.J. Weng, Tuning the AC electric responses of decorated PDA@MWCNT/PVDF nanocomposites, Compos. Sci. Technol., 222 (2022) 109398.
[57] R. Salehiyan, M. Nofar, S.S. Ray, V. Ojijo, Kinetically Controlled Localization of Carbon Nanotubes in Polylactide/Poly(vinylidene fluoride) Blend Nanocomposites and Their Influence on Electromagnetic Interference Shielding, Electrical Conductivity, and Rheological Properties, J. Phys. Chem. C, 123 (2019) 19195-19207.
[58] B. Jóźwiak, G. Dzido, E. Zorębski, A. Kolanowska, R. Jędrysiak, J. Dziadosz, M. Libera, S. Boncel, M. Dzida, Remarkable Thermal Conductivity Enhancement in Carbon-Based Ionanofluids: Effect of Nanoparticle Morphology, ACS Appl. Mater. Interfaces, 12 (2020) 38113-38123.
[59] L. Sun, N. Wu, R. Peng, Negative dielectric permittivity of PVDF nanocomposites induced by carbon nanofibers and polymer crystallization, J. Appl. Polym. Sci, 137 (2020) 49582.
[60] G. Sahoo, N. Sarkar, S.K. Swain, The effect of reduced graphene oxide intercalated hybrid organoclay on the dielectric properties of polyvinylidene fluoride nanocomposite films, Appl. Clay Sci., 162 (2018) 69-82.
[61] A. Gebrekrstos, G. Madras, S. Bose, Piezoelectric Response in Electrospun Poly(vinylidene fluoride) Fibers Containing Fluoro-Doped Graphene Derivatives, ACS Omega, 3 (2018) 5317-5326.
[62] A. Ahmed, Y. Jia, H. Deb, M.F. Arain, H. Memon, K. Pasha, Y. Huang, Q. Fan, J. Shao, Ultra-sensitive all organic PVDF-TrFE E-spun nanofibers with enhanced β-phase for piezoelectric response, J Mater Sci: Mater Electron, 33 (2022) 3965-3981.
[63] A. Anand, D. Meena, K.K. Dey, M.C. Bhatnagar, Enhanced piezoelectricity properties of reduced graphene oxide (RGO) loaded polyvinylidene fluoride (PVDF) nanocomposite films for nanogenerator application, J. Polym. Res., 27 (2020) 358.
[64] M. Kim, V.K. Kaliannagounder, A.R. Unnithan, C.H. Park, C.S. Kim, A. Ramachandra Kurup Sasikala, Development of In-Situ Poled Nanofiber Based Flexible Piezoelectric Nanogenerators for Self-Powered Motion Monitoring, Appl. Sci., 10 (2020) 3493.
[65] M. Haji Abdolrasouli, H. Abdollahi, A. Samadi, PVDF nanofibers containing GO-supported TiO2–Fe3O4 nanoparticle-nanosheets: piezoelectric and electromagnetic sensitivity, J. Mater. Sci: Mater. Electro., 33 (2022) 5970-5982.
[66] M. Khosravi, J. Seyfi, A. Saeidi, H.A. Khonakdar, Spin-coated polyvinylidene fluoride/graphene nanocomposite thin films with improved β-phase content and electrical conductivity, J. Mater. Sci., 55 (2020) 6696-6707.
[67] M. Nasir, R.I. Sugatri, P.P.P. Asri, F. Dara, Ardeniswan, Nanostructure and Surface Characteristic of Electrospun Carbon Black/PVDF Copolymer Nanocomposite, J. Silicate Bas. Compos. Mat., 70 (2018) 209-213.
[68] Z. Liang, J. Wang, Q. Zhang, T. Zhuang, C. Zhao, Y. Fu, Y. Zhang, F. Yang, Composite PVDF ultrafiltration membrane tailored by sandwich-like GO@UiO-66 nanoparticles for breaking the trade-off between permeability and selectivity, Chem. Eng. J., 276 (2021) 119308.
[69] G.-J. Choi, S.-H. Baek, S.-S. Lee, F. Khan, J.H. Kim, I.-K. Park, Performance enhancement of triboelectric nanogenerators based on polyvinylidene fluoride/graphene quantum dot composite nanofibers, J. Alloys Compd., 797 (2019) 945-951.
[70] X. Cai, T. Lei, D. Sun, L. Lin, A critical analysis of the α, β and γ phases in poly(vinylidene fluoride) using FTIR, RSC Adv., 7 (2017) 15382-15389.
[71] X. Li, Q. An, H. Li, Y. Zhang, Z. Chen, K. Teng, J. Zhuang, Y. Zhang, W. Tong, A porous piezoelectric-dielectric flexible energy conversion film for electricity generation from multiple sources, Chem. Phys. Letter., 767 (2021) 138357.

[72] T.T.V. Tran, S.R. Kumar, C.H. Nguyen, J.W. Lee, H.-A. Tsai, C.-H. Hsieh, S.J. Lue, High-permeability graphene oxide and poly(vinyl pyrrolidone) blended poly(vinylidene fluoride) membranes: Roles of additives and their cumulative effects, J. Membr. Sci., 619 (2021) 118773.
[73] C. P.B., N.M. Renukappa, G. Sanjeev, Preparation and crystalline studies of PVDF hybrid composites, AIP Conf Proc, 1942 (2018).
[74] L. Wu, M. Jing, Y. Liu, H. Ning, X. Liu, S. Liu, L. Lin, N. Hu, L. Liu, Power generation by PVDF-TrFE/graphene nanocomposite films, Compos. B. Eng., 164 (2019) 703-709.
[75] T. Xing, L.H. Li, L. Hou, X. Hu, S. Zhou, R. Peter, M. Petravic, Y. Chen, Disorder in ball-milled graphite revealed by Raman spectroscopy, Carbon, 57 (2013) 515-519.
[76] X. Huang, L. Jin, C. Wang, Y. Xu, Z. Peng, M. Xie, C. Zhang, W. Yang, J. Lu, Electrospun luminescent piezo webs as self-powered sensing platform for small accelerations at low frequency, Compos. Commun., 20 (2020) 100348.
[77] Y. Li, J. Tan, K. Liang, Y. Li, J. Sun, H. Zhang, C. Luo, P. Li, J. Xu, H. Jiang, K. Wang, Enhanced piezoelectric performance of multi-layered flexible polyvinylidene fluoride–BaTiO3–rGO films for monitoring human body motions, J. Mater. Sci: Mater. Electron., 33 (2022) 4291-4304.
[78] Y. Du, D. Cao, Y. Liu, Q. Wang, Increased dielectric performance of PVDF-based composites by electrochemical exfoliated graphite additives, Nanotechnology, 32 (2021) 505204.
[79] M.H. Ghajar, M. Mosavi Mashhadi, M. Irannejad, S. Jebril, M. Yavuz, E. Abdel-Rahman, Degree of Crystallinity and Phase Fraction of Polyvinylidene Fluoride Nanocomposites Containing Ionic Liquid and Graphene/Carbon Nanotube, Polym. Compos., 39 (2018) E1208-E1215.
[80] S.C.S.M. dos Santos, B.G. Soares, E.C. Lopes Pereira, T. Indrusiak, A.A. Silva, Impact of phosphonium-based ionic liquids-modified carbon nanotube on the microwave absorbing properties and crystallization behavior of poly(vinylidene fluoride) composites, Mat. Chem. Phys., 280 (2022) 125853.
[81] P. Viswanath, K.K.H. De Silva, H.-H. Huang, M. Yoshimura, Large piezoresponse in ultrathin organic ferroelectric nano lamellae through self-assembly processing, Appl. Surf. Sci., 532 (2020) 147188.
[82] B.T.S. Ramanujam, P.V. Adhyapak, S. Radhakrishnan, R. Marimuthu, Effect of casting solvent on the structure development, electrical, thermal behavior of polyvinylidene fluoride (PVDF)–carbon nanofiber (CNF) conducting binary and hybrid nanocomposites, Polym. Bull., 78 (2021) 1735-1751.
[83] C. Saujanya, S. Radhakrishnan, Structure development and properties of PET fibre filled PP composites, Polymer, 42 (2001) 4537-4548.
[84] S. Song, Z. Zheng, Y. Bi, X. Lv, S. Sun, Improving the electroactive phase, thermal and dielectric properties of PVDF/graphene oxide composites by using methyl methacrylate-co-glycidyl methacrylate copolymers as compatibilizer, J. Mater. Sci., 54 (2019) 3832-3846.
[85] M. Wojtaś, D.V. Karpinsky, M.V. Silibin, S.A. Gavrilov, A.V. Sysa, K.N. Nekludov, Dielectric properties of graphene oxide doped P(VDF-TrFE) films, Polym. Test., 60 (2017) 326-332.
[86] P. Viswanath, K. Kanishka, H. De Silva, M. Yoshimura, Piezoresponse in self oriented ultrathin Poly (vinylidene fluoride) supported on graphene oxide, Jpn. J. Appl. Phys., 59 (2020) SN1006.
[87] S. Mohamadi, N. Sharifi-Sanjani, Effect of thermal annealing on the crystalline structure of PVDF/PMMA-modified graphene nanocomposites, J. Macromol. Sci. A, 55 (2018) 775-784.
[88] R.A. Khan, M. Ashraf, A. Javid, K. Iqbal, A. Rasheed, N. Nasir, Development of self-polarized PVDF films on carbon fabrics for sensing applications, J. Text. Inst., 113 (2022) 2208-2214.
[89] J. Widakdo, T.-H. Huang, T.M. Subrahmanya, H.F.M. Austria, W.-S. Hung, C.-F. Wang, C.-C. Hu, K.-R. Lee, J.-Y. Lai, Tailoring of graphene–organic frameworks membrane to enable reversed electrical-switchable permselectivity in CO2 separation, Carbon, 182 (2021) 545-558.
[90] W.-S. Hung, S.-Y. Ho, Y.-H. Chiao, C.-C. Chan, W.-Y. Woon, M.-J. Yin, C.-Y. Chang, Y.M. Lee, Q.-F. An, Electrical Tunable PVDF/Graphene Membrane for Controlled Molecule Separation, Chem. Mater., 32 (2020) 5750-5758.
[91] J. Widakdo, T.-J. Huang, T.M. Subrahmanya, H.F.M. Austria, H.-L. Chou, W.-S. Hung, C.-F. Wang, C.-C. Hu, K.-R. Lee, J.-Y. Lai, Bioinspired ionic liquid-graphene based smart membranes with electrical tunable channels for gas separation, Appl. Mat. Today, 27 (2022) 101441.
[92] P. Xu, W. Fu, Z. Cui, Y. Ding, Enhancement of polar phase and conductivity relaxation in PIL-modified GO/PVDF composites, Appl. Phys. Letter, 112 (2018).

[93] X. Hu, Z. Ding, L. Fei, Y. Xiang, Y. lin, Wearable piezoelectric nanogenerators based on reduced graphene oxide and in situ polarization-enhanced PVDF-TrFE films, J. Mater. Sci., 54 (2019) 6401-6409.
[94] W. Kim, Y. Oh, J.-Y. Jeon, B.-S. Shin, Y. Eom, D.W. Chae, Combined effect of hybrid carbon fillers on the physical and rheological properties of polyvinylidene fluoride composites, Korea-Aust. Rheol. J., 35 (2023) 137-155.
[95] A.M. Chandran, P.K.S. Mural, Surface silanized MWCNTs doped PVDF nanocomposite with self-organized dipoles: an intrinsic study on the dielectric, piezoelectric, ferroelectric, and energy harvesting phenomenology, Sustain. Energy Fuels, 6 (2022) 1641-1653.
[96] N. Abzan, M. Kharaziha, S. Labbaf, Development of three-dimensional piezoelectric polyvinylidene fluoride-graphene oxide scaffold by non-solvent induced phase separation method for nerve tissue engineering, Mat. Des., 167 (2019) 107636.
[97] C. Shuai, Z. Zeng, Y. Yang, F. Qi, S. Peng, W. Yang, C. He, G. Wang, G. Qian, Graphene oxide assists polyvinylidene fluoride scaffold to reconstruct electrical microenvironment of bone tissue, Mat. Des., 190 (2020) 108564.
[98] S. Song, S. Xia, Y. Wei, X. Lv, S. Sun, Q. Li, Integration of PDA Chemistry and Surface-Initiated ATRP to Prepare Poly(methyl methacrylate)-Grafted Carbon Nanotubes and Its Effect on Poly(vinylidene fluoride)–Carbon Nanotube Composite Properties, Macromol. Mater. Eng., 304 (2019) 1900176.
[99] N.V. Lakshmi, P. Tambe, N.K. Sahu, Giant permittivity of three phase polymer nanocomposites obtained by modifying hybrid nanofillers with polyvinylpyrrolidone, Compos. Inter., 25 (2018) 47-67.
[100] W. Li, A. Zheng, Y. Lin, P. Liu, M. Shen, L. Zhou, H. Liu, J. Yuan, S. Qin, X. Zhang, N. Yang, S. Jiang, G. Zhang, High pyroelectric effect in poly(vinylidene fluoride) composites cooperated with diamond nanoparticles, Mater. Lett., 267 (2020) 127514.
[101] Y. Guo, J. Xu, W. Wu, S. Liu, J. Zhao, E. Pawlikowska, M. Szafran, F. Gao, Ultralight graphene aerogel/PVDF composites for flexible piezoelectric nanogenerators, Compos. Commun., 22 (2020) 100542.
[102] A. Adaval, C.K. Subash, V.H. Shafeeq, M. Aslam, T.W. Turney, G.P. Simon, A.R. Bhattacharyya, A comprehensive investigation on the influence of processing techniques on the morphology, structure, dielectric and piezoelectric properties of poly (vinylidene fluoride)/Graphene oxide nanocomposites, Polymer, 256 (2022) 125239.
[103] J.-h. Yang, X. Xie, Z.-z. He, Y. Lu, X.-d. Qi, Y. Wang, Graphene oxide-tailored dispersion of hybrid barium titanate@polypyrrole particles and the dielectric composites, Chem. Eng. J., 355 (2019) 137-149.
[104] A.M. Ismail, M.I. Mohammed, S.S. Fouad, Optical and structural properties of polyvinylidene fluoride (PVDF) / reduced graphene oxide (RGO) nanocomposites, J. Mol. Struct., 1170 (2018) 51-59.
[105] K. Sabira, M.P. Jayakrishnan, P. Saheeda, S. Jayalekshmi, On the absorption dominated EMI shielding effects in free standing and flexible films of poly(vinylidene fluoride)/graphene nanocomposite, Eur. Pol. J., 99 (2018) 437-444.
[106] K. Roy, D. Mandal, PVDF/rGO hybrid film based efficient piezoelectric energy harvester, AIP Conf. Proc., 2115 (2019).
[107] V.G. Alisha, V. Lalan, G. Subodh, Broadband dielectric response of polyvinylidenefluoride reinforced with carbon nanostructures synthesized from the pith of tapioca stem, AIP Conf. Proc., 2265 (2020).
[108] K. Shi, B. Sun, X. Huang, P. Jiang, Synergistic effect of graphene nanosheet and BaTiO3 nanoparticles on performance enhancement of electrospun PVDF nanofiber mat for flexible piezoelectric nanogenerators, Nano Energy, 52 (2018) 153-162.
[109] T. Huang, S. Yang, P. He, J. Sun, S. Zhang, D. Li, Y. Meng, J. Zhou, H. Tang, J. Liang, G. Ding, X. Xie, Phase-Separation-Induced PVDF/Graphene Coating on Fabrics toward Flexible Piezoelectric Sensors, ACS Appl. Mater. Interfaces, 10 (2018) 30732-30740.
[110] Y. Wang, Q. Jiang, W. Jing, Z. Zhong, W. Xing, Pore structure and surface property design of silicon carbide membrane for water-in-oil emulsification, J. Membr. Sci., 648 (2022) 120347.
[111] A. Sasmal, A. Patra, A. Arockiarajan, Tuning the space charge polarization of PVDF based ternary composite for piezo-tribo hybrid energy harvesting, Appl. Phys. Lett., 121 (2022).

[112] S. Anand, S. Pauline, Electromagnetic Interference Shielding Properties of BaCo2Fe16O27 Nanoplatelets and RGO Reinforced PVDF Polymer Composite Flexible Films, Adv. Mater. Interfaces, 8 (2021) 2001810.
[113] P. Huang, S. Xu, W. Zhong, H. Fu, Y. Luo, Z. Xiao, M. Zhang, Carbon quantum dots inducing formation of β phase in PVDF-HFP to improve the piezoelectric performance, 330 (2021) 112880.
[114] S. Ma, L. Jin, X. Huang, C. Riziotis, R. Huang, C. Zhang, J. Lu, W. Yang, Nanogenerators Begin to Light Up: A Novel Poling-Free Piezoelectric System with Multicolor Photoluminescence as an Efficient Mechatronics Development Platform, Adv. Mater. Interfaces, 5 (2018) 1800587.
[115] B. Prasad, F.S. Gill, V. Panwar, G. Anoop, Development of strain sensor using conductive poly(vinylidene fluoride) (PVDF) nanocomposite membrane reinforced with ionic liquid (IL) & carbon nanofiber (CNF), Compos. B. Eng., 173 (2019) 106990.
[116] L.S. Panwar, V. Panwar, G. Anoop, S. Park, Carbon nanofiber-polyelectrolyte triggered piezoelectric polymer-based hydrophilic nanocomposite for high sensing voltage generation, J. Mater. Res. Technol., 17 (2022) 3246-3261.
[117] W. Guo, Z. Liu, Y. Zhu, L. Li, Fabrication of Poly(Vinylidene Fluoride)/Graphene Nano-Composite Micro-Parts with Increased β-Phase and Enhanced Toughness via Micro-Injection Molding, Polymers, 13 (2021) 3292.
[118] Z. Zhou, Z. Zhang, Q. Zhang, H. Yang, Y. Zhu, Y. Wang, L. Chen, Controllable Core–Shell BaTiO3@Carbon Nanoparticle-Enabled P(VDF-TrFE) Composites: A Cost-Effective Approach to High-Performance Piezoelectric Nanogenerators, ACS Appl. Mater. Interfaces, 12 (2020) 1567-1576.
[119] X. Lin, F. Yu, X. Zhang, W. Li, Y. Zhao, X. Fei, Q. Li, C. Yang, S. Huang, Wearable Piezoelectric Films Based on MWCNT-BaTiO3/PVDF Composites for Energy Harvesting, Sensing, and Localization, ACS Appl. Nano Mater., 6 (2023) 11955-11965.
[120] A. Samadi, R. Ahmadi, S.M. Hosseini, Influence of TiO2-Fe3O4-MWCNT hybrid nanotubes on piezoelectric and electromagnetic wave absorption properties of electrospun PVDF nanocomposites, Org. Electron., 75 (2019) 105405.
[121] Y.A. Barnakov, O. Paul, A. Joaquim, A. Falconer, R. Mu, V.Y. Barnakov, D. Dikin, V.P. Petranovskii, A. Zavalin, A. Ueda, F. Williams, Light intensity-induced phase transitions in graphene oxide doped polyvinylidene fluoride, Opt. Mater. Express, 8 (2018) 2579-2585.
[122] R. Bhunia, R. Dey, S. Das, S. Hussain, R. Bhar, A. Kumar Pal, Enhanced piezo-electric property induced in graphene oxide/polyvinylidene fluoride composite flexible thin films, Polym. Compos., 39 (2018) 4205-4216.
[123] H.Q. Ali, I.E. Tabrizi, R.M. Awais Khan, J. Seyyed Monfared Zanjani, C. Yilmaz, L.H. Poudeh, M. Yildiz, Experimental study on dynamic behavior of woven carbon fabric laminates using in-house piezoelectric sensors, Polym. Compos., 28 (2019) 105004.
[124] M. Hasanzadeh, M.R. Ghahhari, S.M. Bidoki, Enhanced piezoelectric performance of PVDF-based electrospun nanofibers by utilizing in situ synthesized graphene-ZnO nanocomposites, J. Mater. Sci: Mater. Electron., 32 (2021) 15789-15800.
[125] R. Sridhar, V. Amith, S. Aditya, A. Gangadhar, K.A. Vishnumurthy, Electrospun PVDF/Cloisite-30B and PVDF/BaTio3/graphene nanofiber mats for development of nanogenerators, J. Indian Chem. Soc., 99 (2022) 100501.
[126] S. Tiwari, D.K. Dubey, O. Prakash, S. Das, P. Maiti, Effect of functionalization on electrospun PVDF nanohybrid for piezoelectric energy harvesting applications, Energy, 275 (2023) 127492.
[127] B.S. Athira, A. George, K. Vaishna Priya, U.S. Hareesh, E.B. Gowd, K.P. Surendran, A. Chandran, High-Performance Flexible Piezoelectric Nanogenerator Based on Electrospun PVDF-BaTiO3 Nanofibers for Self-Powered Vibration Sensing Applications, ACS Appl. Mater. Interfaces, 14 (2022) 44239-44250.
[128] K. Oflaz, İ. Özaytekin, Analysis of electrospinning and additive effect on β phase content of electrospun PVDF nanofiber mats for piezoelectric energy harvester nanogenerators, Smart Mater. Struct., 31 (2022) 105022.
[129] S. Wang, C.-Y. Tang, J. Jia, X.-J. Zha, J.-H. Liu, X. Zhao, K. Ke, Y. Wang, W. Yang, Liquid Electrolyte-Assisted Electrospinning for Boosting Piezoelectricity of Poly (vinylidene Fluoride) Fiber Mats, Macromolecules, 56 (2023) 7479-7489.
[130] M. Pusty, P.M. Shirage, Insights and perspectives on graphene-PVDF based nanocomposite materials for harvesting mechanical energy, J. Alloys Compd., 904 (2022) 164060.

[131] H. Guo, J. Li, Y. Meng, J. de Claville Christiansen, D. Yu, Z. Wu, S. Jiang, Stretch-induced stable-metastable crystal transformation of PVDF/graphene composites, J. Alloys Compd., 2 (2019) e10079.
[132] M. Seena, J. Husna, V. Prasad, Dielectric properties of hot-pressed Poly(vinylidene fluoride)/Functionalized carbon nanotube composites, Mater. Chem. Phys., 285 (2022) 126134.
[133] B. Lin, G.-D. Chen, F.-A. He, Y. Li, Y. Yang, B. Shi, F.-R. Feng, S.-Y. Chen, K.-H. Lam, Preparation of MWCNTs/PVDF composites with high-content β form crystalline of PVDF and enhanced dielectric constant by electrospinning-hot pressing method, Diam. Relat. Mater., 131 (2023) 109556.
[134] H. Zhang, Y. Zhu, L. Li, Fabrication of PVDF/graphene composites with enhanced β phase via conventional melt processing assisted by solid state shear milling technology, RSC Adv., 10 (2020) 3391-3401.
[135] F.S. Gill, D. Uniyal, B. Prasad, S. Saluja, A. Mishra, R.K. Bachheti, S. Juyal, Investigation of increased electrical conductivity by rGO in rGO/PVDF/PMMA/PTFE nanocomposites, J. Mol. Struct., 1267 (2022) 133541.
[136] Z.-z. He, X. Yu, J.-h. Yang, N. Zhang, T. Huang, Y. Wang, Z.-w. Zhou, Largely enhanced dielectric properties of poly(vinylidene fluoride) composites achieved by adding polypyrrole-decorated graphene oxide, Compos. Part A Appl. Sci. Manuf., 104 (2018) 89-100.
[137] S. Badatya, D.K. Bharti, N. Sathish, A.K. Srivastava, M.K. Gupta, Humidity Sustainable Hydrophobic Poly(vinylidene fluoride)-Carbon Nanotubes Foam Based Piezoelectric Nanogenerator, ACS Appl. Mater. Interfaces, 13 (2021) 27245-27254.
[138] R. Senthil Kumar, T. Sarathi, K.K. Venkataraman, A. Bhattacharyya, Enhanced piezoelectric properties of polyvinylidene fluoride nanofibers using carbon nanofiber and electrical poling, Mater. Letter., 255 (2019) 126515.
[139] M. Fortunato, H.C. Bidsorkhi, C.R. Chandraiahgari, G.D. Bellis, F. Sarto, M.S. Sarto, PFM Characterization of PVDF Nanocomposite Films With Enhanced Piezoelectric Response, IEEE Trans. Nanotechnol., 17 (2018) 955-961.
[140] V. Lalan, A. Puthiyedath Narayanan, K.P. Surendran, S. Ganesanpotti, Room-Temperature Ferromagnetic Sr3YCo4O10+δ and Carbon Black-Reinforced Polyvinylidenefluoride Composites toward High-Performance Electromagnetic Interference Shielding, ACS Omega, 4 (2019) 8196-8206.
[141] J. Zhang, X. Wang, X. Chen, X. Xia, G.J. Weng, Piezoelectricity enhancement in graphene/polyvinylidene fluoride composites due to graphene-induced α → β crystal phase transition, Energy Convers. Manag., 269 (2022) 116121.
[142] Y. Lu, X. Xie, W.-y. Wang, X.-d. Qi, Y.-z. Lei, J.-h. Yang, Y. Wang, ZnO nanoparticles-tailored GO dispersion toward flexible dielectric composites with high relative permittivity, low dielectric loss and high breakdown strength, Compos. Part A Appl. Sci. Manuf., 124 (2019) 105489.
[143] J. Nunes-Pereira, P. Sharma, L.C. Fernandes, J. Oliveira, J.A. Moreira, R.K. Sharma, S. Lanceros-Mendez, Poly(vinylidene fluoride) composites with carbon nanotubes decorated with metal nanoparticles, Compos. B. Eng., 142 (2018) 1-8.
[144] S. Barrau, A. Ferri, A. Da Costa, J. Defebvin, S. Leroy, R. Desfeux, J.-M. Lefebvre, Nanoscale Investigations of α- and γ-Crystal Phases in PVDF-Based Nanocomposites, ACS Appl. Mater. Interfaces, 10 (2018) 13092-13099.
[145] D.M. Nivedhitha, S. Jeyanthi, Polyvinylidene fluoride—An advanced smart polymer for electromagnetic interference shielding applications—A novel review, Polym. Adv. Technol., 34 (2023) 1781-1806.
[146] R.P. Clayton, C.S. Robert, A.S. Mark, Introduction to Electromagnetic Compatibility, Wiley, 2022.
[147] J.-C. Huang, EMI shielding plastics: A review, Adv. Polym. Technol., 14 (1995) 137-150.
[148] Advanced materials for electromagnetic shielding: fundamentals, properties, and applications, Jaroszewski, M.W.

Thomas, S.

Rane, A.V. ed., Wiley, Hoboken, NJ, 2019.
[149] D.D.L. Chung, Electromagnetic interference shielding effectiveness of carbon materials, Carbon, 39 (2001) 279-285.

[150] M.H. Al-Saleh, U. Sundararaj, Electromagnetic interference shielding mechanisms of CNT/polymer composites, Carbon, 47 (2009) 1738-1746.
[151] A. Naseer, M. Mumtaz, M. Raffi, I. Ahmad, S.D. Khan, R.I. Shakoor, S. Shahzada, Reinforcement of Electromagnetic Wave Absorption Characteristics in PVDF-PMMA Nanocomposite by Intercalation of Carbon Nanofibers, Electron. Mater. Lett., 15 (2019) 201-207.
[152] Q. Qin, Y. Hu, S. Guo, Y. Yang, T. Lei, Z. Cui, H. Wang, S. Qin, PVDF-based composites for electromagnetic shielding application: a review, J. Polym. Res., 30 (2023) 130.
[153] Y. Li, J. Sun, S. Li, X. Tian, X. Yao, B. Wang, J. Chen, An experimental study of impulse-current-induced mechanical effects on laminated carbon fibre-reinforced polymer composites, Compos. B. Eng. , 225 (2021) 109245.
[154] Y. Gong, W. Zhou, Z. Wang, L. Xu, Y. Kou, H. Cai, X. Liu, Q. Chen, Z.-M. Dang, Towards suppressing dielectric loss of GO/PVDF nanocomposites with TA-Fe coordination complexes as an interface layer, Journal of Materials Science & Technology, 34 (2018) 2415-2423.
[155] B. Zhao, S. Wang, C. Zhao, R. Li, S.M. Hamidinejad, Y. Kazemi, C.B. Park, Synergism between carbon materials and Ni chains in flexible poly(vinylidene fluoride) composite films with high heat dissipation to improve electromagnetic shielding properties, Carbon, 127 (2018) 469-478.
[156] H.K. Choudhary, R. Kumar, S.P. Pawar, U. Sundararaj, B. Sahoo, Enhancing absorption dominated microwave shielding in Co@C–PVDF nanocomposites through improved magnetization and graphitization of the Co@C-nanoparticles, Phys. Chem. Chem. Phys., 21 (2019) 15595-15608.
[157] P. Rani, M. Basheer Ahamed, K. Deshmukh, Dielectric and electromagnetic interference shielding performance of graphene nanoplatelets and copper oxide nanoparticles reinforced polyvinylidenefluoride/poly(3,4-ethylenedioxythiophene)-block-poly (ethylene glycol) blend nanocomposites, Synthetic Metals, 282 (2021) 116923.
[158] K. Rengaswamy, D.K. Sakthivel, A. Muthukaruppan, B. Natesan, S. Venkatachalam, D. Kannaiyan, Electromagnetic interference (EMI) shielding performance of lightweight metal decorated carbon nanostructures dispersed in flexible polyvinylidene fluoride films, New J. Chem., 42 (2018) 12945-12953.
[159] N.D. Kha Tu, M.-S. Noh, Y. Ko, J.-H. Kim, C.Y. Kang, H. Kim, Enhanced electromechanical performance of P(VDF-TrFE-CTFE) thin films hybridized with highly dispersed carbon blacks, Compos. B. Eng., 152 (2018) 133-138.
[160] S.M. Chougule, A. Twinkle, R. Thomas, M. Balachandran, Quantifying the role of nanocarbon fillers on dielectric properties of poly(vinylidene fluoride) matrix, Polym. Polym. Compos., 30 (2022) 09673911221087597.
[161] X. Yin, K. Wei, W. Wei, G. He, Y. Feng, Improvement of the dielectric properties of polyvinylidene fluoride/reduced graphene oxide composites via in-situ ball milling assisted extensional flow field, J. Appl. Polym. Sci., 139 (2022) e53164.
[162] X. Mei, L. Lu, Y. Xie, W. Wang, Y. Tang, K.S. Teh, An ultra-thin carbon-fabric/graphene/poly(vinylidene fluoride) film for enhanced electromagnetic interference shielding, Nanoscale, 11 (2019) 13587-13599.
[163] E. Kolonelou, E. Loupou, P.A. Klonos, E. Sakellis, D. Valadorou, A. Kyritsis, A.N. Papathanassiou, Thermal and electrical characterization of poly(vinyl)alcohol)/poly(vinylidene fluoride) blends reinforced with nano-graphene platelets, Polymer, 224 (2021) 123731.
[164] Z. Fan, R. Liu, X. Cheng, Nonwoven composite endowed with electromagnetic shielding performance by graphene nanosheets adherence, The Journal of The Textile Institute, 113 (2022) 1411-1417.
[165] B. Zhao, C. Zhao, M. Hamidinejad, C. Wang, R. Li, S. Wang, K. Yasamin, C.B. Park, Incorporating a microcellular structure into PVDF/graphene–nanoplatelet composites to tune their electrical conductivity and electromagnetic interference shielding properties, J. Mater. Chem. C, 6 (2018) 10292-10300.
[166] H. Cheraghi Bidsorkhi, A.G. D'Aloia, A. Tamburrano, G. De Bellis, A. Delfini, P. Ballirano, M.S. Sarto, 3D Porous Graphene Based Aerogel for Electromagnetic Applications, Sci. Rep., 9 (2019) 15719.
[167] W. Tuichai, A. Karaphun, C. Ruttanapun, Improved dielectric properties of PVDF polymer composites filled with Ag nanomaterial deposited reduced graphene oxide (rGO) hybrid particles, Materials Research Bulletin, 145 (2022) 111552.

[168] H. Ma, Z. Xie, Y. Liu, Q. Zhang, P. Gong, F. Meng, Y. Niu, C.B. Park, G. Li, Improved dielectric and electromagnetic interference shielding performance of materials by hybrid filler network design in three-dimensional nanocomposite films, Mater. Des, 226 (2023) 111666.
[169] X.-J. Zha, J.-H. Pu, L.-F. Ma, T. Li, R.-Y. Bao, L. Bai, Z.-Y. Liu, M.-B. Yang, W. Yang, A particular interfacial strategy in PVDF/OBC/MWCNT nanocomposites for high dielectric performance and electromagnetic interference shielding, Compos. - A: Appl. Sci. Manuf., 105 (2018) 118-125.
[170] R. Ram, D. Khastgir, M. Rahaman, Physical properties of polyvinylidene fluoride/multi-walled carbon nanotube nanocomposites with special reference to electromagnetic interference shielding effectiveness, Adv Polym Technol., 37 (2018) 3287-3296.
[171] G.S. Kumar, T.U. Patro, Efficient electromagnetic interference shielding and radar absorbing properties of ultrathin and flexible polymer-carbon nanotube composite films, Mater. Res. Express, 5 (2018) 115304.
[172] S.J. Kazmi, M. Nadeem, M.A. Warsi, S. Manzoor, B. Shabbir, S. Hussain, PVDF/CFO-anchored CNTs ternary composite system with enhanced EMI shielding and EMW absorption properties, J. Alloys Compd., 903 (2022) 163938.
[173] Z. Ertekin, M. Secmen, M. Erol, Electromagnetic shielding effectiveness and microwave properties of expanded graphite-ionic liquid co-doped PVDF, J. Mater. Sci: Mater. Electron., 34 (2023) 43.
[174] Y.a. Bai, X. Wei, D. Dun, S. Bai, H. Zhou, B. Wen, X. Wang, J. Hu, Lightweight poly(vinylidene fluoride) based quaternary nanocomposite foams with efficient and tailorable electromagnetic interference shielding properties, Polym. Compos., 44 (2023) 1951-1966.
[175] Y.-L. Wang, S.-H. Yang, H.-Y. Wang, G.-S. Wang, X.-B. Sun, P.-G. Yin, Hollow porous CoNi/C composite nanomaterials derived from MOFs for efficient and lightweight electromagnetic wave absorber, Carbon, 167 (2020) 485-494.
[176] W. Dong, L. He, C. Chen, J. Kang, H. Niu, J. Zhang, J. Li, K. Li, Preparation and electromagnetic shielding performances of graphene/TPU–PVDF nanocomposites by high-energy ball milling, J Mater Sci: Mater Electron, 33 (2022) 1817-1829.
[177] L.J. Jia, A.D. Phule, Y. Geng, S. Wen, L. Li, Z.X. Zhang, Microcellular Conductive Carbon Black or Graphene/PVDF Composite Foam with 3D Conductive Channel: A Promising Lightweight, Heat-Insulating, and EMI-Shielding Material, Macromol. Mater. Eng., 306 (2021) 2000759.
[178] S. Biswas, T.S. Muzata, B. Krause, P. Rzeczkowski, P. Pötschke, S. Bose, Does the Type of Polymer and Carbon Nanotube Structure Control the Electromagnetic Shielding in Melt-Mixed Polymer Nanocomposites?, J. Compos. Sci., 4 (2020) 9.
[179] K.K. Halder, M. Tomar, V.K. Sachdev, V. Gupta, Carbonized Charcoal-Loaded PVDF Polymer Composite: A Promising EMI Shielding Material, Arab. J. Sci. Eng., 45 (2020) 465-474.
[180] A. Leão, T. Indrusiak, M.F. Costa, B.G. Soares, Exploring the Potential Use of Clean Scrap PVDF as Matrix for Conductive Composites Based on Graphite, Carbon Black and Hybrids: Electromagnetic Interference Shielding Effectiveness (EMI SE), J. Polym. Environ., 28 (2020) 2091-2100.
[181] H. Wang, K. Zheng, X. Zhang, Y. Wang, C. Xiao, L. Chen, X. Tian, Separated poly(vinylidene fluoride)/carbon black composites containing magnetic carbonyl iron particles for efficient electromagnetic interference shielding, Mater. Res. Express, 5 (2018) 125304.
[182] A. Kolanowska, D. Janas, A.P. Herman, R.G. Jędrysiak, T. Giżewski, S. Boncel, From blackness to invisibility – Carbon nanotubes role in the attenuation of and shielding from radio waves for stealth technology, Carbon, 126 (2018) 31-52.
[183] J. Lee, S. Lim, Polarization behavior of polyvinylidene fluoride films with the addition of reduced graphene oxide, Journal of Industrial and Engineering Chemistry, 67 (2018) 478-485.
[184] I. Hussain, S. Yesmin, Enhanced microwave absorption of g-C3N4/poly(vinylidene difluoride)/carbon black composites, Mater. Chem. Phys., 309 (2023) 128337.
[185] H.K. Choudhary, R. Kumar, S.P. Pawar, U. Sundararaj, B. Sahoo, Effect of morphology and role of conductivity of embedded metallic nanoparticles on electromagnetic interference shielding of PVDF-carbonaceous-nanofiller composites, Carbon, 164 (2020) 357-368.
[186] S. Zeng, X. Li, M. Li, J. Zheng, S. E, W. Yang, B. Zhao, X. Guo, R. Zhang, Flexible PVDF/CNTs/Ni@CNTs composite films possessing excellent electromagnetic interference shielding and mechanical properties under heat treatment, Carbon, 155 (2019) 34-43.

[187] K. Dinakaran, K. Narayanasamy, S. Theerthagiri, P. Peethambaram, S. Krishnan, D. Roy, Microwave absorption and dielectric behavior of lead sulfide – graphene composite nanostructure embedded polyvinylidinedilfuoride thin films, International Journal of Polymer Analysis and Characterization, 27 (2022) 277-288.
[188] H.K. Choudhary, R. Kumar, S.P. Pawar, U. Sundararaj, B. Sahoo, Superiority of graphite coated metallic-nanoparticles over graphite coated insulating-nanoparticles for enhancing EMI shielding, New J. Chem., 45 (2021) 4592-4600.
[189] L. Liang, P. Xu, Y. Wang, Y. Shang, J. Ma, F. Su, Y. Feng, C. He, Y. Wang, C. Liu, Flexible polyvinylidene fluoride film with alternating oriented graphene/Ni nanochains for electromagnetic interference shielding and thermal management, Chem. Eng. J., 395 (2020) 125209.
[190] R. Li, S. Wang, P. Bai, B. Fan, B. Zhao, R. Zhang, Enhancement of electromagnetic interference shielding from the synergism between Cu@Ni nanorods and carbon materials in flexible composite films, Mater. Adv., 2 (2021) 718-727.
[191] F. Li, W. Zhan, Y. Su, S.H. Siyal, G. Bai, W. Xiao, A. Zhou, G. Sui, X. Yang, Achieving excellent electromagnetic wave absorption of ZnFe2O4@CNT/polyvinylidene fluoride flexible composite membranes by adjusting processing conditions, Compos. - A: Appl. Sci. Manuf., 133 (2020) 105866.
[192] D. Sharma, A.V. Menon, S. Bose, Graphene templated growth of copper sulphide 'flowers' can suppress electromagnetic interference, Nanoscale Adv., 2 (2020) 3292-3303.
[193] K. Rengaswamy, V.K. Asapu, A. Muthukaruppan, D.K. Sakthivel, S. Venkatachalam, D. Kannaiyan, Enhanced shielding of electromagnetic radiations with flexible, light-weight, and conductive Ag-Cu/MWCNT/rGO architected PVDF nanocomposite films, Polym Adv Technol., 32 (2021) 3759-3769.
[194] P. Zhang, X. Ding, Y. Wang, Y. Gong, K. Zheng, L. Chen, X. Tian, X. Zhang, Segregated double network enabled effective electromagnetic shielding composites with extraordinary electrical insulation and thermal conductivity, Compos. - A: Appl. Sci. Manuf., 117 (2019) 56-64.
[195] K. Sushmita, P. Formanek, B. Krause, P. Pötschke, S. Bose, Distribution of Carbon Nanotubes in Polycarbonate-Based Blends for Electromagnetic Interference Shielding, ACS Appl. Nano Mater., 5 (2022) 662-677.
[196] I. Ahmed, R. Jan, A.N. Khan, I.H. Gul, R. Khan, S. Javed, M.A. Akram, A. Shafqat, H.M. Cheema, I. Ahmad, Graphene-ferrites interaction for enhanced EMI shielding effectiveness of hybrid polymer composites, Mater. Res. Express, 7 (2020) 016304.
[197] B. Zhao, Y. Li, X. Guo, R. Zhang, J. Zhang, H. Hou, T. Ding, J. Fan, Z. Guo, Enhanced electromagnetic wave absorbing nickel (Oxide)-Carbon nanocomposites, Ceram. Int., 45 (2019) 24474-24486.
[198] X. Shen, S.-H. Yang, P.-G. Yin, C.-Q. Li, J.-R. Ye, G.-S. Wang, Enhancement in microwave absorption properties by adjusting the sintering conditions and carbon shell thickness of Ni@C submicrospheres, CrystEngComm, 24 (2022) 765-774.
[199] R. Peymanfar, A. Ahmadi, E. Selseleh-Zakerin, A. Ghaffari, M.M. Mojtahedi, A. Sharifi, Electromagnetic and optical characteristics of wrinkled Ni nanostructure coated on carbon microspheres, Chem. eng. J., 405 (2021) 126985.
[200] W.-G. Cui, X. Zhou, B. Zhao, W. You, Y. Yang, B. Fan, L. Wu, R. Che, 3D porous PVDF foam anchored with ultra-low content of graphene and Ni nanochains towards wideband electromagnetic waves absorption, Carbon, 210 (2023) 118070.
[201] M. Lu, N. Gao, X.-J. Zhang, G.-S. Wang, Reduced graphene oxide decorated with octahedral NiS2/NiS nanocrystals: facile synthesis and tunable high frequency attenuation, RSC Adv., 9 (2019) 5550-5556.
[202] B. Zhao, S. Zeng, X. Li, X. Guo, Z. Bai, B. Fan, R. Zhang, Flexible PVDF/carbon materials/Ni composite films maintaining strong electromagnetic wave shielding under cyclic microwave irradiation, J. Mater. Chem. C, 8 (2020) 500-509.
[203] R. Peymanfar, F. Fazlalizadeh, Fabrication of expanded carbon microspheres/ZnAl2O4 nanocomposite and investigation of its microwave, magnetic, and optical performance, J. Alloys Compd., 854 (2021) 157273.
[204] F. Li, W. Zhan, L. Zhuang, L. Zhou, M. Zhou, G. Bai, A. Zhou, W. Xiao, X. Yang, G. Sui, Acquirement of Strong Microwave Absorption of ZnFe2O4@SiO2@Reduced Graphene Oxide/PVDF

Composite Membranes by Regulating Crystallization Behavior, J. Phys. Chem. C, 124 (2020) 14861-14872.
[205] M. Amini, M. Kamkar, F. Rahmani, A. Ghaffarkhah, F. Ahmadijokani, M. Arjmand, Multilayer Structures of a Zn0.5Ni0.5Fe2O4-Reduced Graphene Oxide/PVDF Nanocomposite for Tunable and Highly Efficient Microwave Absorbers, ACS Appl. Electron. Mater., 3 (2021) 5514-5527.
[206] Q. Qi, L. Ma, B. Zhao, S. Wang, X. Liu, Y. Lei, C.B. Park, An Effective Design Strategy for the Sandwich Structure of PVDF/GNP-Ni-CNT Composites with Remarkable Electromagnetic Interference Shielding Effectiveness, ACS Appl. Mater. Interfaces, 12 (2020) 36568-36577.
[207] Y.-L. Wang, G.-S. Wang, X.-J. Zhang, C. Gao, Porous carbon polyhedrons coupled with bimetallic CoNi alloys for frequency selective wave absorption at ultralow filler loading, J. Mater. Sci. Technol., 103 (2022) 34-41.
[208] S. Anand, S. Pauline, C.J. Prabagar, Zr doped Barium hexaferrite nanoplatelets and RGO fillers embedded Polyvinylidenefluoride composite films for electromagnetic interference shielding applications, Polymer Testing, 86 (2020) 106504.
[209] T. Chakraborty, S. Sharma, T. Debnath, A. Sinha Mahapatra, A. Selvam, S. Chakrabarti, S. Sutradhar, Fabrication of heterostructure composites of Ni-Zn-Cu-Ferrite-C3N4-Poly(vinylidene fluoride) films for the enhancement of electromagnetic interference shielding effectiveness, Chem. Eng. J., 420 (2021) 127683.
[210] Y. Bhattacharjee, S. Bapari, S. Bose, Mechanically robust, UV screener core–double-shell nanostructures provide enhanced shielding for EM radiations over wide angle of incidence, Nanoscale, 12 (2020) 15775-15790.
[211] V. Lalan, S. Ganesanpotti, Broadband Electromagnetic Response and Enhanced Microwave Absorption in Carbon Black and Magnetic Fe3O4 Nanoparticles Reinforced Polyvinylidenefluoride Composites, J. Electron. Mater., 49 (2020) 1666-1676.
[212] H. Cheng, S. Wei, Y. Ji, J. Zhai, X. Zhang, J. Chen, C. Shen, Synergetic effect of Fe3O4 nanoparticles and carbon on flexible poly (vinylidence fluoride) based films with higher heat dissipation to improve electromagnetic shielding, Composites Part A: Applied Science and Manufacturing, 121 (2019) 139-148.
[213] Y. Li, D. Zhang, B. Zhou, C. He, B. Wang, Y. Feng, C. Liu, Synergistically enhancing electromagnetic interference shielding performance and thermal conductivity of polyvinylidene fluoride-based lamellar film with MXene and graphene, Composites Part A: Applied Science and Manufacturing, 157 (2022) 106945.
[214] K. Raagulan, R. Braveenth, H.J. Jang, Y. Seon Lee, C.-M. Yang, B. Mi Kim, J.J. Moon, K.Y. Chai, Electromagnetic Shielding by MXene-Graphene-PVDF Composite with Hydrophobic, Lightweight and Flexible Graphene Coated Fabric, Materials, 11 (2018) 1803.
[215] R. Li, L. Ding, Q. Gao, H. Zhang, D. Zeng, B. Zhao, B. Fan, R. Zhang, Tuning of anisotropic electrical conductivity and enhancement of EMI shielding of polymer composite foam via CO2-assisted delamination and orientation of MXene, Chem. Eng. J., 415 (2021) 128930.
[216] A. Samadi, S.M. Hosseini, M. Mohseni, Investigation of the electromagnetic microwaves absorption and piezoelectric properties of electrospun Fe3O4-GO/PVDF hybrid nanocomposites, Organic Electronics, 59 (2018) 149-155.
[217] Y. Li, Y. Duan, C. Wang, Enhanced Microwave Absorption and Electromagnetic Properties of Si-Modified rGO@Fe3O4/PVDF-co-HFP Composites, Materials, 13 (2020) 933.
[218] S. Ayub, B.H. Guan, K.Y. You, Electromagnetic interference shielding mechanisms of MMG@PVDF composites for a broadband frequency range, Mater. Today Commun., 35 (2023) 106273.
[219] Y. Zhao, J. Hou, Z. Bai, Y. Yang, X. Guo, H. Cheng, Z. Zhao, X. Zhang, J. Chen, C. Shen, Facile preparation of lightweight PE/PVDF/Fe3O4/CNTs nanocomposite foams with high conductivity for efficient electromagnetic interference shielding, Compos. - A: Appl. Sci. Manuf., 139 (2020) 106095.
[220] S. Ayub, B.H. Guan, F. Ahmad, Y.A. Oluwatobi, Z.U. Nisa, M.F. Javed, A. Mosavi, Graphene and Iron Reinforced Polymer Composite Electromagnetic Shielding Applications: A Review, Polymers, 13 (2021) 2580.
[221] C. Liang, M. Hamidinejad, L. Ma, Z. Wang, C.B. Park, Lightweight and flexible graphene/SiC-nanowires/ poly(vinylidene fluoride) composites for electromagnetic interference shielding and thermal management, Carbon, 156 (2020) 58-66.

[222] S. Acharya, S. Datar, Wideband (8–18 GHz) microwave absorption dominated electromagnetic interference (EMI) shielding composite using copper aluminum ferrite and reduced graphene oxide in polymer matrix, J. App. Phys., 128 (2020) 104902
[223] M.A. Darwish, A.T. Morchenko, H.F. Abosheiasha, V.G. Kostishyn, V.A. Turchenko, M.A. Almessiere, Y. Slimani, A. Baykal, A.V. Trukhanov, Impact of the exfoliated graphite on magnetic and microwave properties of the hexaferrite-based composites, J. Alloys Compd., 878 (2021) 160397.
[224] A. Harish Kumar, M.B. Ahamed, K. Deshmukh, M.S. Sirajuddeen, Morphology, Dielectric and EMI Shielding Characteristics of Graphene Nanoplatelets, Montmorillonite Nanoclay and Titanium Dioxide Nanoparticles Reinforced Polyvinylidenefluoride Nanocomposites, J. Inorg. Organomet. Polym., 31 (2021) 2003-2016.
[225] S. Acharya, C.S. Gopinath, P. Alegaonkar, S. Datar, Enhanced microwave absorption property of Reduced Graphene Oxide (RGO)–Strontium Hexaferrite (SF)/Poly (Vinylidene) Fluoride (PVDF), Diamond and Related Materials, 89 (2018) 28-34.
[226] A.G. D'Aloia, H.C. Bidsorkhi, G.D. Bellis, M.S. Sarto, Graphene Based Wideband Electromagnetic Absorbing Textiles at Microwave Bands, IEEE Trans. Electromagn. Compat., 64 (2022) 710-719.
[227] W. Li, Z. Song, J. Qian, Z. Tan, H. Chu, X. Wu, W. Nie, X. Ran, Enhancing conjugation degree and interfacial interactions to enhance dielectric properties of noncovalent functionalized graphene/poly (vinylidene fluoride) composites, Carbon, 141 (2019) 728-738.
[228] M. Li, J. Liu, D. Zheng, M. Zheng, Y. Zhao, M. Hu, G.H. Yue, G. Shan, Enhanced dielectric permittivity and suppressed electrical conductivity in polyvinylidene fluoride nanocomposites filled with 4,4′-oxydiphenol-functionalized graphene, Nanotechnology, 30 (2019) 265705.
[229] Y. Kou, W. Zhou, L. Xu, H. Cai, G. Wang, X. Liu, Q. Chen, Z.-M. Dang, Surface modification of GO by PDA for dielectric material with well-suppressed dielectric loss, High Performance Polymers, 31 (2019) 1183-1194.
[230] L.A. Silva, A.M.S. Batista, T. Serodre, A.T.B. Neto, C.A. Furtado, L.O. Faria, Enhancement of X-ray Shielding Properties of PVDF/BaSO4 Nanocomposites Filled with Graphene Oxide, MRS Advances, 4 (2019) 169-175.
[231] I. Cacciotti, M. Valentini, M. Raio, F. Nanni, Design and development of advanced BaTiO3/MWCNTs/PVDF multi-layered systems for microwave applications, Compos. Struct., 224 (2019) 111075.
[232] S. Chauhan, P. Nikhil Mohan, K.C.J. Raju, S. Ghotia, N. Dwivedi, C. Dhand, S. Singh, P. Kumar, Free-standing polymer/multiwalled carbon nanotubes composite thin films for high thermal conductivity and prominent EMI shielding, Colloids Surf. A: Physicochem. Eng. Asp., 673 (2023) 131811.
[233] S. Habibpour, K. Zarshenas, M. Zhang, M. Hamidinejad, L. Ma, C.B. Park, A. Yu, Greatly Enhanced Electromagnetic Interference Shielding Effectiveness and Mechanical Properties of Polyaniline-Grafted Ti3C2Tx MXene–PVDF Composites, ACS Appl. Mater. Interfaces, 14 (2022) 21521-21534.
[234] D.G. Miller, Thermodynamics of Irreversible Processes. The Experimental Verification of the Onsager Reciprocal Relations, Chem. Rev., 60 (1960) 15-37.
[235] A. Manbachi, R.S.C. Cobbold, Development and Application of Piezoelectric Materials for Ultrasound Generation and Detection, Ultrasound, 19 (2011) 187-196.
[236] IEEE Standard on Piezoelectricity, (1988) 0_1.
[237] J. Yang, An Introduction to the Theory of Piezoelectricity, Springer New York, NY, 2004.
[238] R.S.C. Cobbol, Foundations of Biomedical Ultrasound, Oxford University Press, 2006.
[239] R.S. Dahiya, M. Valle, Robotic Tactile Sensing, Springer Dordrecht, 2012.
[240] K.S. Ramadan, D. Sameoto, S. Evoy, A review of piezoelectric polymers as functional materials for electromechanical transducers, Smart Mater. Struct., 23 (2014) 033001.
[241] Z.L. Wang, J. Song, Piezoelectric Nanogenerators Based on Zinc Oxide Nanowire Arrays, Science, 312 (2006) 242-246.
[242] F.-R. Fan, Z.-Q. Tian, Z. Lin Wang, Flexible triboelectric generator, Nano Energy, 1 (2012) 328-334.
[243] Y. Yang, W. Guo, K.C. Pradel, G. Zhu, Y. Zhou, Y. Zhang, Y. Hu, L. Lin, Z.L. Wang, Pyroelectric Nanogenerators for Harvesting Thermoelectric Energy, Nano Lett., 12 (2012) 2833-2838.

[244] X. Li, W. Wang, W. Cai, H. Liu, H. Liu, N. Han, X. Zhang, MXene/Multiwalled Carbon Nanotube/Polymer Hybrids for Tribopiezoelectric Nanogenerators, ACS Appl. Nano Mater., 5 (2022) 12836-12847.
[245] Ö.F. Ünsal, A. Çelik Bedeloğlu, Correction to "Three-Dimensional Piezoelectric–Triboelectric Hybrid Nanogenerators for Mechanical Energy Harvesting", ACS Appl. Nano Mater., 6 (2023) 17348-17348.
[246] A. Gebrekrstos, T.S. Muzata, S.S. Ray, Nanoparticle-Enhanced β-Phase Formation in Electroactive PVDF Composites: A Review of Systems for Applications in Energy Harvesting, EMI Shielding, and Membrane Technology, ACS Appl. Nano Mater., 5 (2022) 7632-7651.
[247] L. Wu, Z. Jin, Y. Liu, H. Ning, X. Liu, Alamusi, N. Hu, Recent advances in the preparation of PVDF-based piezoelectric materials, Nanotechnol. Rev., 11 (2022) 1386-1407.
[248] L. Lu, W. Ding, J. Liu, B. Yang, Flexible PVDF based piezoelectric nanogenerators, Nano Energy, 78 (2020) 105251.
[249] S.K. Karan, S. Maiti, A.K. Agrawal, A.K. Das, A. Maitra, S. Paria, A. Bera, R. Bera, L. Halder, A.K. Mishra, J.K. Kim, B.B. Khatua, Designing high energy conversion efficient bio-inspired vitamin assisted single-structured based self-powered piezoelectric/wind/acoustic multi-energy harvester with remarkable power density, Nano Energy, 59 (2019) 169-183.
[250] L. Lapčinskis, K. Mālnieks, A. Linarts, J. Blūms, K.n. Šmits, M. Järvekülg, M.r. Knite, A. Šutka, Hybrid Tribo-Piezo-Electric Nanogenerator with Unprecedented Performance Based on Ferroelectric Composite Contacting Layers, ACS Appl. Energy Mater., 2 (2019) 4027-4032.
[251] A. Kaeopisan, H. Wattanasarn, Piezoelectric PVDF/CNT Flexible Applied on Motorcycle, Integr. Ferroelectr., 214 (2021) 166-172.
[252] Ö.F. Ünsal, Y. Altın, A. Çelik Bedeloğlu, Poly(vinylidene fluoride) nanofiber-based piezoelectric nanogenerators using reduced graphene oxide/polyaniline, J. Appl. Polym. Sci., 137 (2020) 48517.
[253] S. Badatya, A. Kumar, C. Sharma, A.K. Srivastava, J.P. Chaurasia, M.K. Gupta, Transparent flexible graphene quantum dot-(PVDF-HFP) piezoelectric nanogenerator, Mat. Lett., 290 (2021) 129493.
[254] J. Vicente, P. Costa, S. Lanceros-Mendez, J.M. Abete, A. Iturrospe, Electromechanical Properties of PVDF-Based Polymers Reinforced with Nanocarbonaceous Fillers for Pressure Sensing Applications, Materials, 12 (2019) 3545.
[255] K.-C. Jung, S.-H. Chang, Performance evaluation of smart grid fabrics comprising carbon dry fabrics and PVDF ribbon sensors for structural health monitoring, Compos. B. Eng., 163 (2019) 690-701.
[256] M.T. Rahman, S.M.S. Rana, M.A. Zahed, S. Lee, E.-S. Yoon, J.Y. Park, Metal-organic framework-derived nanoporous carbon incorporated nanofibers for high-performance triboelectric nanogenerators and self-powered sensors, Nano Energy, 94 (2022) 106921.
[257] W. Zeng, W. Deng, T. Yang, S. Wang, Y. Sun, J. Zhang, X. Ren, L. Jin, L. Tang, W. Yang, Gradient CNT/PVDF piezoelectric composite with enhanced force-electric coupling for soccer training, Nano Res., 16 (2023) 11312-11319.
[258] E.S. Kadir, R.N. Gayen, Graphene oxide incorporated flexible and free-standing PVDF/ZnO composite membrane for mechanical energy harvesting, Sensors and Actuators A: Physical, 333 (2022) 113305.
[259] R. Tao, J. Shi, M. Rafiee, A. Akbarzadeh, D. Therriault, Fused filament fabrication of PVDF films for piezoelectric sensing and energy harvesting applications, Mater. Adv., 3 (2022) 4851-4860.
[260] U. Uyor, A. Popoola, O. Popoola, V. Aigbodion, Enhanced dielectric performance and energy storage density of polymer/graphene nanocomposites prepared by dual fabrication, Journal of Thermoplastic Composite Materials, 33 (2020) 270-285.
[261] D.L. Vu, C.D. Le, K.K. Ahn, Functionalized graphene oxide/polyvinylidene fluoride composite membrane acting as a triboelectric layer for hydropower energy harvesting, Intl J of Energy Research, 46 (2022) 9549-9559.
[262] S. Song, Y. Li, Q. Wang, C. Zhang, Boosting piezoelectric performance with a new selective laser sintering 3D printable PVDF/graphene nanocomposite, Composites Part A: Applied Science and Manufacturing, 147 (2021) 106452.

[263] Y. Yang, J. Chen, Y. Li, D. Shi, B. Lin, S. Zhang, Y. Tang, F. He, K. Lam, Preparation and dielectric properties of composites based on PVDF and PVDF-grafted graphene obtained from electrospinning-hot pressing method, Journal of Macromolecular Science, Part A, 55 (2018) 148-153.
[264] D. Valadorou, A.N. Papathanassiou, E. Kolonelou, E. Sakellis, Boosting the electro-mechanical coupling of piezoelectric polyvinyl alcohol–polyvinylidene fluoride blends by dispersing nano-graphene platelets, J. Phys. D: Appl. Phys., 55 (2022) 295501.
[265] L. Yang, M. Cheng, W. Lyu, M. Shen, J. Qiu, H. Ji, Q. Zhao, Tunable piezoelectric performance of flexible PVDF based nanocomposites from MWCNTs/graphene/MnO2 three-dimensional architectures under low poling electric fields, Composites Part A: Applied Science and Manufacturing, 107 (2018) 536-544.
[266] J. Cai, N. Hu, L. Wu, Y. Liu, Y. Li, H. Ning, X. Liu, L. Lin, Preparing carbon black/graphene/PVDF-HFP hybrid composite films of high piezoelectricity for energy harvesting technology, Composites Part A: Applied Science and Manufacturing, 121 (2019) 223-231.
[267] R. Li, Q. Guo, Z. Shi, J. Pei, Effects of conductive carbon black on PZT/PVDF composites, Ferroelectrics, 526 (2018) 176-186.
[268] Y. Li, W. Tong, J. Yang, Z. Wang, D. Wang, Q. An, Y. Zhang, Electrode-free piezoelectric nanogenerator based on carbon black/polyvinylidene fluoride–hexafluoropropylene composite achieved via interface polarization effect, Chem. Eng. J., 457 (2023) 141356.
[269] D. Xu, H. Zhang, L. Pu, L. Li, Fabrication of Poly(vinylidene fluoride)/Multiwalled carbon nanotube nanocomposite foam via supercritical fluid carbon dioxide: Synergistic enhancement of piezoelectric and mechanical properties, Compos. Sci. Technol., 192 (2020) 108108.
[270] Y. Li, L. Zheng, L. Song, Y. Han, Y. Yang, C. Tan, Toward Balanced Piezoelectric and Mechanical Performance: 3D Printed Polyvinylidene Fluoride/Carbon Nanotube Energy Harvester with Hierarchical Structure, Ind. Eng. Chem. Res., 61 (2022) 13063-13071.
[271] B. Zhao, M. Hamidinejad, C. Zhao, R. Li, S. Wang, Y. Kazemi, C.B. Park, A versatile foaming platform to fabricate polymer/carbon composites with high dielectric permittivity and ultra-low dielectric loss, J. Mater. Chem. A, 7 (2019) 133-140.
[272] K. Li, X. Liu, Y. Liu, X. Wang, A piezoelectric generator based on PVDF/GO nanofiber membrane, J. Phys.: Conf. Ser., 1052 (2018) 012110.
[273] S.P. Muduli, S. Parida, S. Nayak, S.K. Rout, Effect of Graphene Oxide loading on ferroelectric and dielectric properties of hot pressed poly(vinylidene fluoride) matrix composite film, Polym. Compos., 41 (2020) 2855-2865.
[274] I.O. Pariy, A.A. Ivanova, V.V. Shvartsman, D.C. Lupascu, G.B. Sukhorukov, T. Ludwig, A. Bartasyte, S. Mathur, M.A. Surmeneva, R.A. Surmenev, Piezoelectric Response in Hybrid Micropillar Arrays of Poly(Vinylidene Fluoride) and Reduced Graphene Oxide, Polymers, 11 (2019) 1065.
[275] Y. Dai, X. Zhong, T. Xu, Y. Li, Y. Xiong, S. Zhang, High-Performance Triboelectric Nanogenerator Based on Electrospun Polyvinylidene Fluoride-Graphene Oxide Nanosheet Composite Nanofibers, Energy Technol., 11 (2023) 2300426.
[276] J. Kang, T. Liu, Y. Lu, L. Lu, K. Dong, S. Wang, B. Li, Y. Yao, Y. Bai, W. Fan, Polyvinylidene fluoride piezoelectric yarn for real-time damage monitoring of advanced 3D textile composites, Compos. B. Eng., 245 (2022) 110229.
[277] M. Abbasipour, R. Khajavi, A.A. Yousefi, M.E. Yazdanshenas, F. Razaghian, A. Akbarzadeh, Improving piezoelectric and pyroelectric properties of electrospun PVDF nanofibers using nanofillers for energy harvesting application, Polym Adv Technol, 30 (2019) 279-291.
[278] Z. Zhu, Y. Liu, J. Ge, Z. Hu, G. Zeng, X. Peng, W. Xu, X. Peng, High dielectric polymer composites from thermal-induced in-situ formation of conjugated structures and reduced graphene oxide, Materials Chemistry and Physics, 262 (2021) 124276.
[279] Z. Peng, X. Zhang, C. Zhao, C. Gan, C. Zhu, Hydrophobic and stable MXene/ reduced graphene oxide/polymer hybrid materials pressure sensors with an ultrahigh sensitive and rapid response speed pressure sensor for health monitoring, Materials Chemistry and Physics, 271 (2021) 124729.
[280] Y. Chen, W. Tong, X. Wang, P. Zhang, S. Wang, Y. Zhang, MXene effectively enhances the electron-withdrawing (EW) ability and dielectric properties of PVDF-TrFE nanofibers for triboelectric nanogenerators, Colloids Surf. A: Physicochem. Eng. Asp., 664 (2023) 131172.
[281] Y. Wang, P. He, F. Li, Graphene-improved dielectric property of CCTO/PVDF composite film, Ferroelectrics, 540 (2019) 154-161.

[282] M. Yang, C. Hu, H. Zhao, P. Haghi-Ashtiani, D. He, Y. Yang, J. Yuan, J. Bai, Core@double-shells nanowires strategy for simultaneously improving dielectric constants and suppressing losses of poly(vinylidene fluoride) nanocomposites, Carbon, 132 (2018) 152-156.
[283] F. Kong, M. Chang, Z. Wang, Comprehensive Analysis of Mechanical Properties of CB/SiO2/PVDF Composites, Polymers, 12 (2020) 146.
[284] X. Zheng, H. Yu, S. Yue, R. Xing, Q. Zhang, Y. Liu, B. Zhang, Functionalization of Graphene and Dielectric Property Relationships in PVDF/graphene Nanosheets Composites, Int. J. Electrochem. Sci., (2018).
[285] M. Zeyrek Ongun, S. Oguzlar, E.C. Doluel, U. Kartal, M. Yurddaskal, Enhancement of piezoelectric energy-harvesting capacity of electrospun β-PVDF nanogenerators by adding GO and rGO, J Mater Sci: Mater Electron, 31 (2020) 1960-1968.
[286] T. Aravinda, S. Rao, V. Kumbla, J. Pattar, Improved dielectric performance of polyvinylidene fluoride (PVDF) - Carbon dots composites, Physica E Low Dimens. Syst. Nanostruct., 147 (2023) 115589.
[287] M. Yang, H. Zhao, C. Hu, P. Haghi-Ashtiani, D. He, Z.-M. Dang, J. Bai, Largely enhanced dielectric constant of PVDF nanocomposites through a core–shell strategy, Phys. Chem. Chem. Phys., 20 (2018) 2777-2786.
[288] Z. Chen, H. Li, G. Xie, K. Yang, Core–shell structured Ag@C nanocables for flexible ferroelectric polymer nanodielectric materials with low percolation threshold and excellent dielectric properties, RSC Adv., 8 (2018) 1-9.
[289] T. Patodia, K.B. Sharma, S. Dixit, N. Srivastava, S. Katyayan, S.K. Jain, B. Tripathi, A comparative study on dielectric and structural properties of graphene Oxide(GO)/Reduced graphene Oxide(rGO)/Multiwall carbon nanotubes (MWNT) based polyvinylidene Fluoride(PVDF) nanocomposites, AIP Conf. Procced., 2265 (2020).
[290] C. Zhang, H. Sun, Q. Zhu, Study on flexible large-area Poly(vinylidene fluoride)-based piezoelectric films prepared by extrusion-casting process for sensors and microactuators, Materials Chemistry and Physics, 275 (2022) 125221.
[291] S. Moharana, T. Yadav, P.A. Alvi, A. Pathak, R.N. Mahaling, Enhanced dielectric and electrical properties of tri-phase percolative PVDF–BiFeO3–Carbon Black (CB) composite film, J. Mater. Sci: Mater. Electron., 32 (2021) 6038-6046.
[292] N. Maity, A. Mandal, K. Roy, A.K. Nandi, Physical and dielectric properties of poly(vinylidene fluoride)/polybenzimidazole functionalized graphene nanocomposites, J. Polym. Sci. Part B: Polym. Phys., 57 (2019) 189-201.
[293] J. Liu, M. Zhang, L. Guan, C. Wang, L. Shi, Y. Jin, C. Han, J. Wang, Z. Han, Preparation of BT/GNP/PS/PVDF composites with controllable phase structure and dielectric properties, Polymer Testing, 100 (2021) 107236.
[294] F. Mokhtari, G.M. Spinks, S. Sayyar, J. Foroughi, Dynamic Mechanical and Creep Behaviour of Meltspun PVDF Nanocomposite Fibers, Nanomaterials, 11 (2021) 2153.
[295] E.A. Bakar, M.A. Mohamed, P.C. Ooi, M.F.M.R. Wee, C.F. Dee, B.Y. Majlis, Fabrication of indium-tin-oxide free, all-solution-processed flexible nanogenerator device using nanocomposite of barium titanate and graphene quantum dots in polyvinylidene fluoride polymer matrix, Organic Electronics, 61 (2018) 289-295.
[296] P. Xu, Z. Cui, H. Chen, Y. Ding, Synergistic enhanced dielectric properties of PVDF nanocomposites containing γ-oxo-pyrenebutyric acid functionalized graphene and BaTiO3 nanofillers, Composites Communications, 13 (2019) 63-69.
[297] M. Silibin, D. Karpinsky, V. Bystrov, D. Zhaludkevich, M. Bazarova, P.M. Vaghefi, P.A.A.P. Marques, B. Singh, I. Bdikin, Preparation, Stability and Local Piezoelectrical Properties of P(VDF-TrFE)/Graphene Oxide Composite Fibers, C, 5 (2019) 48.
[298] N.A. Shepelin, P.C. Sherrell, E. Goudeli, E.N. Skountzos, V.C. Lussini, G.W. Dicinoski, J.G. Shapter, A.V. Ellis, Printed recyclable and self-poled polymer piezoelectric generators through single-walled carbon nanotube templating, Energy Environ. Sci., 13 (2020) 868-883.
[299] S. Lee, Y. Lim, Generating Power Enhancement of Flexible PVDF Generator by Incorporation of CNTs and Surface Treatment of PEDOT:PSS Electrodes, Macromol. Mater. Eng., 303 (2018) 1700588.

[300] S. Sukumaran, P.K. Szewczyk, J. Knapczyk-Korczak, U. Stachewicz, Optimizing Piezoelectric Coefficient in PVDF Fibers: Key Strategies for Energy Harvesting and Smart Textiles, Adv. Electron. Mater., 9 (2023) 2300404.
[301] D. Sarkar, N. Das, M.M. Saikh, P. Biswas, S. Roy, S. Paul, N.A. Hoque, R. Basu, S. Das, High β-crystallinity comprising nitrogenous carbon dot/PVDF nanocomposite decorated self-powered and flexible piezoelectric nanogenerator for harvesting human movement mediated energy and sensing weights, Ceram. Int., 49 (2023) 5466-5478.
[302] Y. Jin, N. Chen, Y. Li, Q. Wang, The selective laser sintering of a polyamide 11/BaTiO3/graphene ternary piezoelectric nanocomposite, RSC Adv., 10 (2020) 20405-20413.
[303] K. Silakaew, P. Thongbai, Significantly improved dielectric properties of multiwall carbon nanotube-BaTiO3/PVDF polymer composites by tuning the particle size of the ceramic filler, RSC Adv., 9 (2019) 23498-23507.
[304] D. Ponnamma, A. Erturk, H. Parangusan, K. Deshmukh, M.B. Ahamed, M. Al Ali Al-Maadeed, Stretchable quaternary phasic PVDF-HFP nanocomposite films containing graphene-titania-SrTiO3 for mechanical energy harvesting, emergent mater., 1 (2018) 55-65.
[305] L.-y. Li, S.-l. Li, Y. Shao, R. Dou, B. Yin, M.-b. Yang, PVDF/PS/HDPE/MWCNTs/Fe3O4 nanocomposites: Effective and lightweight electromagnetic interference shielding material through the synergetic effect of MWCNTs and Fe3O4 nanoparticles, Curr. Appl. Phys., 18 (2018) 388-396.
[306] S. Moharana, R.N. Mahaling, Enhancement investigations on dielectric and electrical properties of niobium pentoxide (Nb2O5) reinforced poly(vinylidene fluoride) (PVDF)- graphene oxide (GO) nanocomposite films, J. Asian Ceram. Soc., 9 (2021) 1183-1193.
[307] G. Baek, S.-C. Yang, Effect of the Two-Dimensional Magnetostrictive Fillers of CoFe2O4-Intercalated Graphene Oxide Sheets in 3-2 Type Poly(vinylidene fluoride)-Based Magnetoelectric Films, Polymers, 13 (2021) 1782.
[308] H. Zhang, Y. Chen, J. Xiao, F. Song, C. Wang, H. Wang, Fabrication of enhanced dielectric PVDF nanocomposite based on the conjugated synergistic effect of ionic liquid and graphene, Materials Today: Proceedings, 16 (2019) 1512-1517.
[309] A. Karaphun, W. Tuichai, N. Chanlek, C. Sriwong, C. Ruttanapun, Dielectric and electrochemical properties of hybrid Pt nanoparticles deposited on reduced graphene oxide nanoparticles /poly (vinylidene fluoride) nanocomposites, Materials Today Communications, 27 (2021) 102232.
[310] Y. Li, B. Wang, B. Zhang, X. Ge, C. Bulin, R. Xing, Hydrophilic Fluoro-Functionalized Graphene Oxide / Polyvinylidene Fluoride Composite Films with High Dielectric Constant and Low Dielectric Loss, ChemistrySelect, 4 (2019) 570-575.
[311] S. Shetty, A.M. Shanmugharaj, S. Anandhan, Physico-chemical and piezoelectric characterization of electroactive nanofabrics based on functionalized graphene/talc nanolayers/PVDF for energy harvesting, J Polym Res, 28 (2021) 419.
[312] E. Kar, N. Bose, B. Dutta, N. Mukherjee, S. Mukherjee, MWCNT@SiO2 Heterogeneous Nanofiller-Based Polymer Composites: A Single Key to the High-Performance Piezoelectric Nanogenerator and X-band Microwave Shield, ACS Appl. Nano Mater., 1 (2018) 4005-4018.
[313] C. Xu, L. Jin, L. Zhang, C. Wang, X. Huang, X. He, Y. Xu, R. Huang, C. Zhang, W. Yang, J. Lu, Pressure-crystallized piezopolymer/ionomer/graphene quantum dot composites: A novel poling-free dynamic hybrid electret with enhanced energy harvesting properties, Composites Science and Technology, 164 (2018) 282-289.
[314] S. Bairagi, S.W. Ali, Investigating the role of carbon nanotubes (CNTs) in the piezoelectric performance of a PVDF/KNN-based electrospun nanogenerator, Soft Matter, 16 (2020) 4876-4886.
[315] Y. Li, D. Zhang, S. Wang, Y. Zhan, J. Yin, X. Tao, X. Ge, S.C. Tjong, H.-Y. Liu, Y.W. Mai, Fe3O4 decorated graphene/poly(vinylidene fluoride) nanocomposites with high dielectric constant and low dielectric loss, Composites Science and Technology, 171 (2019) 152-161.
[316] M. Sekkarapatti Ramasamy, A. Rahaman, B. Kim, Effect of phenyl-isocyanate functionalized graphene oxide on the crystalline phases, mechanical and piezoelectric properties of electrospun PVDF nanofibers, Ceramics International, 47 (2021) 11010-11021.
[317] A. Kumar, S. Jaiswal, R. Joshi, S. Yadav, A. Dubey, D. Sharma, D. Lahiri, I. Lahiri, Energy harvesting by piezoelectric polyvinylidene fluoride/zinc oxide/carbon nanotubes composite under cyclic uniaxial tensile deformation, Polym Compos., 44 (2023) 4746-4756.

[318] A.R. Chowdhury, J. Jaksik, I. Hussain, P. Tran, S. Danti, M.J. Uddin, Surface-Modified Nanostructured Piezoelectric Device as a Cost-Effective Transducer for Energy and Biomedicine, Energy Technol., 7 (2019) 1800767.
[319] R. Barstugan, M. Barstugan, I. Ozaytekin, PBO/graphene added β-PVDF piezoelectric composite nanofiber production, Compos Part B: Eng, 158 (2019) 141-148.
[320] Z. Xu, Y. Liu, L. Dong, A.B. Closson, N. Hao, M. Oglesby, G.P. Escobar, S. Fu, X. Han, C. Wen, J. Liu, M.D. Feldman, Z. Chen, J.X.J. Zhang, Tunable Buckled Beams with Mesoporous PVDF-TrFE/SWCNT Composite Film for Energy Harvesting, ACS Appl. Mater. Interfaces, 10 (2018) 33516-33522.
[321] Y. Han, C. Jiang, H. Lin, C. Luo, R. Qi, H. Peng, Piezoelectric Nanogenerators Based on Helical Carbon Materials and Polyvinyledenedifluoride–Trifluoroethylene Hybrids with Enhanced Energy-Harvesting Performance, 8 (2020) 1901249.
[322] J.P.F. Santos, B. de Melo Carvalho, R.E. Suman Bretas, Remarkable change in the broadband electrical behavior of poly(vinylidene fluoride)–multiwalled carbon nanotube nanocomposites with the use of different processing routes, 136 (2019) 47409.
[323] S. Badatya, A. Kumar, A.K. Srivastava, M.K. Gupta, Flexible Interconnected Cu-Ni Nanoalloys Decorated Carbon Nanotube-Poly(vinylidene fluoride) Piezoelectric Nanogenerator, 7 (2022) 2101281.
[324] M. Shoorangiz, Z. Sherafat, E. Bagherzadeh, CNT loaded PVDF-KNN nanocomposite films with enhanced piezoelectric properties, Ceram.Int., 48 (2022) 15180-15188.
[325] C. Zhao, Y. Hong, X. Chu, Y. Dong, Z. Hu, X. Sun, S. Yan, Enhanced ferroelectric properties of P(VDF-TrFE) thin film on single-layer graphene simply adjusted by crystallization condition, Materials Today Energy, 20 (2021) 100678.
[326] J. Tu, H. Li, Z. Cai, J. Zhang, X. Hu, J. Huang, C. Xiong, M. Jiang, L. Huang, Phase change-induced tunable dielectric permittivity of poly(vinylidene fluoride)/polyethylene glycol/graphene oxide composites, Compos Part B: Eng, 173 (2019) 106920.
[327] E.O. Taha, H.A. Alyousef, A.M. Dorgham, O.M. Hemeda, H.M.H. Zakaly, P. Noga, M.M. Abdelhamied, M.M. Atta, Electron beam irradiation and carbon nanotubes influence on PVDF-PZT composites for energy harvesting and storage applications: Changes in dynamic-mechanical and dielectric properties, Inorg. Chem. Commun., 151 (2023) 110624.
[328] Z. Wang, Y. Zhao, J. Ji, T. Yao, H. Zhang, A tactile skin based on the piezoelectric effect of PVDF and room temperature vulcanised silicone rubber, Materials Technology, 37 (2022) 2123-2131.
[329] Z. Sun, Z. Yan, R. Yin, X. Huang, K. Yue, A. Li, L. Qian, Homogeneous poly(vinylidenefluoride)-graphene films with secondary percolation behaviors towards negative permittivity properties, Composites Communications, 17 (2020) 18-21.
[330] K. Sun, W. Duan, Y. Lei, Z. Wang, J. Tian, P. Yang, Q. He, M. Chen, H. Wu, Z. Zhang, R. Fan, Flexible multi-walled carbon nanotubes/polyvinylidene fluoride membranous composites with weakly negative permittivity and low frequency dispersion, 156 (2022) 106854.
[331] K. Krishnamoorthy, V.K. Mariappan, P. Pazhamalai, S. Sahoo, S.-J. Kim, Mechanical energy harvesting properties of free-standing carbyne enriched carbon film derived from dehydrohalogenation of polyvinylidene fluoride, Nano Energy, 59 (2019) 453-463.
[332] F.F. Hatta, M.A.S. Mohammad Haniff, M.A. Mohamed, A review on applications of graphene in triboelectric nanogenerators, Intl J of Energy Research, 46 (2022) 544-576.
[333] Z. Yu, Y. Zhang, Y. Wang, J. Zheng, Y. Fu, D. Chen, G. Wang, J. Cui, S. Yu, L. Zheng, H. Zhou, D. Li, Integrated piezo-tribo hybrid acoustic-driven nanogenerator based on porous MWCNTs/PVDF-TrFE aerogel bulk with embedded PDMS tympanum structure for broadband sound energy harvesting, Nano Energy, 97 (2022) 107205.
[334] E.C. Statharas, K. Yao, M. Rahimabady, A.M. Mohamed, F.E.H. Tay, Polyurethane/poly(vinylidene fluoride)/MWCNT composite foam for broadband airborne sound absorption, J. Appl. Polym. Sci., 136 (2019) 47868.
[335] K. Kumar Reddy Bannuru, A. Raj Plamootil Mathai, P. Valdivia y Alvarado, H. Yee Low, Freestanding 3D piezoelectric PVDF sensors via electroprinting, Mater. Today: Proceed., 70 (2022) 447-453.
[336] J. Cong, J. Jing, C. Chen, Z. Dai, Development of a PVDF Sensor Array for Measurement of the Dynamic Pressure Field of the Blade Tip in an Axial Flow Compressor, Sensors, 19 (2019) 1404.

[337] C. Xiu, J. Hou, Y. Zang, G. Xu, C. Liu, Synchronous Control of Hysteretic Creep Chaotic Neural Network, IEEE Access, 4 (2016) 8617-8624.
[338] J.X. Lin, H.W. Hu, J. Luo, L. Miao, Z.H. Yang, M. Chen, M. Zhang, J.Z. Ou, Micro/nanoarrays and their applications in flexible sensors: A review, Mater. Today Nano, 19 (2022) 100224.
[339] Y.J. Dias, T.C. Gimenes, S.A.P.V. Torres, J.A. Malmonge, A.J. Gualdi, F.R. de Paula, PVDF/Ni fibers synthesis by solution blow spinning technique, J. Mater. Sci: Mater. Electron., 29 (2018) 514-518.
[340] C. Su, X. Huang, L. Zhang, Y. Zhang, Z. Yu, C. Chen, Y. Ye, S. Guo, Robust superhydrophobic wearable piezoelectric nanogenerators for self-powered body motion sensors, Nano Energy, 107 (2023) 108095.
[341] Y. Luo, L. Zhao, G. Luo, M. Li, X. Han, Y. Xia, Z. Li, Q. Lin, P. Yang, L. Dai, G. Niu, X. Wang, J. Wang, D. Lu, Z. Jiang, All electrospun fabrics based piezoelectric tactile sensor, Nanotechnol., 33 (2022) 415502.
[342] H. Le Xuan, E. Haentzsche, A. Nocke, N.H.A. Tran, I. Kruppke, C. Cherif, Development of fiber-based piezoelectric sensors for the load monitoring of dynamically stressed fiber-reinforced composites, Smart Mater. Sctructur., 32 (2023) 045013.
[343] D. Hernández-Rivera, G. Rodríguez-Roldán, R. Mora-Martínez, E. Suaste-Gómez, A Capacitive Humidity Sensor Based on an Electrospun PVDF/Graphene Membrane, Sensors, 17 (2017) 1009.
[344] C.-A. Ku, C.-K. Chung, Advances in Humidity Nanosensors and Their Application: Review, Sensors, 23 (2023) 2328.
[345] Q. Wang, Z. Wu, J. Li, J. Wei, J. Guo, M. Yin, Spontaneous and Continuous Actuators Driven by Fluctuations in Ambient Humidity for Energy-Harvesting Applications, ACS Appl. Mater. Interfaces, 14 (2022) 38972-38980.
[346] J.P. Lee, J.W. Lee, J.M. Baik, The Progress of PVDF as a Functional Material for Triboelectric Nanogenerators and Self-Powered Sensors, Micromachines, 9 (2018) 532.
[347] D.-L. Vu, K.-K. Ahn, Triboelectric Enhancement of Polyvinylidene Fluoride Membrane Using Magnetic Nanoparticle for Water-Based Energy Harvesting, Polymers, 14 (2022) 1547.
[348] K. Roy, S.K. Ghosh, A. Sultana, S. Garain, M. Xie, C.R. Bowen, K. Henkel, D. Schmeiβer, D. Mandal, A Self-Powered Wearable Pressure Sensor and Pyroelectric Breathing Sensor Based on GO Interfaced PVDF Nanofibers, ACS Appl. Nano Mater., 2 (2019) 2013-2025.
[349] Y. Li, J. Sun, P. Li, X. Li, J. Tan, H. Zhang, T. Li, J. Liang, Y. Zhou, Z. Hai, J. Zhang, High-performance piezoelectric nanogenerators based on hierarchical ZnO@CF/PVDF composite film for self-powered meteorological sensor, J. Mater. Chem. A, 11 (2023) 13708-13719.
[350] J. Tartière, M. Arrigoni, A. Nême, H. Groeneveld, S. Van Der Veen, PVDF Based Pressure Sensor for the Characterisation of the Mechanical Loading during High Explosive Hydro Forming of Metal Plates, Sensors, 21 (2021) 4429.
[351] J.S. Lee, K.-Y. Shin, O.J. Cheong, J.H. Kim, J. Jang, Highly Sensitive and Multifunctional Tactile Sensor Using Free-standing ZnO/PVDF Thin Film with Graphene Electrodes for Pressure and Temperature Monitoring, Sci. Rep., 5 (2015) 7887.
[352] O.Y. Kweon, S.J. Lee, J.H. Oh, Wearable high-performance pressure sensors based on three-dimensional electrospun conductive nanofibers, NPG Asia Mater., 10 (2018) 540-551.
[353] R.K. Singh, S.W. Lye, J. Miao, PVDF Nanofiber Sensor for Vibration Measurement in a String, Sensors, 19 (2019) 3739.
[354] B. Marchiori, S. Regal, Y. Arango, R. Delattre, S. Blayac, M. Ramuz, PVDF-TrFE-Based Stretchable Contact and Non-Contact Temperature Sensor for E-Skin Application, Sensors, 20 (2020) 623.
[355] K. Maity, S. Garain, K. Henkel, D. Schmeißer, D. Mandal, Self-Powered Human-Health Monitoring through Aligned PVDF Nanofibers Interfaced Skin-Interactive Piezoelectric Sensor, ACS Appl. Polym. Mater., 2 (2020) 862-878.
[356] B. Mahanty, K. Maity, S. Sarkar, D. Mandal, Human Skin Interactive Self-powered Piezoelectric e-skin Based on PVDF/MWCNT Electrospun Nanofibers for Non-invasive Health Care Monitoring, 21 (2020) 1964-1968.
[357] Y. Mao, W. Yue, T. Zhao, M. Shen, B. Liu, S. Chen, A Self-Powered Biosensor for Monitoring Maximal Lactate Steady State in Sport Training, Biosensors, 10 (2020) 75.

[358] G. Chen, G. Chen, L. Pan, D. Chen, Electrospun flexible PVDF/GO piezoelectric pressure sensor for human joint monitoring, Diam. Relat. Mater., 129 (2022) 109358.
[359] N. Shehata, A.H. Hassanin, E. Elnabawy, R. Nair, S.A. Bhat, I. Kandas, Acoustic Energy Harvesting and Sensing via Electrospun PVDF Nanofiber Membrane, Sensors, 20 (2020) 3111.
[360] M.M. Alam, A. Sultana, D. Mandal, Biomechanical and Acoustic Energy Harvesting from TiO2 Nanoparticle Modulated PVDF Nanofiber Made High Performance Nanogenerator, ACS Appl. Energy Mater., 1 (2018) 3103-3112.
[361] M. Shehzad, Y. Wang, PVDF based piezoelectric condenser loudspeaker and microphone, 346 (2022) 113861.
[362] J. Gutiérrez, A. Lasheras, P. Martins, N. Pereira, J.M. Barandiarán, S. Lanceros-Mendez, Metallic Glass/PVDF Magnetoelectric Laminates for Resonant Sensors and Actuators: A Review, Sensors, 17 (2017) 1251.
[363] C. Merlini, R.d.S. Almeida, M.A. D'Ávila, W.H. Schreiner, G.M.d.O. Barra, Development of a novel pressure sensing material based on polypyrrole-coated electrospun poly(vinylidene fluoride) fibers, Mat. Sci. Eng. B, 179 (2014) 52-59.
[364] Y. Miao, P. Li, S. Cheng, Q. Zhou, M. Cao, J. Yi, H. Zhang, Preparation of multi-axial compressible 3D PVDF nanofibre/graphene wearable composites sensor sponge and application of integrated sensor, Sens. Actuator A Phys., 342 (2022) 113648.
[365] C. Mangone, W. Kaewsakul, M.K. Gunnewiek, L.A.E.M. Reuvekamp, J.W.M. Noordermeer, A. Blume, Design and performance of flexible polymeric piezoelectric energy harvesters for battery-less tyre sensors, Smart Mater. Structur., 31 (2022) 095034.
[366] S. Sharafkhani, M. Kokabi, Enhanced sensing performance of polyvinylidene fluoride nanofibers containing preferred oriented carbon nanotubes, Adv. Compos. Hybrid Mater., 5 (2022) 3081-3093.
[367] S. Sharafkhani, M. Kokabi, Coaxially oriented PVDF/MWCNT nanofibers as a high-performance piezoelectric actuator, Polym Compos., 44 (2023) 8780-8791.
[368] S. Choi, J. Lim, H. Park, H.S. Kim, A Flexible Piezoelectric Device for Frequency Sensing from PVDF/SWCNT Composite Fibers, Polymers, 14 (2022) 4773.
[369] Y. Khazani, E. Rafiee, A. Samadi, Piezoelectric fibers-based PVDF-ZnS-carbon nano onions as a flexible nanogenerator for energy harvesting and self-powered pressure sensing, Colloids Surf. A: Physicochem. Eng. Asp., 675 (2023) 132004.
[370] L. Chen, M. Song, J. Guan, Y. Shu, D. Jin, G. Fan, Q. Xu, X.-Y. Hu, A highly-specific photoelectrochemical platform based on carbon nanodots and polymers functionalized organic-inorganic perovskite for cholesterol sensing, Talanta, 225 (2021) 122050.
[371] H. Kim, B.R. Wilburn, E. Castro, C.A. Garcia Rosales, L.A. Chavez, T.-L.B. Tseng, Y. Lin, Multifunctional SENSING using 3D printed CNTs/BaTiO3/PVDF nanocomposites, J. Compos. Mater., 53 (2019) 1319-1328.
[372] P. Zhang, S. Lei, W. Fu, J. Niu, G. Liu, J. Qian, J. Sun, The effects of agglomerate on the piezoresistivity of conductive carbon nanotube/polyvinylidene fluoride composites, 281 (2018) 176-184.
[373] R. Salehiyan, M. Nofar, D. Makwakwa, S.S. Ray, Shear-Induced Carbon Nanotube Migration and Morphological Development in Polylactide/Poly(vinylidene fluoride) Blend Nanocomposites and Their Impact on Dielectric Constants and Rheological Properties, J Phys Chem C, 124 (2020) 9536-9547.
[374] X. Tang, P. Pötschke, J. Pionteck, Y. Li, P. Formanek, B. Voit, Tuning the Piezoresistive Behavior of Poly(Vinylidene Fluoride)/Carbon Nanotube Composites Using Poly(Methyl Methacrylate), ACS Appl. Mater. Interfaces, 12 (2020) 43125-43137.
[375] Y.R. Lee, J. Park, Y. Jeong, J.S. Park, Improved Mechanical and Electrical Properties of Carbon Nanotube Yarns by Wet Impregnation and Multi-ply Twisting, Fibers Polym., 19 (2018) 2478-2482.
[376] C. Zhao, W. Yuan, Y. Zhao, N. Hu, B. Gu, H. Liu, Alamusi, Unified equivalent circuit model for carbon nanotube-based nanocomposites, Nanotechnology, 29 (2018) 305503.
[377] C. Zhao, W. Yuan, H. Liu, B. Gu, N. Hu, Alamusi, Y. Ning, F. Jia, Equivalent circuit model for the strain sensing characteristics of multi-walled carbon nanotube/polyvinylidene fluoride films in alternating current circuit, Carbon, 129 (2018) 585-591.
[378] J.R. Dios, C. Garcia-Astrain, S. Gonçalves, P. Costa, S. Lanceros-Méndez, Piezoresistive performance of polymer-based materials as a function of the matrix and nanofiller content to walking detection application, Compos. Sci. Technol., 181 (2019) 107678.

[379] S. Aziz, S.-H. Chang, Smart-fabric sensor composed of single-walled carbon nanotubes containing binary polymer composites for health monitoring, Compos. Sci. Technol, 163 (2018) 1-9.
[380] M. Chaturvedi, V. Panwar, B. Prasad, Piezoresistive sensitivity tuning using polyelectrolyte as interface linker in carbon based polymer composites, 312 (2020) 112151.
[381] M.A. Rahman, F. Rubaiya, N. Islam, K. Lozano, A. Ashraf, Graphene-Coated PVDF/PAni Fiber Mats and Their Applications in Sensing and Nanogeneration, ACS Appl. Mater. Interfaces, 14 (2022) 38162-38171.
[382] C.-H. Huang, J.-X. Huang, Y.-H. Chiao, C.-M. Chang, W.-S. Hung, S.J. Lue, C.-F. Wang, C.-C. Hu, K.-R. Lee, H.-H. Pan, J.-Y. Lai, Tailoring of a Piezo-Photo-Thermal Solar Evaporator for Simultaneous Steam and Power Generation, Adv. Funct. Mater., 31 (2021) 2010422.
[383] S. Subburaj, B. Patel, C.-H. Yeh, T.-H. Huang, C.-Y. Chang, W.-S. Hung, P.T. Lin, Design and fabrication of curved sensor based on polyvinylidene fluoride/graphene composite film with a self-assembling mechanism for monitoring of human body parts movement, Sens. Actuator A Phys., 356 (2023) 114360.
[384] S. Liu, X. Liu, Z. Zhu, S. Wang, Y. Gu, F. Shan, Y. Zou, Improved flexible ZnO/CsPbBr3/Graphene UV photodetectors with interface optimization by solution process, Materials Research Bulletin, 130 (2020) 110956.
[385] K. Miao, H. Su, M. Li, H. Yao, Y. Zhang, H. Wang, Y. Zhang, L. Yang, H. Zheng, The interaction mechanism of photogenerated carriers and piezoelectric charges of a photoactive piezoelectric nanogenerator, Appl. Phys. Lett., 121 (2022).
[386] E.S. Kadir, R.N. Gayen, M.P. Chowdhury, Enhanced photodetection properties of GO incorporated flexible PVDF membranes under solar spectrum, J. Pol. Res., 29 (2022) 529.
[387] B. Mondal, H.K. Mishra, D. Sengupta, A. Kumar, A. Babu, D. Saini, V. Gupta, D. Mandal, Lead-Free Perovskite Cs3Bi2I9-Derived Electroactive PVDF Composite-Based Piezoelectric Nanogenerators for Physiological Signal Monitoring and Piezo-Phototronic-Aided Strain Modulated Photodetectors, Langmuir, 38 (2022) 12157-12172.
[388] S. Veeralingam, S. Badhulika, Low-density, stretchable, adhesive PVDF-polypyrrole reinforced gelatin based organohydrogel for UV photodetection, tactile and strain sensing applications, 150 (2022) 111779.
[389] B.-D. On, G.-J. Choi, S.-S. Lee, I.-K. Park, Rollable Ultraviolet Photodetector Based on ZnAl-Layered Double Hydroxide/Polyvinylidene Fluoride Membrane, Adv. Mater. Interfaces, 9 (2022) 2201052.
[390] T. Jiang, J. Wang, L. Xie, C. Bai, M. Wang, Y. Wu, F. Zhang, Y. Zhao, B. Chen, Y. Wang, In Situ Fabrication of Lead-Free Cs3Cu2I5 Nanostructures Embedded in Poly(Vinylidene Fluoride) Electrospun Fibers for Polarized Emission, ACS Appl. Nano Mater., 5 (2022) 508-516.
[391] K.A. Ahmad, M.F. Abdullah, N. Abdullah, Design and Characterization of an Interdigitated Electrode PVDF based Energy Harvesting Device, in: 2019 9th IEEE International Conference on Control System, Computing and Engineering (ICCSCE), 2019, pp. 172-177.
[392] Western Blotting Humana New York, NY, 2015.
[393] C.J.G. Nielsen, D. Tian, K. Wang, A. Preumont, Adaptive Deployable Thin Spherical Shell Reflectors, Actuators, 11 (2022) 198.
[394] D. Khatua, M. Padhy, R.K. Singh, R.N.P. Choudhary, P.G.R. Achary, Investigation of electrical and thermal properties of poly (vinylidene fluoride)/strontium hexaferrite polymer composites, Mater. Sci: Mater Electron., 31 (2020) 22687-22698.
[395] Y. Lu, M. Amabili, J. Wang, F. Yang, H. Yue, Y. Xu, H. Tzou, Active vibration control of a polyvinylidene fluoride laminated membrane plate mirror, J. Vib. Control, 25 (2019) 2611-2626.
[396] H. Peng, Z. Cheng, L. Zeng, X. Ji, Photoacoustic microscopy based on transparent piezoelectric ultrasound transducers, J. Innov. Opt. Health Sci., 0 2330001.
[397] T. Zhao, M. Zhang, S. Ourselin, W. Xia, Wavefront Shaping-Assisted Forward-Viewing Photoacoustic Endomicroscopy Based on a Transparent Ultrasound Sensor, App. Sci., 12 (2022) 12619.
[398] S. Li, Y. Zhai, X. Pan, Y. Jin, M. Wang, High sensitive luminescence thermometer using Ruphen-based temperature sensitive paint, Infrared Phys. Technol., 123 (2022) 104151.
[399] R.-T. Jung, N.M.T. Naing, Impulsive forces of two spark-generated cavity bubbles with phase differences, Ultrason Sonochem., 86 (2022) 106042.

[400] B. Zeng, J. Yang, Z. Hang, Y. Fan, C. Feng, J. Yang, High pyroelectric performances of graphene nanoplatelet reinforced polyvinylidene fluoride composite film, Polym Compos., 44 (2023) 5148-5158.
[401] A.A. Shakaty, J.K. Hmood, B.R. Mahdi, R.I. Mahdi, A.A. Al-Azzawi, Q-switched erbium-doped fiber laser based on nanodiamond saturable absorber, Opt. Laser Technol., 146 (2022) 107569.
[402] N. Ali, A.A. Babar, Y. Zhang, N. Iqbal, X. Wang, J. Yu, B. Ding, Porous, flexible, and core-shell structured carbon nanofibers hybridized by tin oxide nanoparticles for efficient carbon dioxide capture, J. Colloid Interf. Sci., 560 (2020) 379-387.
[403] H.-Y. Park, C.H. Lee, D.-W. Cho, C.-H. Lee, J.-H. Park, Synthesis of porous carbon derived from poly(vinylidenefluoride) and its adsorption characteristics for CO2 and CH4, Microporous Mesoporous Mater., 299 (2020) 110121.
[404] S. Singh, N. Shauloff, C.P. Sharma, R. Shimoni, C.J. Arnusch, R. Jelinek, Carbon dot-polymer nanoporous membrane for recyclable sunlight-sterilized facemasks, J. Colloid Interf. Sci., 592 (2021) 342-348.
[405] C.N. Kumar, M.N. Prabhakar, S. Jung-il, PVDF green nanofibers as potential carriers for improving self-healing and mechanical properties of carbon fiber/epoxy prepregs, Nanotechnol. Rev., 11 (2022) 1890-1900.
[406] D. Hernández-Rivera, S.D. Torres-Landa, M. Rangel-Ayala, V. Agarwal, Fluorescent films based on PVDF doped with carbon dots for evaluation of UVA protection of sunscreens and fabrication of cool white LEDs, RSC Adv., 11 (2021) 32604-32614.
[407] L. Dong, Z. Xiong, X. Liu, D. Sheng, Y. Zhou, Y. Yang, Synthesis of carbon quantum dots to fabricate ultraviolet-shielding poly(vinylidene fluoride) films, J. Appl. Polym. Sci., 136 (2019) 47555.
[408] M.C. Bertolini, S. Dul, G.M.O. Barra, A. Pegoretti, Poly(vinylidene fluoride)/thermoplastic polyurethane flexible and 3D printable conductive composites, J. Appl. Polym. Sci., 138 (2021) 50305.
[409] G. Jiang, Z. Liu, J. Hu, Superhydrophobic and Photothermal PVDF/CNTs Durable Composite Coatings for Passive Anti-Icing/Active De-Icing, Adv. Mater. Interfaces, 9 (2022) 2101704.
[410] E. Korczeniewski, P. Bryk, S. Koter, P. Kowalczyk, M. Zięba, M. Łępicka, K.J. Kurzydłowski, K.H. Markiewicz, A.Z. Wilczewska, W. Kujawski, S. Boncel, S. Al-Gharabli, M. Świdziński, D.J. Smoliński, K. Kaneko, J. Kujawa, A.P. Terzyk, Are nanohedgehogs thirsty? Toward new superhydrophobic and anti-icing carbon nanohorn-polymer hybrid surfaces, Chem. Eng. J., 446 (2022) 137126.
[411] S. Zeng, Q. Su, L.-Z. Zhang, Molecular-level evaluation and manipulation of thermal conductivity, moisture diffusivity and hydrophobicity of a GO-PVP/PVDF composite membrane, International Journal of Heat and Mass Transfer, 152 (2020) 119508.
[412] C. Muzzi, A. Gotzias, E. Fontananova, E. Tocci, Stability of Graphene Oxide Composite Membranes in an Aqueous Environment from a Molecular Point of View, Applied Sciences, 12 (2022) 3460.
[413] C. Li, Q. Han, Semi-analytical wave characteristics analysis of graphene-reinforced piezoelectric polymer nanocomposite cylindrical shells, International Journal of Mechanical Sciences, 186 (2020) 105890.
[414] J.A. Krishnaswamy, F.C. Buroni, E. García-Macías, R. Melnik, L. Rodriguez-Tembleque, A. Saez, Design of nano-modified PVDF matrices for lead-free piezocomposites: Graphene vs carbon nanotube nano-additions, Mechanics of Materials, 142 (2020) 103275.
[415] X. Xia, G.J. Weng, D. Hou, W. Wen, Tailoring the frequency-dependent electrical conductivity and dielectric permittivity of CNT-polymer nanocomposites with nanosized particles, Int. J. Eng. Sci., 142 (2019) 1-19.
[416] H. Cho, A. Shakil, A.A. Polycarpou, S. Kim, Enabling Selectively Tunable Mechanical Properties of Graphene Oxide/Silk Fibroin/Cellulose Nanocrystal Bionanofilms, ACS Nano, 15 (2021) 19546-19558.
[417] X. Chen, K. Fan, Y. Liu, Y. Li, X. Liu, W. Feng, X. Wang, Recent Advances in Fluorinated Graphene from Synthesis to Applications: Critical Review on Functional Chemistry and Structure Engineering, Adv. Mat., 34 (2022) 2101665.
[418] X. Li, Y. Wang, Y. Zhao, J. Zhang, L. Qu, Graphene Materials for Miniaturized Energy Harvest and Storage Devices, Small Struct., 3 (2022) 2100124.

[419] Y. Feng, Q. Deng, C. Peng, Q. Wu, High dielectric and breakdown properties achieved in ternary BaTiO3/MXene/PVDF nanocomposites with low-concentration fillers from enhanced interface polarization, Ceram. Int., 45 (2019) 7923-7930.
[420] S. Wang, H.-Q. Shao, Y. Liu, C.-Y. Tang, X. Zhao, K. Ke, R.-Y. Bao, M.-B. Yang, W. Yang, Boosting piezoelectric response of PVDF-TrFE via MXene for self-powered linear pressure sensor, Compos. Sci. Technol., 202 (2021) 108600.
[421] Y. Gogotsi, B. Anasori, The Rise of MXenes, ACS Nano, 13 (2019) 8491-8494.
[422] S. Bhunia, S. Chandel, S.K. Karan, S. Dey, A. Tiwari, S. Das, N. Kumar, R. Chowdhury, S. Mondal, I. Ghosh, A. Mondal, B.B. Khatua, N. Ghosh, C.M. Reddy, Autonomous self-repair in piezoelectric molecular crystals, Science, 373 (2021) 321-327.
[423] P. Zhou, Q. Zhu, X. Sun, L. Liu, Z. Cai, J. Xu, Recent advances in MXene-based membrane for solar-driven interfacial evaporation desalination, Chem. Eng. J., 464 (2023) 142508.
[424] S. Kumar Singh, A. Kumar Tiwari, H.K. Paliwal, A holistic review of MXenes for solar device applications: Synthesis, characterization, properties and stability, FlatChem, 39 (2023) 100493.
[425] C. Naga Kumar, M.N. Prabhakar, J.-i. Song, Synthesis of vinyl ester resin-carrying PVDF green nanofibers for self-healing applications, Sci. Rep., 11 (2021) 908.
[426] N.K. C, P. M. N, M.T. Hassim, J.-i. Song, Development of Self-Healing Carbon/Epoxy Composites with Optimized PAN/PVDF Core–Shell Nanofibers as Healing Carriers, ACS Omega, 7 (2022) 42396-42407.
[427] S. Wang, M.W. Urban, Self-healing polymers, Nat. Rev. Mater., 5 (2020) 562-583.
[428] A. Nag, R.B.V.B. Simorangkir, S. Sapra, J.L. Buckley, B. O'Flynn, Z. Liu, S.C. Mukhopadhyay, Reduced Graphene Oxide for the Development of Wearable Mechanical Energy-Harvesters: A Review, IEEE Sensors J., 21 (2021) 26415-26425.
[429] H. Kim, T. Fernando, M. Li, Y. Lin, T.-L.B. Tseng, Fabrication and characterization of 3D printed BaTiO3/PVDF nanocomposites, J. Compos. Mat., 52 (2018) 197-206.
[430] X. Liu, Y. Shang, J. Zhang, C. Zhang, Ionic Liquid-Assisted 3D Printing of Self-Polarized β-PVDF for Flexible Piezoelectric Energy Harvesting, ACS Appl. Mater. Interfaces, 13 (2021) 14334-14341.
[431] Y. Guo, C. Liu, H. Liu, W. Wang, H. Li, C. Zhang, Influences of gamma-ray irradiation on PVDF membrane behavior: An experimental study based on simulation and numerical analysis, Polym. Degrad. Stab., 193 (2021) 109722.
[432] W. Sun, G. Ji, J. Chen, D. Sui, J. Zhou, J. Huber, Enhancing the acoustic-to-electrical conversion efficiency of nanofibrous membrane-based triboelectric nanogenerators by nanocomposite composition, Nano Energy, 108 (2023) 108248.
[433] X. Yang, G. Liu, Q. Guo, H. Wen, R. Huang, X. Meng, J. Duan, Q. Tang, Triboelectric sensor array for internet of things based smart traffic monitoring and management system, Nano Energy, 92 (2022) 106757.
[434] A. Aboelmakarem Farag, The Story of NEOM City: Opportunities and Challenges, in: New Cities and Community Extensions in Egypt and the Middle East Springer, Cham, 2018.
[435] B.H. AlDoaies, H. Almagwashi, Exploitation of the Promising Technology: Using BlockChain to Enhance the Security of IoT, in: 2018 21st Saudi Computer Society National Computer Conference (NCC), 2018, pp. 1-6.
[436] F. Alharbi, Integrating internet of things in electrical engineering education, Int. J. Electr. Eng. Educ., 0 0020720920903422.
[437] X. Zhao, H. Askari, J. Chen, Nanogenerators for smart cities in the era of 5G and Internet of Things, Joule, 5 (2021) 1391-1431.
[438] B. Meena, M. Kumar, S. Gupta, L. Sinha, P. Subramanyam, C. Subrahmanyam, Rational design of TiO2/BiSbS3 heterojunction for efficient solar water splitting, Sustain. Energy Technol. Assess., 49 (2022) 101775.
[439] M. Verma, L. Sinha, P.M. Shirage, Electrodeposited nanostructured flakes of cobalt, manganese and nickel-based sulfide (CoMnNiS) for electrocatalytic alkaline oxygen evolution reaction (OER), J. Mater. Sci: Mater. Electron., 32 (2021) 12292-12307.
[440] Y. Wang, Y. Gui, S. He, J. Yang, MOF-derived NiO/N-MWCNTs@PVDF film and MXene/Carbon/Ecoflex electrode for enhanced hybrid nanogenerator towards environmental monitoring, Compos. Part A Appl. Sci., 173 (2023) 107692.